\documentclass[11pt]{article}
\usepackage[T1]{fontenc}
\usepackage[utf8]{inputenc}
\usepackage{lmodern}
\usepackage[margin=1in]{geometry}
\usepackage{amsmath,amssymb,amsthm,mathtools}
\usepackage{microtype}
\usepackage{booktabs}
\usepackage{array}
\usepackage{enumitem}
\usepackage{xurl}
\usepackage{graphicx}
\usepackage[font=small,labelfont=bf]{caption}
\usepackage{tikz}
\usetikzlibrary{arrows.meta,positioning,calc,fit,backgrounds}
\usepackage{pgfplots}
\pgfplotsset{compat=1.18}
\usepgfplotslibrary{groupplots}
\usepackage[colorlinks=true,linkcolor=blue,citecolor=blue,urlcolor=blue]{hyperref}
\usepackage{aliascnt}
\usepackage[capitalise,nameinlink,noabbrev]{cleveref}
\numberwithin{equation}{section}
\newtheorem{theorem}{Theorem}[section]
\newaliascnt{lemma}{theorem}
\newtheorem{lemma}[lemma]{Lemma}
\aliascntresetthe{lemma}
\crefname{lemma}{Lemma}{Lemmas}
\Crefname{lemma}{Lemma}{Lemmas}
\newaliascnt{proposition}{theorem}
\newtheorem{proposition}[proposition]{Proposition}
\aliascntresetthe{proposition}
\crefname{proposition}{Proposition}{Propositions}
\Crefname{proposition}{Proposition}{Propositions}
\newaliascnt{corollary}{theorem}
\newtheorem{corollary}[corollary]{Corollary}
\aliascntresetthe{corollary}
\crefname{corollary}{Corollary}{Corollaries}
\Crefname{corollary}{Corollary}{Corollaries}
\theoremstyle{definition}
\newaliascnt{definition}{theorem}
\newtheorem{definition}[definition]{Definition}
\aliascntresetthe{definition}
\crefname{definition}{Definition}{Definitions}
\Crefname{definition}{Definition}{Definitions}
\newaliascnt{example}{theorem}
\newtheorem{example}[example]{Example}
\aliascntresetthe{example}
\crefname{example}{Example}{Examples}
\Crefname{example}{Example}{Examples}
\theoremstyle{remark}
\newaliascnt{remark}{theorem}
\newtheorem{remark}[remark]{Remark}
\aliascntresetthe{remark}
\crefname{remark}{Remark}{Remarks}
\Crefname{remark}{Remark}{Remarks}
\crefname{figure}{Figure}{Figures}
\Crefname{figure}{Figure}{Figures}
\crefname{table}{Table}{Tables}
\Crefname{table}{Table}{Tables}

\DeclareMathOperator{\Har}{Har}
\DeclareMathOperator{\Tr}{Tr}
\DeclareMathOperator{\rank}{rank}
\DeclareMathOperator{\supp}{supp}

\DeclareMathOperator{\sgn}{sgn}
\newcommand{\id}{I}
\newcommand{\ket}[1]{\lvert #1\rangle}
\newcommand{\bra}[1]{\langle #1\rvert}
\newcommand{\proj}[1]{\ket{#1}\!\bra{#1}}

\newcommand{\ov}{m}
\newcommand{\Dtest}[1]{D_{\mathrm H}^{#1}}
\newcommand{\Dhat}[1]{\widehat D_{\mathrm H}^{#1}}
\newcommand{\Dlo}[1]{\underline D^{#1}}
\newcommand{\Dup}[1]{\overline D^{#1}}
\newcommand{\rhot}{\rho_{1}\otimes\cdots\otimes\rho_{n}}
\newcommand{\sigt}{\sigma_{1}\otimes\cdots\otimes\sigma_{n}}
\newcommand{\ee}{\varepsilon}
\newcommand{\bb}{\beta}
\setlist{itemsep=2pt,topsep=4pt}

\newcommand{\ran}{\operatorname{range}}
\newcommand{\cH}{\mathcal{H}}
\newcommand{\cS}{\mathcal{S}}
\newcommand{\eps}{\varepsilon}
\newcommand{\cN}{\mathcal{N}}
\newcommand{\cL}{\mathcal{L}}
\newcommand{\cD}{\mathcal{D}}
\newcommand{\Dmax}{D_{\max}}
\newcommand{\NS}{\mathrm{NS}}
\definecolor{vizblue}{HTML}{2A78D6}
\definecolor{vizorange}{HTML}{EB6834}
\definecolor{vizaqua}{HTML}{1BAF7A}
\definecolor{vizgray}{HTML}{52514E}
\definecolor{vizlight}{HTML}{D9D8D2}

\pgfplotsset{
  vizaxis/.style={
    grid=major, grid style={vizlight, very thin},
    axis line style={vizgray}, tick style={vizgray},
    tick label style={font=\footnotesize}, label style={font=\small},
    title style={font=\small},
    legend style={font=\footnotesize, draw=vizlight, fill=white},
    legend cell align=left,
    every axis plot/.append style={line width=1.1pt},
  },
}
\tikzset{
  rmbox/.style={draw=vizgray, rounded corners=3pt, align=center, font=\footnotesize, inner sep=3pt, fill=white, text width=#1},
  rmbox/.default=2.9cm,
  rmarrow/.style={-{Stealth[length=2mm]}, vizgray, line width=0.6pt},
}
\title{Super-Multiplicativity of the Noncommutative Minimum and Applications to Quantum Privacy and Hypothesis Testing
}

\author{
  Theshani Nuradha\textsuperscript{1,2}
  \thanks{Email: \texttt{theshani.gallage@gmail.com}.}, \ 
  Sujeet Bhalerao\textsuperscript{1},
  Christoph Hirche\textsuperscript{3}, 
  Mark M.~Wilde\textsuperscript{4},
  Mingze Xu\textsuperscript{5}\\[0.75em]
  {\small \textsuperscript{1}Department of Mathematics,}
  {\small University of Illinois Urbana-Champaign, Urbana, IL 61801, USA}\\[0.25em]
  {\small \textsuperscript{2}Illinois Quantum Information Science and Technology (IQUIST) Center,}\\
  {\small University of Illinois Urbana-Champaign, Urbana, IL 61801, USA}\\[0.25em]
  {\small \textsuperscript{3}Institute for Information Processing (tnt/L3S),}
  {\small Leibniz Universit\"at Hannover, Germany}\\[0.25em]
  {\small \textsuperscript{4}School of Electrical and Computer Engineering,}
  {\small Cornell University, Ithaca, NY 14850, USA}\\[0.25em]
  {\small \textsuperscript{5}Department of Electrical and Computer Engineering,}\\
  {\small University of Illinois Urbana-Champaign, Urbana, IL 61801, USA}
}
\date{}

\begin{document}

\maketitle
\begin{abstract}
The hockey-stick divergence is a fundamental distinguishability measure in quantum information theory. It characterizes quantum differential privacy and determines the optimal error probability of binary quantum hypothesis testing with arbitrary priors. In this paper, we study how the hockey-stick divergence behaves on tensor-product states. To this end, we consider the noncommutative minimum $m(A,B)\coloneqq(\Tr[A]+\Tr[B]-\lVert A-B\rVert_1)/2$ of positive semidefinite operators, which reduces to $\sum_x\min\{a_x,b_x\}$ when $A$ and $B$ commute, and we prove that it is super-multiplicative, i.e., $m(A\otimes C,B\otimes D)\ge m(A,B)\,m(C,D)$. 
As a consequence, we establish a sub-additivity property of the hockey-stick divergence on product states, in which the parameters of the individual systems are multiplied, and no additional prefactor appears. We show that this bound is tight, also for noncommuting states, and that it is strict for faithful states whenever at least two of the divergences are positive.
Using this, we obtain sub-additivity of the Datta--Leditzky divergences, a parallel composition theorem for quantum differentially private channels under product neighboring inputs with at most linear degradation of the privacy parameters,
and a sharp sub-additivity property of the hypothesis-testing relative entropy normalized by the acceptance probability. Finally, we apply the same tools to symmetric binary quantum hypothesis testing. We determine the exact prefactor of the quantum Chernoff bound for every pair of noncommuting states satisfying mild conditions including a nonlattice condition and, under a quantitative version of this condition, the full asymptotic expansion to every fixed order in $1/n$, with explicitly computable coefficients. As a consequence, we determine the sample complexity of symmetric binary quantum hypothesis testing up to one sample, as the error probability tends to zero.
\end{abstract}

\tableofcontents

\section{Introduction}
\label{sec:intro}

Distinguishability measures between quantum states are central to quantum information theory, as they quantify the ultimate performance of statistical tasks such as hypothesis testing, and they are used to formalize security and privacy guarantees~\cite{wilde2017book}. A basic question about such a measure is how it behaves when the states are tensor products, i.e., when independent systems are combined. For several measures, the answer is simple. For instance, the quantum relative entropy and the Petz R\'enyi divergences are additive on tensor-product states. However, this is not the case for the hockey-stick divergence and for other one-shot quantities obtained from it. The difficulty is already visible in the trace norm $\lVert A\otimes C-B\otimes D\rVert_1$, which does not factorize, because the optimal measurement for distinguishing $A\otimes C$ from $B\otimes D$ can be a collective (entangled) measurement across the two systems.

The quantum hockey-stick divergence is defined for $\gamma\ge1$ and states $\rho$ and $\sigma$  as~\cite{SW12}: 
\begin{equation}
E_\gamma(\rho\Vert\sigma)\coloneqq\Tr[(\rho-\gamma\sigma)_+].
\label{eq:HS-def-intro}
\end{equation}
It plays several roles. First, $(\varepsilon,\delta)$-quantum differential privacy (QDP) of a quantum channel is equivalent to an upper bound on the hockey-stick divergence of its outputs, with $\gamma=e^\varepsilon$~\cite{hirche2023quantum}. Second, the optimal error probability of binary quantum hypothesis testing with priors $p$ and $1-p$~\cite{helstrom1969quantum,holevo1973statistical} is a function of $E_\gamma$ with $\gamma=(1-p)/p$ for $p\le1/2$, and with $\gamma=p/(1-p)$ and the roles of the two states exchanged for $p\ge1/2$. Third, the Datta--Leditzky information-spectrum divergences~\cite{datta2014second} and the hypothesis-testing relative entropy~\cite{wang2012one} are obtained from the hockey-stick divergence through a one-dimensional optimization. Furthermore, contraction properties of the hockey-stick divergence under privacy constraints were used to study private quantum hypothesis testing~\cite{nuradha2024contraction,Christoph2024sample}. Also, linear and non-linear strong data-processing inequalities for the quantum hockey-stick divergence have been established to study information contraction and mixing times of noisy quantum systems~\cite{nuradha2025nonLinearSDPI,dasgupta2025quantum}.  

The behavior of these quantities under composition has direct operational consequences. In the classical setting, the composition of differentially private mechanisms is well understood~\cite{dwork2006calibrating,dwork2014algorithmic,kairouz2017composition}. In particular, if $\mathcal A_1$ and $\mathcal A_2$ satisfy $(\varepsilon_1,\delta_1)$- and $(\varepsilon_2,\delta_2)$-differential privacy, then their parallel composition satisfies $(\varepsilon_1+\varepsilon_2,\delta_1+\delta_2)$-differential privacy. In the quantum setting, an adversary may perform a collective measurement on the joint output of $\mathcal A_1\otimes\mathcal A_2$. The parallel composition results known for such general measurements contain additional terms~\cite{hirche2023quantum,nuradha_QPP}, which become large in the low privacy regime, {whereas, for $\delta_1,\delta_2>0$, a composition rule of the classical type was established only for restricted classes of measurements~\cite{nuradha_QPP,alabi2026quantum}.}
{Thus it has remained unclear whether the classical parameters $(\varepsilon_1+\varepsilon_2,\delta_1+\delta_2)$ remain valid when collective measurements are allowed.}

At the level of the hockey-stick divergence, the parallel composition question reduces to bounding $E_{\gamma_1\gamma_2}(\rho_1\otimes\rho_2\Vert\sigma_1\otimes\sigma_2)$ in terms of $E_{\gamma_1}(\rho_1\Vert\sigma_1)$ and $E_{\gamma_2}(\rho_2\Vert\sigma_2)$. Previously, the main known sub-additivity property of the quantum hockey-stick divergence for general states was~\cite[Proposition~II.5]{hirche2023quantum}: for $\gamma_1,\gamma_2\ge1$ and states $\rho_1,\sigma_1,\rho_2,\sigma_2$,
\begin{equation}
E_{\gamma_1\gamma_2}(\rho_1\otimes\rho_2\Vert\sigma_1\otimes\sigma_2)\le\min\!\left\{E_{\gamma_1}(\rho_1\Vert\sigma_1)+\gamma_1E_{\gamma_2}(\rho_2\Vert\sigma_2),\ E_{\gamma_1}(\rho_2\Vert\sigma_2)+\gamma_1E_{\gamma_2}(\rho_1\Vert\sigma_1)\right\},
\label{eq:intro-HRF}
\end{equation}
which follows from a triangle inequality for the hockey-stick divergence together with its stability under tensoring with a common state. 
Moreover, iterating~\eqref{eq:intro-HRF} over $n$ systems leads to bounds in which the hockey-stick divergences of the later systems are multiplied by products of the earlier parameters $\gamma_i$.

A related question arises in quantum hypothesis testing. For $n$ independent copies, the optimal error probability of symmetric binary hypothesis testing decays exponentially with the exponent given by the quantum Chernoff bound~\cite{audenaert2007chernoff,nussbaum2009chernoff}, and the asymmetric setting is governed by the quantum Stein lemma~\cite{hiai1991proper,ogawa2000strong} and its second-order refinements~\cite{li2014second,tomamichel2013hierarchy}. In the classical setting, much more precise statements are known. Indeed, the exact asymptotics of large-deviation probabilities, including the prefactor of the exponential term, go back to Bahadur and Ranga Rao~\cite{bahadur1960deviations}, and higher-order expansions are available under suitable nonlattice conditions~\cite{petrov1975sums,chaganty1993strong,angst2017weak,fernando2021higher}. In the quantum setting, the precise prefactor of the quantum Chernoff bound has remained open. Recently, Cheng and Liu~\cite{cheng2026multiple} compared the optimal quantum error with a harmonic-mean quantity associated with the Nussbaum--Szko{\l}a distributions, and they determined the prefactor up to a factor of two in each direction.

\subsection{Contributions}

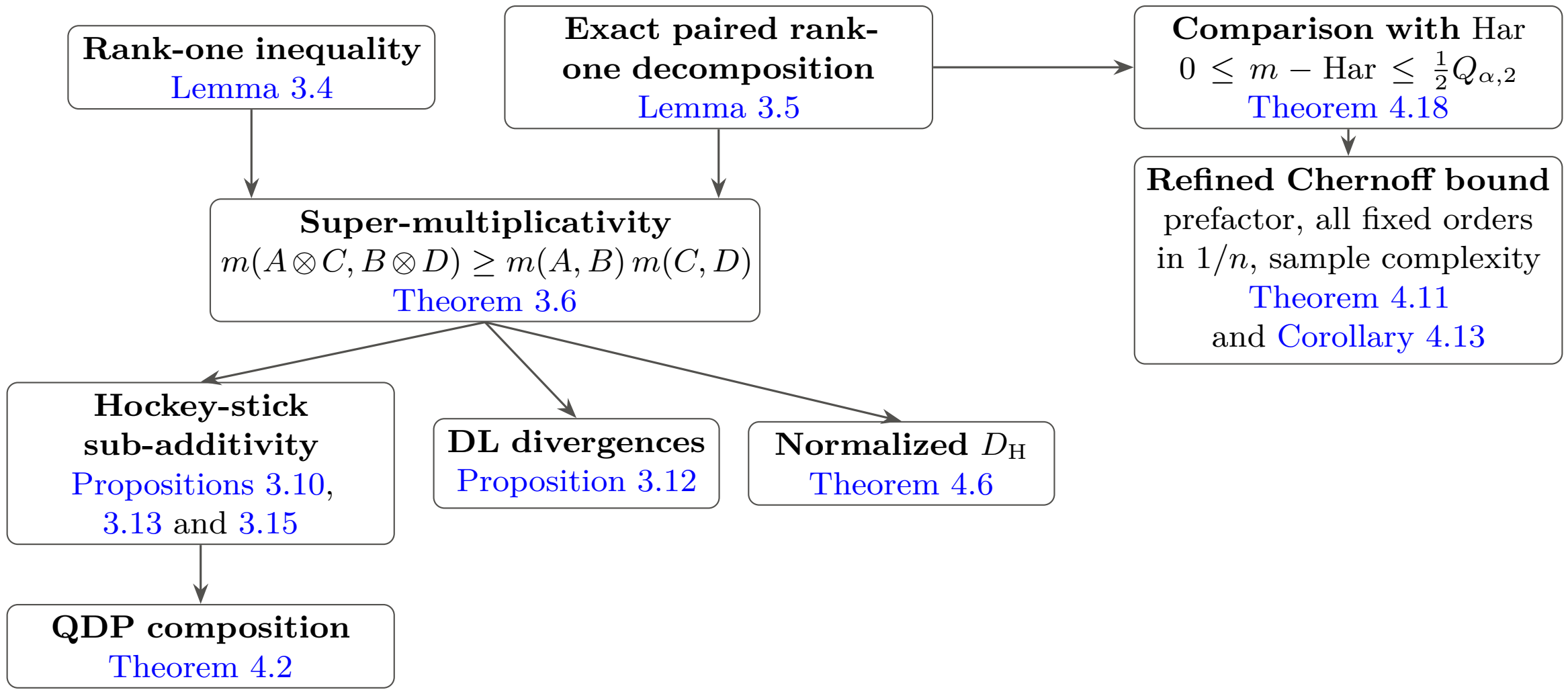
\begin{figure}[t]
\centering
\begin{tikzpicture}
\node[rmbox=3.4cm] (r1) at (0,0) {\textbf{Rank-one inequality}\\ \cref{lem:rank-one}};
\node[rmbox=4.0cm] (dec) at (4.6,0) {\textbf{Exact paired rank-one decomposition}\\ \cref{lem:decomposition_general_to_rank}};
\node[rmbox=4.0cm] (cmp) at (10.8,0) {\textbf{Comparison with $\Har$}\\ $0\le m-\Har\le\frac12 Q_{\alpha,2}$\\ \cref{thm:ch-three-bounds}};
\node[rmbox=5.2cm] (ov) at (2.3,-1.9) {\textbf{Super-multiplicativity}\\ $m(A\otimes C,B\otimes D)\ge m(A,B)\,m(C,D)$\\ \cref{thm:overlap}};
\node[rmbox=4.0cm] (ch) at (10.8,-1.9) {\textbf{Refined Chernoff bound}\\ prefactor, all fixed orders in $1/n$, sample complexity\\ \cref{thm:ch-main,cor:ch-sample}};
\node[rmbox=3.6cm] (hs) at (-0.5,-3.9) {\textbf{Hockey-stick sub-additivity}\\ \cref{Prop:n_HS_subADD,prop:optimality,prop:strict}};
\node[rmbox=2.6cm] (dl) at (3.2,-3.9) {\textbf{DL divergences}\\ \cref{prop:n_DL}};
\node[rmbox=2.8cm] (ht) at (6.4,-3.9) {\textbf{Normalized $D_{\mathrm H}$}\\ \cref{thm:ht}};
\node[rmbox=3.6cm] (qdp) at (-0.5,-5.7) {\textbf{QDP composition}\\ \cref{thm:parallel_composability_QDP}};
\draw[rmarrow] (r1.south) -- (ov.north -| r1.south);
\draw[rmarrow] (dec.south) -- (ov.north -| dec.south);
\draw[rmarrow] (dec) -- (cmp);
\draw[rmarrow] (cmp) -- (ch);
\draw[rmarrow] (ov.south) -- (hs.north);
\draw[rmarrow] (ov.south) -- (dl.north);
\draw[rmarrow] (ov.south) -- (ht.north);
\draw[rmarrow] (hs) -- (qdp);
\end{tikzpicture}
\caption{Structure of the paper. The super-multiplicativity of the noncommutative minimum $m$ yields the sub-additivity results for divergences and their applications to quantum differential privacy and hypothesis testing. The exact rank-one decomposition also yields the comparison of $m$ with the harmonic overlap $\Har$, which leads to the refined Chernoff asymptotics.}
\label{fig:roadmap}
\end{figure}

In this paper, we address the above questions through a single structural result. For positive semidefinite operators $A$ and $B$, we consider the noncommutative minimum
\begin{equation}
m(A,B)\coloneqq\frac{\Tr[A]+\Tr[B]-\lVert A-B\rVert_1}{2}.
\label{eq:intro-overlap}
\end{equation}
If $A$ and $B$ commute, with eigenvalues $a_x$ and $b_x$ in a common eigenbasis, then $m(A,B)=\sum_x\min\{a_x,b_x\}$. For states and $\gamma\ge1$, $m(\rho,\gamma\sigma)=1-E_\gamma(\rho\Vert\sigma)$, and $m(p\rho,(1-p)\sigma)$ is the optimal error probability of binary hypothesis testing with priors $p$ and $1-p$. \Cref{fig:roadmap} shows how the results of this paper follow from this single quantity.

First, we prove that the noncommutative minimum is super-multiplicative, i.e., $m(A\otimes C,B\otimes D)\ge m(A,B)\,m(C,D)$ for all positive semidefinite operators (\cref{thm:overlap}). The key step is an exact paired rank-one decomposition (\cref{lem:decomposition_general_to_rank}): every pair $(A,B)$ can be written as $A=\sum_iA_i$ and $B=\sum_iB_i$ with operators of rank at most one, such that $m(A,B)=\sum_im(A_i,B_i)$. This reduces the problem to a scalar inequality for rank-one operators (\cref{lem:rank-one}).

Second, as a consequence, we establish the sub-additivity of the hockey-stick divergence. For every $n\ge2$, states $\rho_1,\ldots,\rho_n$ and $\sigma_1,\ldots,\sigma_n$, and $\gamma_1,\ldots,\gamma_n\ge1$ with $\Gamma\coloneqq\prod_{i=1}^n\gamma_i$, we have that
\begin{equation}
E_{\Gamma}(\rho_1\otimes\cdots\otimes\rho_n\Vert\sigma_1\otimes\cdots\otimes\sigma_n)\le1-\prod_{i=1}^n\left(1-E_{\gamma_i}(\rho_i\Vert\sigma_i)\right)\le\sum_{i=1}^nE_{\gamma_i}(\rho_i\Vert\sigma_i)
\label{eq:intro-HS-n}
\end{equation}
(\cref{Cor:sub_additivity_HS,Prop:n_HS_subADD}). In contrast to~\eqref{eq:intro-HRF}, no factor $\gamma_i$ appears in front of the divergences on the right-hand side, and the bound is symmetric in the $n$ systems. We show that~\eqref{eq:intro-HS-n} is tight for every $n$ and every value of the divergences $E_{\gamma_i}(\rho_i\Vert\sigma_i)$, and that equality is attained also by noncommuting states (\cref{prop:optimality,ex:noncommuting-equality}). For faithful states, the first inequality in~\eqref{eq:intro-HS-n} is strict whenever $E_{\gamma_i}(\rho_i\Vert\sigma_i)>0$ for at least two indices $i$ (\cref{prop:strict}). Moreover, sub-additivity fails if the parameter $\gamma$ is kept fixed instead of being multiplied (\cref{ex:same-error}). We also derive sub-additivity of the Datta--Leditzky divergences for $n$ product states (\cref{Cor:sub_additivity_DL,prop:n_DL}).

Third, we apply these results to quantum differential privacy. We prove that the parallel composition $\mathcal A_1\otimes\cdots\otimes\mathcal A_n$ of $(\varepsilon_i,\delta_i)$-QDP channels satisfies $(\sum_i\varepsilon_i,1-\prod_i(1-\delta_i))$-QDP for product neighboring inputs, even when the adversary performs collective measurements (\cref{thm:parallel_composability_QDP}). Thus the privacy parameters degrade at most linearly, as in the classical setting, and the parameter $1-\prod_i(1-\delta_i)$ cannot be improved, already for classical mechanisms (\cref{rem:QDP-tight}). In particular, for channels that are QDP with respect to all measurements, the parameter $\delta_1+\delta_2$, previously established for product measurements and a class of separable measurements~\cite[Theorem~1 and the discussion after it]{nuradha_QPP}, also suffices for global measurements, which answers the question raised in~\cite[Section~5.2]{alabi2026quantum}. 

Fourth, we study the hypothesis-testing relative entropy. We show that the type-II error normalized by the acceptance probability is super-multiplicative when the type-I error parameters are multiplied (\cref{thm:ht}). This statement is sharp, and it fails if the error parameter is kept fixed (\cref{ex:same-error_HT}). The normalized divergence has a simple operational meaning: it is the logarithm of the largest likelihood ratio of an acceptance event (\cref{lem:ht-operational}). Thus, \cref{thm:ht} implies the following. Consider a global test, possibly based on an entangled measurement, with type-I error at most $\prod_i\ee_i$. Then the logarithm of its likelihood ratio is at most $\sum_i\Dhat{\ee_i}$, which is the value achieved by the product test with the AND rule, at the larger type-I error $1-\prod_i(1-\ee_i)$ (\cref{sec:ht-operational}). For many copies, \cref{thm:ht} also gives a converse that holds for every $n$ (see~\eqref{eq:ht-many-copies}).

Finally, in \cref{sec:chernoff}, we study refined asymptotics of symmetric binary quantum hypothesis testing with $n$ copies. Our main result there is \cref{thm:ch-main}: for every noncommuting pair with an interior Chernoff minimizer $\alpha_*\in(0,1)$, positive tilted variance $V_*=\Phi''(\alpha_*)>0$, and a nonlattice log-likelihood ratio (\cref{def:ch-regular}), and for each fixed $p\in(0,1)$, the optimal error probability satisfies
\begin{equation}
P_{e,n}^*(p)=A_{\mathrm q}\,\frac{e^{-n\xi}}{\sqrt n}\,[1+o(1)],
\qquad
A_{\mathrm q}\coloneqq\frac{\pi\,p^{\alpha_*}(1-p)^{1-\alpha_*}}{\sin(\pi\alpha_*)\sqrt{2\pi V_*}},
\label{eq:intro-prefactor}
\end{equation}
where $\xi$ is the quantum Chernoff exponent, $\alpha_*$ is the optimal order, and $V_*$ is a variance. This removes the remaining constant-factor uncertainty in~\cite{cheng2026multiple}. Under the quantitative nonlattice condition of \cref{def:ch-QN}, we further obtain
\[
-\ln P_{e,n}^*(p)=n\xi+\frac12\ln n-\ln A_{\mathrm q}+\sum_{j=1}^{M}\frac{C_j(p)}{n^j}+O(n^{-M-1}),
\]
for every fixed integer $M\ge1$. Appendix~\ref{sec:ch-coefficients} gives a finite rule for every coefficient and explicit formulas for $C_1(p)$, $C_2(p)$, and $C_3(p)$. As a consequence, the minimum number $n^*(p,\delta)$ of copies needed to achieve error probability at most $\delta$ satisfies $-o(1)\le n^*(p,\delta)-x_\delta\le1+o(1)$ as $\delta\to0$ (\cref{cor:ch-sample}), where $x_\delta$ is the unique positive solution of~\eqref{eq:ch-sample-root} and
\[
x_\delta=\frac1\xi\left[\ln\frac1\delta-\frac12\ln\!\left(\frac1\xi\ln\frac1\delta\right)+\ln A_{\mathrm q}\right]+o(1).
\]
Thus the sample complexity is determined up to one sample asymptotically, whereas the general bounds of~\cite[Theorem~II.7]{cheng2024invitation} determine it up to a constant factor.

We show that every pair of distinct, noncommuting faithful states satisfies all assumptions of \cref{thm:ch-main} except, possibly, the nonlattice condition, and that this condition holds for generic noisy states (\cref{prop:ch-regular-pairs}). We compare these predictions with numerical evaluations of the optimal error for up to $200$ copies of a pair of noisy qubit states (\cref{ex:ch-numerics,fig:chernoff-numerics,tab:sample}).

Our main tool is the harmonic overlap $\Har(A,B)$. Cheng and Liu established the comparison $\Har(A,B)\le m(A,B)\le2\Har(A,B)$~\cite[Theorem~4.2]{cheng2026multiple}. Using the exact paired rank-one decomposition, we also show that, for $A,B\ge0$,
\[
0\le m(A,B)-\Har(A,B)\le\frac12Q_{\alpha,2}(A,B),\qquad\alpha\in(0,1),
\]
where $Q_{\alpha,2}$ is the $\alpha$-$z$ R\'enyi quasi-entropy with $z=2$ (\cref{thm:ch-three-bounds}), so that, for $n$ copies of noncommuting faithful states, the optimal error is given by a classical quantity of the Nussbaum--Szko{\l}a distributions up to an exponentially smaller term. As a separate application, the same harmonic-overlap comparison determines the exact bounded correction for the integral R\'enyi divergence of~\cite{hirche2024quantum} for noncommuting states, without a nonlattice assumption (\cref{sec:ch-integral}).

\subsection{Further Comparisons with Prior Work}
\label{sec:intro-comparisons}

Lee and Lami's one-shot Chernoff bound~\cite[Theorem~4]{lee2026sharp}, specialized to two hypotheses, gives a lower bound of the form $n\xi+O(1)$ on $-\ln P_{e,n}^*(p)$, without the $\frac12\ln n$ term. This scale already follows from the standard binary Chernoff bound~\cite{audenaert2007chernoff}; the main result of~\cite{lee2026sharp} is a sharp reduction of multiple hypotheses to binary errors. The third-order converse of Cheng and Liu~\cite[Proposition~6.7]{cheng2026multiple} concerns testing with a fixed type-I error, with the scales $n$, $\sqrt n$, and $\ln n$, and it does not determine the Chernoff prefactor at a fixed prior.

The acceptance normalization of $\Dhat\ee$ already appears in~\cite{dupuis2013generalized}; our contribution in \cref{thm:ht} is its sharp tensor-product law with multiplied error parameters. Finally, \cref{thm:ch-integral-exact} refines the known regularization of the integral R\'enyi divergence to the Petz divergence for $0<\alpha<1$~\cite{hirche2024quantum} by determining the exact bounded correction and an exponentially decaying remainder for noncommuting states.

\subsection{Organization}

The rest of our paper is organized as follows. In \cref{sec:background}, we introduce notation and the quantum divergences used in this work. In \cref{sec:supermult}, we prove the super-multiplicativity of the noncommutative minimum and derive the sub-additivity of the hockey-stick and Datta--Leditzky divergences, together with their optimality. In \cref{sec:applications}, we present applications to quantum differential privacy, to the hypothesis-testing relative entropy, and to refined quantum Chernoff bounds and the sample complexity of symmetric hypothesis testing. We conclude in \cref{sec:conclusion} with a summary and future directions. The appendices contain the proofs of \cref{lem:rank-one,lem:decomposition_general_to_rank}, as well as auxiliary results and the computation of the coefficients for \cref{sec:chernoff}.

\section{Background}
\label{sec:background}

We collect the operator notation and quantum divergences used throughout the paper.

\subsection{Definitions and Notation}
\label{sec:defs}

For a Hermitian operator $X$, its Jordan decomposition and absolute value are
\[
X=X_+-X_-,
\qquad
|X|=X_++X_-,
\]
where $X_+,X_-\ge0$ have orthogonal supports. The trace norm and the Hilbert--Schmidt norm of a linear operator $X$ are
\[
\lVert X\rVert_1\coloneqq\Tr|X|,
\qquad
\lVert X\rVert_2\coloneqq\sqrt{\Tr[X^\dagger X]}.
\]
For a Hermitian operator $H$ with spectral decomposition
\[
H=\sum_j \lambda_j\,u_ju_j^\dagger,
\]
we define its sign operator by
\begin{equation}
\sgn(H)
\coloneqq
\sum_j \sgn(\lambda_j)\,u_ju_j^\dagger,
\label{eq:sign-def}
\end{equation}
where
\[
\sgn(\lambda)=
\begin{cases}
1, & \lambda>0,\\
0, & \lambda=0,\\
-1, & \lambda<0.
\end{cases}
\]

Throughout, we consider only finite dimensional Hilbert spaces. For a linear operator $T$ on a Hilbert space $\cH$, we write
\[
\ran(T)\coloneqq\{Tx:x\in\cH\},
\qquad
\ker(T)\coloneqq\{x\in\cH:T x=0\}
\]
for its range and kernel, respectively. For a positive semidefinite operator $T\ge0$, we define its support by $\supp(T)\coloneqq(\ker T)^\perp.$
Since $T$ is Hermitian and the Hilbert space is finite dimensional,
\begin{equation}
\ran(T)=\supp(T)=(\ker T)^\perp.
\label{eq:range-support}
\end{equation}
If $T$ is singular, we never use an inverse on the whole Hilbert space. Rather, whenever an inverse of $T$ is considered, we first restrict $T$ to $\supp(T)$, where it is strictly positive definite. Thus
$
\left(T|_{\supp(T)}\right)^{-1}$ and $
\left(T|_{\supp(T)}\right)^{-1/2}
$
denote the ordinary inverse and inverse square root of the restricted operator $T|_{\supp(T)}$.

We write $\cD(A)$ for the set of quantum states (density operators) on a system $A$, $\cL(A)$ for the linear operators on $A$, and $\pi_A\coloneqq I_A/\dim A$ for the maximally mixed state. A state is called faithful if it is positive definite. A channel $\cN\colon\cL(A)\to\cL(B)$ is a completely positive and trace-preserving map. Throughout, $\ln$ denotes the natural logarithm; only the hypothesis-testing relative entropies in \cref{sec:HT} use the logarithm to base~$2$.

\subsection{Quantum Divergences}

For a state $\rho$, a positive semidefinite operator $\sigma$, and $\gamma\ge1$, define the hockey-stick divergence~\cite{SW12} by
\begin{equation}
    E_{\gamma}(\rho\| \sigma) \coloneqq  \Tr\!\left[(\rho-\gamma \sigma)_{+}\right],
    \label{eq:HS-def}
\end{equation}
where $(\cdot)_+$ denotes the positive part in the Jordan decomposition in \cref{sec:defs}. For states $\rho$ and $\sigma$, the case $\gamma=1$ gives the trace distance $T(\rho,\sigma)\coloneqq\frac12\lVert\rho-\sigma\rVert_1=E_1(\rho\Vert\sigma)$. For weights $\gamma\in(0,1)$, we do not use the hockey-stick divergence and work instead with the noncommutative minimum $m(\rho,\gamma\sigma)$ introduced in \cref{def:m}. We also use the max-relative entropy $\Dmax(\rho\Vert\sigma)\coloneqq\inf\{\lambda\in\mathbb R:\rho\le e^\lambda\sigma\}$, so that, for $\lambda\ge0$, $E_{e^\lambda}(\rho\Vert\sigma)=0$ if and only if $\lambda\ge\Dmax(\rho\Vert\sigma)$.

Fix $\delta \in [0,1]$, a state $\rho$, and a positive semidefinite operator~$\sigma$. The Datta--Leditzky information-spectrum divergences are defined as follows~\cite{datta2014second}:
\begin{subequations}
\begin{align}
\underline{D}^{\delta}(\rho\Vert\sigma) &  \coloneqq  \sup\left\{  \gamma
\in\mathbb{R}:\operatorname{Tr}[\left(  \rho-e^{\gamma}\sigma\right)
_{+}]\geq1-\delta\right\} ,
\label{eq:1st-DL-def} \\
\overline{D}^{\delta}(\rho\Vert\sigma) &  \coloneqq  \inf\left\{  \gamma
\in\mathbb{R}:\operatorname{Tr}[\left(  \rho-e^{\gamma}\sigma\right)
_{+}]\leq\delta\right\}.
\label{eq:2nd-DL-def}
\end{align}\label{eq:DL_div}%
\end{subequations}
Hereafter we abbreviate these divergences as DL~divergences.
For $0<\delta<1$ and $\supp\rho\subseteq\supp\sigma$, the equality stated in~\cite[Proposition~4.3(ii)]{datta2014second} holds:\footnote{The support assumption is absent from the cited statement. Without it, the function $\gamma\mapsto\Tr[(\rho-e^\gamma\sigma)_+]$ may have a positive plateau, so the two generalized inverses need not coincide.}
\begin{equation}
\underline{D}^{\delta}(\rho\Vert\sigma)=\overline{D}%
^{1-\delta}(\rho\Vert\sigma). \label{eq:inf-spec-equality}
\end{equation}
Under these assumptions, it suffices to consider $\overline{D}^{\delta}$ from~\eqref{eq:2nd-DL-def}.

\section{Super-Multiplicativity and Sub-Additivity}
\label{sec:supermult}

We now introduce the quantity on which all results of our paper are based.

\begin{definition}[Noncommutative Minimum]
\label{def:m}
For positive semidefinite operators $A$ and $B$, define
\begin{equation}
m(A,B)
\coloneqq
\frac{\Tr A+\Tr B-\lVert A-B\rVert_1}{2}.
\label{eq:m-def}
\end{equation}
\end{definition}

If $A=\sum_xa_x\proj{x}$ and $B=\sum_xb_x\proj{x}$ commute, then $\lVert A-B\rVert_1=\sum_x|a_x-b_x|$, and since $(a+b-|a-b|)/2=\min\{a,b\}$, we have that
\begin{equation}
m(A,B)=\sum_x\min\{a_x,b_x\}.
\label{eq:m-commuting}
\end{equation}
This is why we call $m(A,B)$ the noncommutative minimum of $A$ and $B$; it is denoted by $\Tr[A\wedge B]$ in~\cite{cheng2026multiple}. The first observation is that it is exactly the complement of the hockey-stick divergence when the first argument has unit trace, and that it is the minimal weighted error of a binary test between $A$ and $B$.

\begin{lemma}[Basic Properties of the Noncommutative Minimum]
\label{lem:overlap-identities}
For $A,B\ge0$, the noncommutative minimum in~\eqref{eq:m-def} satisfies
\begin{align}
m(A,B)
&=
\Tr A-\Tr(A-B)_+,
\label{eq:m-positive-part}\\
0\le m(A,B)
&\le \min\{\Tr A,\Tr B\},
\label{eq:m-bounds}\\
m(A,B)&=\min_{0\le T\le\id}\left\{\Tr[AT]+\Tr[B(\id-T)]\right\}.
\label{eq:m-variational}
\end{align}
In particular, for a state $\rho$, a positive semidefinite operator $\sigma$, and $\gamma\ge1$,
\begin{equation}
m(\rho,\gamma\sigma)
=
1-E_\gamma(\rho\Vert\sigma).
\label{eq:m-hsd}
\end{equation}
\end{lemma}

\begin{proof}
Equation~\eqref{eq:m-variational} is the Helstrom--Holevo formula; see~\cite[Eqs.~(13)--(14), (17)--(18)]{cheng2026multiple}. Equation~\eqref{eq:m-positive-part} follows from~\cite[Eq.~(II.2)]{hirche2023quantum}, and~\eqref{eq:m-hsd} is its unit-trace specialization with $B=\gamma\sigma$. The bounds in~\eqref{eq:m-bounds} follow from the nonnegative objective in~\eqref{eq:m-variational} and the feasible choices $T=0$ and $T=\id$.
\end{proof}

\begin{lemma}[Joint Super-Additivity]
\label{lem:superadd}
For finite families $A_i,B_i\ge0$, the noncommutative minimum in~\eqref{eq:m-def} satisfies
\begin{equation}
m\!\left(\sum_iA_i,\sum_iB_i\right)
\ge
\sum_i m(A_i,B_i).
\label{eq:superadd}
\end{equation}
\end{lemma}

\begin{proof}
By \eqref{eq:m-def} and the triangle inequality for trace norm,
\begin{align*}
m\!\left(\sum_iA_i,\sum_iB_i\right)
&=
\frac{
\sum_i\Tr A_i+\sum_i\Tr B_i
-\left\|\sum_i(A_i-B_i)\right\|_1
}{2}\\
&\ge
\frac{
\sum_i\Tr A_i+\sum_i\Tr B_i
-\sum_i\|A_i-B_i\|_1
}{2}\\
&=
\sum_i m(A_i,B_i),
\end{align*}
thus concluding the proof.
\end{proof}

\begin{lemma}[Rank-One Super-Multiplicativity]
\label{lem:rank-one}
Let $A,B,C,D\ge0$ have rank at most one. Then the noncommutative minimum in~\eqref{eq:m-def} satisfies
\begin{equation}
m(A\otimes C,B\otimes D)
\ge
m(A,B)m(C,D).
\label{eq:rank-one}
\end{equation}
Moreover, the inequality is strict if $A$ and $B$ are nonzero and not proportional to each other, satisfying $m(A,B)>0$, and the same conditions hold for $C$ and $D$.
\end{lemma}
\begin{proof}
The proof is given in Appendix~\ref{Sec:Lemma_rank_one}.
\end{proof}

We now state the structural result needed to go from rank-one pairs to general operators.

\begin{lemma}[Exact Paired Rank-One Decomposition]
\label{lem:decomposition_general_to_rank}
For every finite-dimensional pair $A,B\ge0$, there exist finitely many pairs $(A_i,B_i)$ such that
\[
A=\sum_iA_i,
\qquad
B=\sum_iB_i,
\qquad
\operatorname{rank}A_i,\operatorname{rank}B_i\le1,
\]
and the noncommutative minimum in~\eqref{eq:m-def} satisfies
\begin{equation}
m(A,B)=\sum_i m(A_i,B_i).
\label{eq:exact-overlap}
\end{equation}
Moreover, if $(A-B)_+\nleq A$, the pairs can be chosen so that, for some index $i$, $A_i$ and $B_i$ are nonzero and not proportional to each other, satisfying $m(A_i, B_i)>0$.
\end{lemma}
\begin{proof}
The proof is given in Appendix~\ref{sec:proof_lemma_decomposition_general_to_rank}.
\end{proof}

The idea behind the proof of \cref{lem:decomposition_general_to_rank} is as follows. We first construct a common positive part $C_0\le A,B$ by normalizing with $(A+B)^{1/2}$ on its support. After removing this part, the ranges of $A-C_0$ and $B-C_0$ intersect only in the zero vector. For the remaining pair, we construct rank-one pairs $(A_i,B_i)$ such that the positive part of every difference $A_i-B_i$ lies in the positive spectral subspace of $A-B$ and the negative part lies in the negative spectral subspace of $A-B$. Then the differences cannot cancel each other in trace norm, i.e., $\sum_i\lVert A_i-B_i\rVert_1=\lVert A-B\rVert_1$, and substituting this into~\eqref{eq:m-def} gives~\eqref{eq:exact-overlap}. Equivalently, by~\eqref{eq:m-variational}, the projection onto the positive spectral subspace of $B-A$ is an optimal test for every pair $(A_i,B_i)$ simultaneously. The degenerate case occurs when $(A-B)_+\leq A$. In this case, set $C\coloneqq A-(A-B)_+\geq0$, and we have $A=C+(A-B)_+$ and $B=C+(A-B)_-$. After removing this common positive part $C$, the residual operators have orthogonal supports.

With these tools, we now prove the main result of this section.

\begin{theorem}[Super-Multiplicativity of the Noncommutative Minimum]
\label{thm:overlap}
For positive semidefinite operators $A,B,C,$ and $D$, the noncommutative minimum in~\eqref{eq:m-def} satisfies
\begin{equation}
m(A\otimes C,B\otimes D)
\ge
m(A,B)m(C,D).
\label{eq:overlap-supermult}
\end{equation}
\end{theorem}

\begin{proof}
First by applying \cref{lem:decomposition_general_to_rank} separately to $(A,B)$ and $(C,D)$, we have that
\[
A=\sum_iA_i,
\qquad
B=\sum_iB_i,
\qquad
C=\sum_jC_j,
\qquad
D=\sum_jD_j,
\]
with all summands of rank at most one and
\[
\sum_i m(A_i,B_i)=m(A,B),
\qquad
\sum_j m(C_j,D_j)=m(C,D).
\]
Then
\[
A\otimes C=\sum_{i,j}A_i\otimes C_j,
\qquad
B\otimes D=\sum_{i,j}B_i\otimes D_j.
\]
Using \cref{lem:superadd} and then \cref{lem:rank-one}, we obtain
\begin{align}
m(A\otimes C,B\otimes D)
&\ge
\sum_{i,j}
m(A_i\otimes C_j,B_i\otimes D_j)\label{eq:supermult-step1}\\
&\ge
\sum_{i,j}
m(A_i,B_i)m(C_j,D_j)\label{eq:supermult-step2}\\
&=
\left(\sum_i m(A_i,B_i)\right)
\left(\sum_j m(C_j,D_j)\right)\notag\\
&=
m(A,B)m(C,D),\notag
\end{align}
concluding the proof.
\end{proof}

\begin{remark}[Operational Meaning]
\label{rem:overlap-operational}
By~\eqref{eq:m-variational}, $m(A,B)$ is the minimal weighted error of a binary test between the weights $A$ and $B$. For commuting operators, \cref{thm:overlap} reduces by~\eqref{eq:m-commuting} to the elementary inequality $\min\{a_xc_y,b_xd_y\}\ge\min\{a_x,b_x\}\min\{c_y,d_y\}$, summed over $x$ and $y$. In the noncommutative case, the optimal test for the product pair is in general an entangled projection that does not factorize, and \cref{thm:overlap} shows that such a collective measurement cannot push the weighted error below the product of the individual weighted errors.
\end{remark}

\subsection{Sub-Additivity of Hockey-Stick Divergence and Datta--Leditzky Divergence}

We now translate the super-multiplicativity of the noncommutative minimum into sub-additivity bounds for the hockey-stick and Datta--Leditzky divergences.

\begin{corollary}[Sub-Additivity of Hockey-Stick Divergence] \label{Cor:sub_additivity_HS}
    Let $\rho_i$ and $\sigma_i$ be quantum states and $\gamma_i \geq 1$ for $i\in\{1,2\}$. Then, the hockey-stick divergence in~\eqref{eq:HS-def} satisfies
    \begin{align}
        E_{\gamma_1 \gamma_2}(\rho_1 \otimes \rho_2 \Vert \sigma_1 \otimes \sigma_2) &\leq 1- \left( 1- E_{\gamma_1}(\rho_1 \Vert \sigma_1)\right) \left( 1- E_{\gamma_2}(\rho_2 \Vert \sigma_2)\right) \label{eq:HS-two}\\
        &= E_{\gamma_1}(\rho_1 \Vert \sigma_1) + E_{\gamma_2}(\rho_2 \Vert \sigma_2)-E_{\gamma_1}(\rho_1 \Vert \sigma_1)E_{\gamma_2}(\rho_2 \Vert \sigma_2).\notag
    \end{align}
\end{corollary}
\begin{proof}
Apply \cref{thm:overlap} with $A=\rho_1,
B=\gamma_1\sigma_1,
C=\rho_2,
D=\gamma_2\sigma_2.$
By \eqref{eq:m-hsd}, $m(\rho_i,\gamma_i\sigma_i)
=
1-E_{\gamma_i}(\rho_i\Vert \sigma_i)$
and
\[
m(\rho_1\otimes\rho_2,
\gamma_1\gamma_2\,\sigma_1\otimes\sigma_2)
=
1-
E_{\gamma_1\gamma_2}
(\rho_1\otimes\rho_2\Vert\sigma_1\otimes\sigma_2).
\]
Hence
\[
1-
E_{\gamma_1\gamma_2}
(\rho_1\otimes\rho_2\Vert\sigma_1\otimes\sigma_2)
\ge
\left( 1- E_{\gamma_1}(\rho_1 \Vert \sigma_1)\right) \left( 1- E_{\gamma_2}(\rho_2 \Vert \sigma_2)\right).
\]
Rearranging the above inequality gives the desired inequality.
\end{proof}

The bound of \cref{Cor:sub_additivity_HS} has the form of the probability that at least one of two independent events occurs: if $E_{\gamma_i}(\rho_i\Vert\sigma_i)$ is read as the probability that the $i$-th pair leaks at level $\gamma_i$, then the product pair leaks at level $\gamma_1\gamma_2$ with probability at most $1-(1-E_{\gamma_1}(\rho_1\Vert\sigma_1))(1-E_{\gamma_2}(\rho_2\Vert\sigma_2))$. In \cref{sec:optimality}, we make this interpretation precise.

\begin{corollary}[Sub-Additivity of Datta--Leditzky Divergence] \label{Cor:sub_additivity_DL}
Let $\rho_i$ be quantum states, $\sigma_i$ be positive semidefinite operators, and $\delta_i\in[0,1)$ and $\eps_i\in(0,1)$ for $i\in\{1,2\}$. Then the DL divergences in~\eqref{eq:DL_div} satisfy
    \begin{align}
 \Dup{1-(1-\delta_1)(1-\delta_2)}(\rho_1 \otimes \rho_2 \Vert \sigma_1 \otimes \sigma_2)
 &\leq \Dup{\delta_1}(\rho_1\|\sigma_1) + \Dup{\delta_2}(\rho_2\|\sigma_2) ,\\
 \Dlo{\eps_1 \eps_2}(\rho_1 \otimes \rho_2 \Vert \sigma_1 \otimes \sigma_2)
 &\leq\Dlo{\ee_1}(\rho_1\|\sigma_1) + \Dlo{\ee_2}(\rho_2\|\sigma_2).
\end{align}
\end{corollary}
\begin{proof}
For a state $\rho$ and a positive semidefinite operator $\sigma$, the function $\gamma\mapsto\Tr[(\rho-e^\gamma\sigma)_+]$ on $\mathbb R$, which appears in~\eqref{eq:DL_div}, is nonincreasing. By~\eqref{eq:m-positive-part}, $m(\rho,e^\gamma\sigma)=1-\Tr[(\rho-e^\gamma\sigma)_+]$ for every $\gamma\in\mathbb R$. Hence \cref{thm:overlap} with $A=\rho_1$, $B=e^{\gamma_1}\sigma_1$, $C=\rho_2$, and $D=e^{\gamma_2}\sigma_2$ gives, for all $\gamma_1,\gamma_2\in\mathbb R$,
\begin{equation}
1-\operatorname{Tr}\!\left[
  \left(\rho_1\otimes\rho_2
  -e^{\gamma_1+\gamma_2}\sigma_1\otimes\sigma_2\right)_+
\right]
\ge
\left(1-\operatorname{Tr}\!\left[(\rho_1-e^{\gamma_1}\sigma_1)_+\right]\right)
\left(1-\operatorname{Tr}\!\left[(\rho_2-e^{\gamma_2}\sigma_2)_+\right]\right).
\label{eq:DL-key}
\end{equation}

For the first inequality, if either term on the right-hand side is $+\infty$, the claim is immediate. Otherwise, consider $\gamma_1,\gamma_2\in\mathbb R$ such that
\begin{equation}
    \Tr[(\rho_i-e^{\gamma_i} \sigma_i)_+] \leq \delta_i\qquad\text{for }i\in\{1,2\}.
\end{equation}
Then~\eqref{eq:DL-key} gives
\begin{equation}
     \Tr\!\left[\left(\rho_1 \otimes \rho_2-e^{\gamma_1+\gamma_2}\sigma_1 \otimes \sigma_2\right)_+\right] \leq 1-(1-\delta_1) (1-\delta_2),
\end{equation}
so that $\gamma_1+\gamma_2$ is a feasible value for the optimization in~\eqref{eq:2nd-DL-def}. Thus
\begin{equation}
     \Dup{1-(1-\delta_1)(1-\delta_2)}(\rho_1 \otimes \rho_2 \Vert \sigma_1 \otimes \sigma_2)
     \leq \gamma_1+\gamma_2.
\end{equation}
Taking the infimum over all such $\gamma_1$ and $\gamma_2$, we arrive at
\begin{equation}
    \Dup{1-(1-\delta_1)(1-\delta_2)}(\rho_1 \otimes \rho_2 \Vert \sigma_1 \otimes \sigma_2) \leq \Dup{\delta_1}(\rho_1\|\sigma_1) + \Dup{\delta_2}(\rho_2\|\sigma_2),
\end{equation}
concluding the proof of the first inequality.

For the second inequality, write $d_i\coloneqq\Dlo{\eps_i}(\rho_i\Vert\sigma_i)$. Since $\eps_i>0$, each $d_i>-\infty$. If either $d_i=+\infty$, the claim is immediate. Otherwise, choose any $R_i>d_i$. By~\eqref{eq:1st-DL-def},
\[
\Tr[(\rho_i-e^{R_i}\sigma_i)_+]<1-\eps_i\qquad\text{for }i\in\{1,2\},
\]
and~\eqref{eq:DL-key} gives
\[
\Tr\!\left[\left(\rho_1\otimes\rho_2-e^{R_1+R_2}\sigma_1\otimes\sigma_2\right)_+\right]<1-\eps_1\eps_2.
\]
Since the function $\gamma\mapsto\Tr[(\rho_1\otimes\rho_2-e^\gamma\sigma_1\otimes\sigma_2)_+]$ is nonincreasing, no $R\ge R_1+R_2$ is feasible in the definition~\eqref{eq:1st-DL-def} of $\Dlo{\eps_1\eps_2}(\rho_1\otimes\rho_2\Vert\sigma_1\otimes\sigma_2)$. Hence
\[
\Dlo{\eps_1\eps_2}(\rho_1\otimes\rho_2\Vert\sigma_1\otimes\sigma_2)\le R_1+R_2.
\]
Letting $R_i\downarrow d_i$ completes the proof without any support assumption.
\end{proof}

\subsection{\texorpdfstring{$n$}{n} Product States}

Iterating the two-factor inequalities from the previous subsection gives bounds for an arbitrary number of product states.

\begin{proposition}\label{Prop:n_HS_subADD}
    Let $\rho_i$ and $\sigma_i$ be states, $\gamma_i\ge1$ for all $i \in\{1,\ldots,n\}$, and set $\Gamma\coloneqq\prod_i\gamma_i$. Then the hockey-stick divergence in~\eqref{eq:HS-def} satisfies
\begin{equation}
 E_{\Gamma}\!\left(\rhot\middle\|\sigt\right)
 \leq 1-\prod_{i=1}^n\left(1-E_{\gamma_i}(\rho_i\|\sigma_i)\right)\le\sum_{i=1}^nE_{\gamma_i}(\rho_i\|\sigma_i).
 \label{eq:subadd_HS}
\end{equation}
\end{proposition}

\begin{proof}
For $n=2$, the first inequality follows by~\cref{Cor:sub_additivity_HS}.
Assuming that the first inequality holds for $n-1$ factors, and together with the $n=2$ case, we have the following by considering $n$ tensor products as two tensor products of $n-1$ and one state:
\begin{align}
     &E_{\Gamma}\!\left(\rhot\middle\|\sigt\right) \nonumber \\
     & \leq 1- (1-E_{\gamma_n}(\rho_n \Vert \sigma_n)) \left(1- E_{\prod_{i=1}^{n-1}\gamma_i}(\rho_1 \otimes \cdots \otimes \rho_{n-1} \Vert \sigma_1 \otimes \cdots \otimes \sigma_{n-1})  \right) \\
     &\leq 1- (1-E_{\gamma_n}(\rho_n \Vert \sigma_n)) \left( 1-\left(1-\prod_{i=1}^{n-1}\left(1-E_{\gamma_i}(\rho_i\|\sigma_i)\right) \right) \right) \\
     &=  1-\prod_{i=1}^n\left(1-E_{\gamma_i}(\rho_i\|\sigma_i)\right).
\end{align}
This concludes the proof of the first inequality for all $n\in \mathbb{N}$ by induction. The second inequality follows from $1-\prod_i(1-x_i)\le\sum_ix_i$ for $x_i\in[0,1]$.
\end{proof}

\begin{figure}[ht]
\centering
\begin{tikzpicture}
\begin{groupplot}[group style={group size=2 by 2, horizontal sep=15mm, vertical sep=23mm}, vizaxis, width=0.48\textwidth, height=0.31\textwidth, xmin=0.5, xmax=10.5, xtick={2,4,...,10}, ylabel={trace distance}]
\nextgroupplot[ymin=0, ymax=1.05, xtick={1,2,...,10}, xlabel={number of copies $n$}, title={(a) $T(\rho^{\otimes n},\sigma^{\otimes n})$}]
\addplot[vizorange, densely dashed, mark=triangle*, mark size=1.6pt] coordinates {(1,0.26926) (2,0.53852) (3,0.80777) (4,1.00000) (5,1.00000) (6,1.00000) (7,1.00000) (8,1.00000) (9,1.00000) (10,1.00000)};
\addplot[vizaqua, mark=square*, mark size=2.2pt] coordinates {(1,0.26926) (2,0.46602) (3,0.60980) (4,0.71486) (5,0.79164) (6,0.84774) (7,0.88874) (8,0.91870) (9,0.94059) (10,0.95659)};
\addplot[vizblue, mark=*, mark size=1.4pt] coordinates {(1,0.26926) (2,0.46602) (3,0.60980) (4,0.71486) (5,0.79164) (6,0.84774) (7,0.88874) (8,0.91870) (9,0.94059) (10,0.95659)};
\addplot[vizgray, only marks, mark=diamond*, mark size=2pt] coordinates {(1,0.26926) (2,0.46057) (3,0.60067) (4,0.70410) (5,0.78067) (6,0.83743) (7,0.87951) (8,0.91071) (9,0.93384) (10,0.95099)};
\nextgroupplot[ymin=0, ymax=1.05, xlabel={number of systems $n=2k$}, title={(b) $T(\rho^{\otimes k}\otimes\sigma^{\otimes k},\sigma^{\otimes k}\otimes\rho^{\otimes k})$}, legend to name=tdlegend, legend columns=2, legend style={/tikz/every even column/.append style={column sep=5mm}}]
\addplot[vizorange, densely dashed, mark=triangle*, mark size=1.6pt] coordinates {(2,0.53852) (4,1.00000) (6,1.00000) (8,1.00000) (10,1.00000)};
\addlegendentry{triangle inequality: $\min\{1,\sum_iT_i\}$}
\addplot[vizblue, mark=*, mark size=1.4pt] coordinates {(2,0.46602) (4,0.71486) (6,0.84774) (8,0.91870) (10,0.95659)};
\addlegendentry{\eqref{eq:trace-distance-subadd} ($\gamma_i=1$): $1-\prod_i(1-T_i)$}
\addplot[vizaqua, mark=square*, mark size=1.4pt] coordinates {(2,0.27885) (4,0.47994) (6,0.62495) (8,0.72953) (10,0.80495)};
\addlegendentry{\cref{cor:profile}, optimized over $\gamma_1,\ldots,\gamma_n$}
\addplot[vizgray, only marks, mark=diamond*, mark size=2pt] coordinates {(2,0.27821) (4,0.47821) (6,0.62230) (8,0.72630) (10,0.80148)};
\addlegendentry{exact}
\nextgroupplot[ymin=0, ymax=0.42, ytick={0,0.1,0.2,0.3,0.4}, xtick={1,2,...,10}, xlabel={number of copies $n$}, ylabel={bound $-$ exact}, title={(c) random pair, $n$ copies}]
\addplot[draw=none, fill=vizorange, fill opacity=0.15, forget plot] coordinates {(1,0.0000) (2,0.3233) (3,0.3817) (4,0.3784) (5,0.3575) (6,0.3293) (7,0.3071) (8,0.2841) (9,0.2644) (10,0.2468) (10,0.0000) (9,0.0000) (8,0.0000) (7,0.0000) (6,0.0106) (5,0.0320) (4,0.0648) (3,0.1094) (2,0.1307) (1,0.0000)} -- cycle;
\addplot[draw=none, fill=vizblue, fill opacity=0.15, forget plot] coordinates {(1,0.0000) (2,0.1505) (3,0.1895) (4,0.2116) (5,0.2205) (6,0.2212) (7,0.2174) (8,0.2115) (9,0.2044) (10,0.1964) (10,0.0000) (9,0.0000) (8,0.0009) (7,0.0110) (6,0.0237) (5,0.0381) (4,0.0530) (3,0.0650) (2,0.0513) (1,0.0000)} -- cycle;
\addplot[draw=none, fill=vizaqua, fill opacity=0.15, forget plot] coordinates {(1,0.0000) (2,0.1265) (3,0.1751) (4,0.2008) (5,0.2124) (6,0.2150) (7,0.2127) (8,0.2078) (9,0.2015) (10,0.1941) (10,0.0000) (9,0.0000) (8,0.0008) (7,0.0105) (6,0.0224) (5,0.0355) (4,0.0485) (3,0.0577) (2,0.0481) (1,0.0000)} -- cycle;
\addplot[vizorange, densely dashed, mark=triangle*, mark size=1.6pt] coordinates {(1,0.0000) (2,0.2270) (3,0.2456) (4,0.2216) (5,0.1948) (6,0.1699) (7,0.1502) (8,0.1333) (9,0.1190) (10,0.1070)};
\addplot[vizblue, mark=*, mark size=1.4pt] coordinates {(1,-0.0000) (2,0.1009) (3,0.1273) (4,0.1323) (5,0.1293) (6,0.1224) (7,0.1142) (8,0.1062) (9,0.0980) (10,0.0905)};
\addplot[vizaqua, mark=square*, mark size=1.4pt] coordinates {(1,0.0000) (2,0.0873) (3,0.1164) (4,0.1246) (5,0.1240) (6,0.1187) (7,0.1116) (8,0.1043) (9,0.0967) (10,0.0895)};
\nextgroupplot[ymin=0, ymax=0.42, ytick={0,0.1,0.2,0.3,0.4}, xtick={1,2,...,10}, xlabel={number of pairs $n$}, ylabel={bound $-$ exact}, title={(d) $n$ independent random pairs}]
\addplot[draw=none, fill=vizorange, fill opacity=0.15, forget plot] coordinates {(1,0.0000) (2,0.3147) (3,0.3400) (4,0.2642) (5,0.1907) (6,0.1493) (7,0.1169) (8,0.0879) (9,0.0696) (10,0.0527) (10,0.0061) (9,0.0087) (8,0.0127) (7,0.0203) (6,0.0319) (5,0.0504) (4,0.0743) (3,0.1253) (2,0.1590) (1,0.0000)} -- cycle;
\addplot[draw=none, fill=vizblue, fill opacity=0.15, forget plot] coordinates {(1,0.0000) (2,0.1309) (3,0.1611) (4,0.1587) (5,0.1398) (6,0.1210) (7,0.1014) (8,0.0800) (9,0.0654) (10,0.0505) (10,0.0066) (9,0.0097) (8,0.0144) (7,0.0224) (6,0.0336) (5,0.0488) (4,0.0652) (3,0.0774) (2,0.0566) (1,0.0000)} -- cycle;
\addplot[draw=none, fill=vizaqua, fill opacity=0.15, forget plot] coordinates {(1,0.0000) (2,0.0850) (3,0.1098) (4,0.1124) (5,0.1058) (6,0.0952) (7,0.0826) (8,0.0681) (9,0.0572) (10,0.0453) (10,0.0066) (9,0.0092) (8,0.0129) (7,0.0191) (6,0.0260) (5,0.0339) (4,0.0397) (3,0.0400) (2,0.0262) (1,0.0000)} -- cycle;
\addplot[vizorange, densely dashed, mark=triangle*, mark size=1.6pt] coordinates {(1,0.0000) (2,0.2369) (3,0.2326) (4,0.1692) (5,0.1206) (6,0.0906) (7,0.0686) (8,0.0503) (9,0.0392) (10,0.0294)};
\addplot[vizblue, mark=*, mark size=1.4pt] coordinates {(1,0.0000) (2,0.0937) (3,0.1193) (4,0.1120) (5,0.0943) (6,0.0773) (7,0.0619) (8,0.0472) (9,0.0375) (10,0.0286)};
\addplot[vizaqua, mark=square*, mark size=1.4pt] coordinates {(1,0.0000) (2,0.0556) (3,0.0749) (4,0.0761) (5,0.0699) (6,0.0606) (7,0.0508) (8,0.0405) (9,0.0332) (10,0.0260)};
\end{groupplot}
\end{tikzpicture}

\vspace{2mm}
\ref{tdlegend}
\caption[Upper bounds on the trace distance of product states]{Upper bounds on the trace distance $T(\bigotimes_i\rho_i,\bigotimes_i\sigma_i)$, where $T_i\coloneqq T(\rho_i,\sigma_i)$; the exact values are obtained by diagonalization. (a) $n$ copies of the qubit pair $(\rho,\sigma)$ given in~\eqref{eq:rho_sigma_pair_eg}. 
(b) $k$ pairs $(\rho,\sigma)$ and $k$ pairs $(\sigma,\rho)$ of the same states. 
(c) $n$ copies of one random pair. (d) $n$ independent random pairs. In (c) and~(d), each qubit state is drawn independently from the Hilbert--Schmidt measure, i.e., $GG^\dagger/\Tr[GG^\dagger]$ for a $2\times2$ matrix $G$ with independent standard complex Gaussian entries. The curves show the average over $300$ samples, and the shaded regions show the average $\pm$ one standard deviation, cut off at $0$ since each bound is at least the exact value. 
}
\label{fig:trace-distance}
\end{figure}
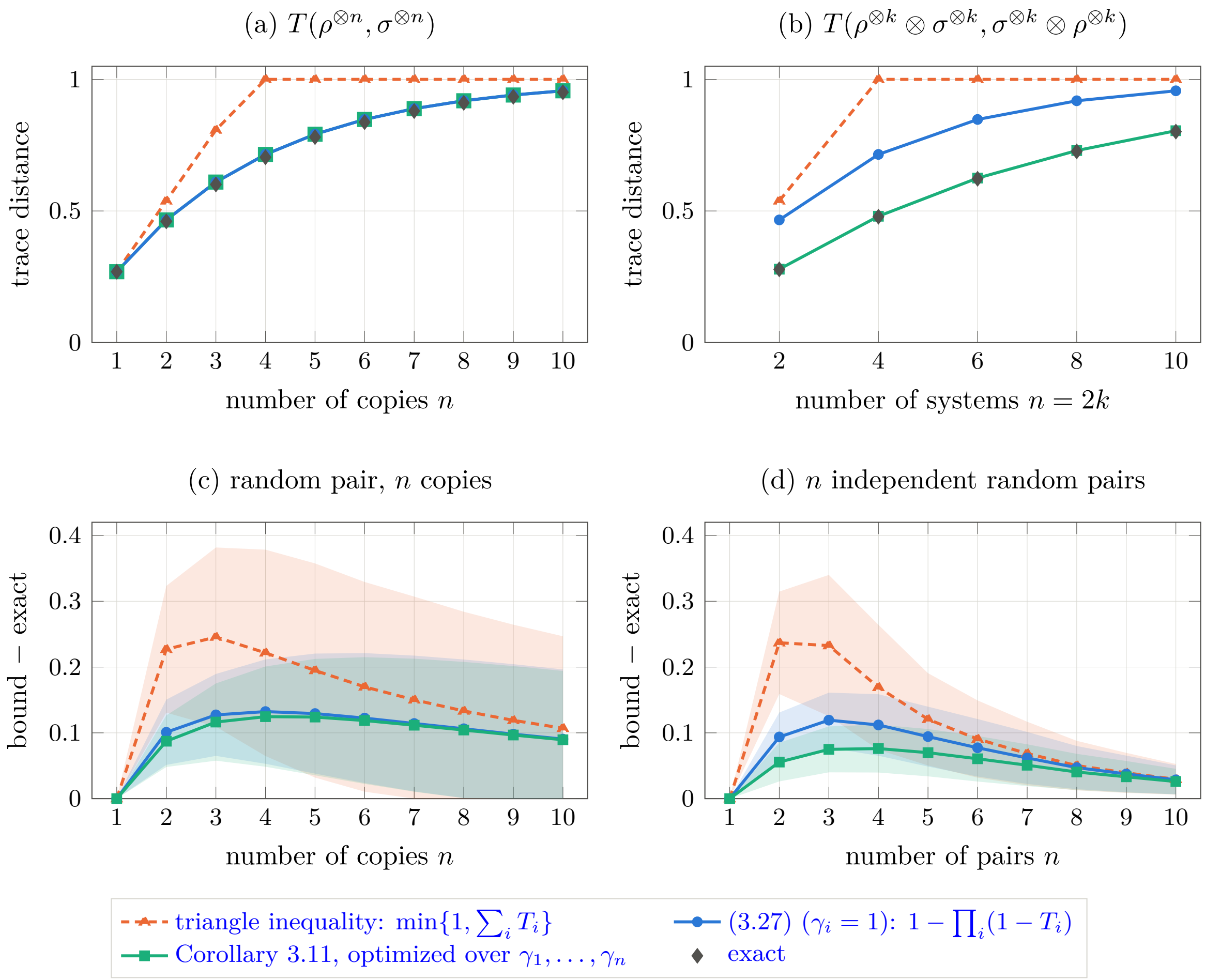

Since \cref{thm:overlap} holds for arbitrary positive weights, the factors $\gamma_i$ in \cref{Prop:n_HS_subADD} need not be at least one, and they can be chosen freely subject to $\prod_i\gamma_i=\Gamma$. Since the hockey-stick divergence is defined only for $\gamma\ge1$, we express the factors through the noncommutative minimum. This gives the following strengthening.

\begin{corollary}[Optimized Sub-Additivity]
\label{cor:profile}
Let $\rho_i$ and $\sigma_i$ be states for $i\in\{1,\ldots,n\}$, and let $\Gamma>0$. Then the noncommutative minimum in~\eqref{eq:m-def} satisfies
\begin{equation}
m\!\left(\rhot,\Gamma\,\sigt\right)\ge\sup\left\{\prod_{i=1}^nm(\rho_i,\gamma_i\sigma_i):\gamma_1,\ldots,\gamma_n>0,\ \prod_{i=1}^n\gamma_i=\Gamma\right\}.
\label{eq:profile}
\end{equation}
In particular, for $\Gamma\ge1$, the hockey-stick divergence in~\eqref{eq:HS-def} satisfies
\begin{equation}
E_\Gamma\!\left(\rhot\middle\|\sigt\right)\le1-\sup\left\{\prod_{i=1}^nm(\rho_i,\gamma_i\sigma_i):\gamma_1,\ldots,\gamma_n>0,\ \prod_{i=1}^n\gamma_i=\Gamma\right\},
\label{eq:profile-HS}
\end{equation}
and, for $\Gamma=1$ and $\gamma_i=1$, the trace distance satisfies
\begin{equation}
T\!\left(\rhot,\sigt\right)\le1-\prod_{i=1}^n\left(1-T(\rho_i,\sigma_i)\right)\le\sum_{i=1}^nT(\rho_i,\sigma_i).
\label{eq:trace-distance-subadd}
\end{equation}
\end{corollary}

\begin{proof}
Fix $\gamma_1,\ldots,\gamma_n>0$ with $\prod_i\gamma_i=\Gamma$. Since $\Gamma\,\sigt=\bigotimes_i(\gamma_i\sigma_i)$, applying \cref{thm:overlap} $n-1$ times gives
\[
m\!\left(\rhot,\Gamma\,\sigt\right)=m\!\left(\textstyle\bigotimes_i\rho_i,\bigotimes_i(\gamma_i\sigma_i)\right)\ge\prod_{i=1}^nm(\rho_i,\gamma_i\sigma_i).
\]
Taking the supremum over $(\gamma_i)_i$ proves~\eqref{eq:profile}. For $\Gamma\ge1$, \eqref{eq:m-hsd} gives $m(\rhot,\Gamma\,\sigt)=1-E_\Gamma(\rhot\Vert\sigt)$, which proves~\eqref{eq:profile-HS}. Finally, \eqref{eq:trace-distance-subadd} follows from~\eqref{eq:profile-HS} with $\gamma_i=1$, since $m(\rho_i,\sigma_i)=1-T(\rho_i,\sigma_i)$ by~\eqref{eq:m-hsd}, and from $1-\prod_i(1-x_i)\le\sum_ix_i$ for $x_i\in[0,1]$.
\end{proof}

The factors $\gamma_i<1$ in~\eqref{eq:profile-HS} can help. For example, for $\rho=\operatorname{diag}(2/3,1/3)$ and $\sigma=\operatorname{diag}(1/3,2/3)$, the choice $\gamma_1=\gamma_2=1$ gives $T(\rho^{\otimes2},\sigma^{\otimes2})\le5/9$, whereas $\gamma_1=2$ and $\gamma_2=1/2$ give $m(\rho,2\sigma)=1$ and $m(\rho,\frac12\sigma)=\frac12$ by~\eqref{eq:m-commuting}, and hence the better bound $1-1\cdot\frac12=\frac12$. \Cref{fig:trace-distance} compares these bounds with the bound $\sum_iT(\rho_i,\sigma_i)$ from the triangle inequality and with the exact trace distance. In \cref{fig:trace-distance}(a) and~(b), we use the noncommuting qubit states 
\begin{equation}
    \rho=\frac1{10}\begin{pmatrix}9&3\\3&1\end{pmatrix}, \quad 
\sigma=\frac1{20}\begin{pmatrix}13&4\\4&7\end{pmatrix}, \label{eq:rho_sigma_pair_eg}
\end{equation}
where $\rho=|\psi\rangle\!\langle\psi|$ is pure with $|\psi\rangle=(3|0\rangle+|1\rangle)/\sqrt{10}$, and $T\coloneqq T(\rho,\sigma)=\sqrt{29}/20$. For $n$ copies of this pair, the best choice in~\eqref{eq:profile} is $\gamma_i=1$ (numerically), so that the optimized bound coincides with~\eqref{eq:trace-distance-subadd}. For the states $\rho^{\otimes k}\otimes\sigma^{\otimes k}$ and $\sigma^{\otimes k}\otimes\rho^{\otimes k}$ on $n=2k$ systems, we choose $\gamma_i=\gamma$ for the $k$ pairs $(\rho,\sigma)$ and $\gamma_i=1/\gamma$ for the $k$ pairs $(\sigma,\rho)$. Since $m$ is symmetric and homogeneous, $m(\sigma,\rho/\gamma)=m(\rho,\gamma\sigma)/\gamma$, and~\eqref{eq:profile} gives $T\!\left(\rho^{\otimes k}\otimes\sigma^{\otimes k},\sigma^{\otimes k}\otimes\rho^{\otimes k}\right)\le1-\left(\frac{m(\rho,\gamma\sigma)^2}{\gamma}\right)^k.$
The best choice is $\gamma=\langle\psi|\sigma^{-1}|\psi\rangle=104/75$, the smallest $\gamma$ with $\rho\le\gamma\sigma$, for which $m(\rho,\gamma\sigma)=1$ and the bound becomes $1-(75/104)^k$. Numerically, this choice is optimal among all $\gamma_1,\ldots,\gamma_n$, and the resulting bound is much better than~\eqref{eq:trace-distance-subadd} and close to the exact trace distance.

\Cref{fig:trace-distance}(c) and~(d) show the difference between each bound and the exact trace distance for random qubit states, with its average and standard deviation. There, the optimized bound in~\eqref{eq:profile} is evaluated numerically, by maximizing $\sum_i\ln m(\rho_i,\gamma_i\sigma_i)$ subject to $\sum_i\ln\gamma_i=0$ over a grid of $\ln\gamma_i\in[-6,6]$ with step size $0.02$. Restricting the $\gamma_i$ to a grid can only weaken the bound, so the plotted values are still valid upper bounds. In both cases, the bound~\eqref{eq:trace-distance-subadd} is much better than the triangle inequality, which grows linearly in $n$ until it reaches $1$, and it is close to the optimized bound. Since~\eqref{eq:trace-distance-subadd} only requires the values $T(\rho_i,\sigma_i)$ and no optimization over $\gamma_1,\ldots,\gamma_n$, it is a good choice when this optimization is hard.

\begin{proposition}\label{prop:n_DL}
    Let $\rho_i$ be states and $\sigma_i$ be positive semidefinite operators for all $i \in\{1,\ldots,n\}$, and let $\delta_i\in[0,1)$ and $\eps_i\in(0,1)$. Then the DL divergences in~\eqref{eq:DL_div} satisfy
\begin{align}
 \Dup{1-\prod_i(1-\delta_i)}\!\left(\rhot\middle\|\sigt\right)
 &\leq\sum_{i=1}^n\Dup{\delta_i}(\rho_i\|\sigma_i),\label{eq:intro-upper}\\
 \Dlo{\prod_i\ee_i}\!\left(\rhot\middle\|\sigt\right)
 &\leq\sum_{i=1}^n\Dlo{\ee_i}(\rho_i\|\sigma_i).\label{eq:intro-lower}
\end{align}
\end{proposition}
\begin{proof}
    The proof follows by repeatedly applying~\cref{Cor:sub_additivity_DL}. In particular, by starting with
    \begin{align}
         &\Dup{1-(1-\delta_1) \prod_{i=2}^n(1-\delta_i)}\!\left(\rhot\middle\|\sigt\right) \nonumber \\
 &\leq\Dup{\delta_1}(\rho_1 \Vert \sigma_1) + \Dup{1- \prod_{i=2}^n(1-\delta_i)}\!\left(\rho_2 \otimes \cdots \otimes \rho_{n} \middle\|\sigma_2 \otimes \cdots \otimes \sigma_{n} \right)
    \end{align}
and
 \begin{align}
         &\Dlo{\eps_1 \prod_{i=2}^n \eps_i}\!\left(\rhot\middle\|\sigt\right) \nonumber \\
 &\leq\Dlo{\eps_1}(\rho_1 \Vert \sigma_1) + \Dlo{\prod_{i=2}^n\eps_i}\!\left(\rho_2 \otimes \cdots \otimes \rho_{n} \middle\|\sigma_2 \otimes \cdots \otimes \sigma_{n} \right)
    \end{align}
and decomposing the tensor product terms one by one by applying~\cref{Cor:sub_additivity_DL} at each step concludes the proof.
\end{proof}

\subsection{Optimality}
\label{sec:optimality}

We first show that the bound in \cref{Prop:n_HS_subADD} cannot be improved as a function of the values $E_{\gamma_i}(\rho_i\Vert\sigma_i)$, and that equality can be attained by noncommuting states.

\begin{proposition}[Equality Cases]\label{prop:optimality}
For $i\in\{1,\ldots,n\}$, let $\sigma_i$ be a state, $\gamma_i\ge1$, $e_i\in[0,1]$, and let $\omega_i$ and $\tau_i$ be states such that
\begin{equation}
\supp(\tau_i)\subseteq\ker(\sigma_i),
\qquad
(1-e_i)\,\omega_i\le\gamma_i\sigma_i.
\label{eq:equality-conditions}
\end{equation}
Set $\rho_i\coloneqq(1-e_i)\omega_i+e_i\tau_i$. Then $E_{\gamma_i}(\rho_i\Vert\sigma_i)=e_i$ for every $i$, and equality holds in the first inequality of~\eqref{eq:subadd_HS}. In particular, for every $n$, all $\gamma_i\ge1$, and all values $e_i\in[0,1]$, equality holds for the commuting states $\sigma_i=\proj0$ and $\rho_i=(1-e_i)\proj0+e_i\proj1$.
\end{proposition}
\begin{proof}
By the definition of $\rho_i$,
\[
\rho_i-\gamma_i\sigma_i=\big[(1-e_i)\omega_i-\gamma_i\sigma_i\big]+e_i\tau_i.
\]
The first operator is negative semidefinite by~\eqref{eq:equality-conditions} and supported in $\supp(\sigma_i)$, since $\supp(\omega_i)\subseteq\supp(\sigma_i)$ if $e_i<1$, and the operator is $-\gamma_i\sigma_i$ if $e_i=1$. The second operator is positive semidefinite and supported in $\ker(\sigma_i)$. Since the two supports are orthogonal, the positive part of $\rho_i-\gamma_i\sigma_i$ is $e_i\tau_i$, and $E_{\gamma_i}(\rho_i\Vert\sigma_i)=e_i$.

For the tensor product, let $[n]\coloneqq\{1,\ldots,n\}$. Expanding $\rhot$ gives
\[
\rhot-\Gamma\,\sigt=\left[\bigotimes_{i=1}^n(1-e_i)\omega_i-\bigotimes_{i=1}^n\gamma_i\sigma_i\right]+\sum_{\varnothing\neq S\subseteq[n]}\left(\prod_{i\in S}e_i\right)\bigotimes_{i\notin S}(1-e_i)\omega_i\otimes\bigotimes_{i\in S}\tau_i,
\]
where the tensor factors are ordered according to $i$. The operator in the square brackets is negative semidefinite, because $0\le X_i\le Y_i$ for all $i$ implies $\bigotimes_iX_i\le\bigotimes_iY_i$, and it is supported in $\bigotimes_i\supp(\sigma_i)$. For every $S\neq\varnothing$, the corresponding term is positive semidefinite and supported in a subspace in which at least one factor lies in $\ker(\sigma_i)$, so it is orthogonal to $\bigotimes_i\supp(\sigma_i)$. Therefore, the positive part of $\rhot-\Gamma\,\sigt$ is the sum over $S\neq\varnothing$, whose trace is $\prod_i[(1-e_i)+e_i]-\prod_i(1-e_i)=1-\prod_i(1-e_i)$. This proves the equality. For the last claim, choose $\omega_i=\sigma_i=\proj0$ and $\tau_i=\proj1$, which satisfy~\eqref{eq:equality-conditions} since $1-e_i\le1\le\gamma_i$.
\end{proof}

\begin{example}[Equality for Noncommuting States]
\label{ex:noncommuting-equality}
On $\mathbb C^3$, let $\sigma=\operatorname{diag}(2/3,1/3,0)$, $\omega=\proj{+}$ with $\ket{+}\coloneqq(\ket0+\ket1)/\sqrt2$, and $\tau=\proj2$. Then $\langle+|(\sigma|_{\supp(\sigma)})^{-1}|+\rangle=9/4$, so that $(1-e)\omega\le\sigma$ for every $e\in[5/9,1]$. By \cref{prop:optimality}, for all such $e_1,\ldots,e_n$ and all $\gamma_i\ge1$, the states $\rho_i=(1-e_i)\omega+e_i\tau$ and $\sigma_i=\sigma$ attain equality in~\eqref{eq:subadd_HS}, and $[\rho_i,\sigma_i]\neq0$ whenever $e_i<1$.
\end{example}

The pairs in \cref{prop:optimality} have the following structure: the part $e_i\tau_i$ of $\rho_i$ that is responsible for $E_{\gamma_i}(\rho_i\Vert\sigma_i)$ is supported outside the support of $\sigma_i$, so that it is detected without error, while the remaining part satisfies the privacy constraint $(1-e_i)\omega_i\le\gamma_i\sigma_i$. This is the quantum version of the classical mechanisms that fail with probability $\delta$ and are otherwise private (see \cref{rem:QDP-tight}). The next result shows that, for faithful states, equality is impossible as soon as at least two of the values $E_{\gamma_i}(\rho_i\Vert\sigma_i)$ are positive. Together with \cref{prop:optimality}, it describes when equality can and cannot hold, but it does not give a complete characterization of the equality cases.

\begin{proposition}[Strict Inequality for Faithful States]
\label{prop:strict}
Let $\rho_1,\rho_2,\sigma_1,\sigma_2$ be faithful states and $\gamma_1,\gamma_2\ge1$ such that $E_{\gamma_1}(\rho_1\Vert\sigma_1)>0$ and $E_{\gamma_2}(\rho_2\Vert\sigma_2)>0$. Then the inequality in~\eqref{eq:HS-two} is strict. More generally, if $\rho_1,\ldots,\rho_n,\sigma_1,\ldots,\sigma_n$ are faithful and $E_{\gamma_i}(\rho_i\Vert\sigma_i)>0$ for at least two indices $i$, then the first inequality in~\eqref{eq:subadd_HS} is strict.
\end{proposition}

\begin{proof}
Let $A\coloneqq\rho_1$, $B\coloneqq\gamma_1\sigma_1$, $C'\coloneqq\rho_2$, and $D\coloneqq\gamma_2\sigma_2$, and suppose that equality holds in~\eqref{eq:HS-two}, i.e., $m(A\otimes C',B\otimes D)=m(A,B)\,m(C',D)$. Apply \cref{lem:decomposition_general_to_rank} to both pairs.
Suppose that $(A-B)_+\nleq A$ and $(C'-D)_+\nleq C'$. We can choose exact decompositions of $(A, B)$ and $(C',D)$, each containing a pair of nonzero, nonproportional operators with positive noncommutative minimum. Applying the strict part of \cref{lem:rank-one} to these two pairs makes~\eqref{eq:supermult-step2} strict, a contradiction. Thus at least one of $(A-B)_+\leq A$ and $(C'-D)_+\leq C'$ holds. By symmetry, we may assume that $(C'-D)_+\leq C'$. Set $X=(\rho_2-\gamma_2\sigma_2)_+, Y=(\rho_2-\gamma_2\sigma_2)_-$, and $C=C'-X\geq0$. Then $C'=C+X$ and $D=C+Y$.
Since $X$ and $Y$ have orthogonal supports, $\lVert X-Y\rVert_1=\Tr X+\Tr Y$, so that $m(C',D)=\Tr C$. Moreover,
\[
A\otimes C'-B\otimes D=(A-B)\otimes C+A\otimes X-B\otimes Y,
\]
and the triangle inequality gives $\lVert A\otimes C'-B\otimes D\rVert_1\le\lVert A-B\rVert_1\Tr C+\Tr A\Tr X+\Tr B\Tr Y$, which is equivalent to $m(A\otimes C',B\otimes D)\ge m(A,B)\Tr C$. Thus equality requires equality in this triangle inequality.

For Hermitian operators $T_1,T_2,T_3$, the equality $\lVert\sum_kT_k\rVert_1=\sum_k\lVert T_k\rVert_1$ holds if and only if the supports of all the positive parts $(T_k)_+$ are orthogonal to the supports of all the negative parts $(T_k)_-$. Indeed, with $J\coloneqq\sgn(\sum_kT_k)$, we have that $\lVert\sum_kT_k\rVert_1=\sum_k\Tr[JT_k]$ and $\Tr[JT_k]\le\lVert T_k\rVert_1$, with equality if and only if $J$ acts as the identity on $\supp((T_k)_+)$ and as minus the identity on $\supp((T_k)_-)$. Moreover, two subspaces of the form $U_1\otimes V_1$ and $U_2\otimes V_2$ with nonzero $U_k,V_k$ are orthogonal if and only if $U_1\perp U_2$ or $V_1\perp V_2$.

Now, $E_{\gamma_1}(\rho_1\Vert\sigma_1)>0$ gives $(A-B)_+\neq0$, and $(A-B)_-\neq0$, since otherwise $\rho_1\ge\gamma_1\sigma_1$ together with $\Tr\rho_1=1\le\gamma_1$ would give $\rho_1=\sigma_1$ and $E_{\gamma_1}(\rho_1\Vert\sigma_1)=0$. Similarly, $X\neq0$. Furthermore, $C\neq0$, since otherwise $\rho_2=X$ and $\gamma_2\sigma_2=Y$ would have orthogonal supports, which is impossible for faithful states. Applying the above criterion to $T_1=(A-B)\otimes C$, $T_2=A\otimes X$, and $T_3=-B\otimes Y$, and using that $\supp(A)$ and $\supp(B)$ are the whole space, we find that equality requires $\supp(X)\perp\supp(C)$ (from $T_2$ and $(T_1)_-$) and, if $Y\neq0$, $\supp(Y)\perp\supp(C)$ (from $T_3$ and $(T_1)_+$). Since also $\supp(X)\perp\supp(Y)$, it follows that $\supp(X)$ is orthogonal to $\supp(C+Y)=\supp(\gamma_2\sigma_2)$, which is the whole space. Hence $X=0$, a contradiction.

For the general statement, let $E_{\gamma_a}(\rho_a\Vert\sigma_a)>0$ and $E_{\gamma_b}(\rho_b\Vert\sigma_b)>0$ for some $a\neq b$. Applying \cref{Prop:n_HS_subADD} with the pair $(\rho_a\otimes\rho_b,\sigma_a\otimes\sigma_b)$ treated as a single system and the strict inequality for this pair gives the claim, because $1-E_{\gamma_i}(\rho_i\Vert\sigma_i)=m(\rho_i,\gamma_i\sigma_i)>0$ for faithful states by~\eqref{eq:m-hsd}, as $m(\rho_i,\gamma_i\sigma_i)=0$ would require $\rho_i$ and $\sigma_i$ to have orthogonal supports.
\end{proof}

\begin{remark}[Faithful States and Tensor Powers]
\label{rem:tightness-faithful}
Although equality with at least two positive values $E_{\gamma_i}(\rho_i\Vert\sigma_i)$ requires states that are not faithful (\cref{prop:strict}), the bound in~\eqref{eq:subadd_HS} cannot be improved as a function of the values $E_{\gamma_i}(\rho_i\Vert\sigma_i)$ even if we restrict to faithful states. Indeed, fix $\gamma_i\ge1$ and $e_i\in[0,1)$, let $\pi\coloneqq\id_2/2$, and, for $\eta\in(0,1)$, set $e_i'\coloneqq(e_i+\eta(\gamma_i-1)/2)/(1-\eta)$, which lies in $[0,1]$ for all sufficiently small $\eta$. Consider the depolarized versions
\[
\rho_i^{\eta}\coloneqq(1-\eta)\left[(1-e_i')\proj0+e_i'\proj1\right]+\eta\pi,
\qquad
\sigma^{\eta}\coloneqq(1-\eta)\proj0+\eta\pi
\]
of the commuting pairs in \cref{prop:optimality}. These states are faithful and diagonal, and $\rho_i^\eta-\gamma_i\sigma^\eta=\operatorname{diag}(1-\gamma_i-e_i,\,e_i)$, so that $E_{\gamma_i}(\rho_i^\eta\Vert\sigma^\eta)=e_i$ for every such $\eta$. As $\eta\to0$, these pairs converge to the pairs of \cref{prop:optimality} with the values $e_i$, for which~\eqref{eq:subadd_HS} is an equality. Since the hockey-stick divergence is continuous in its arguments, $E_{\Gamma}(\bigotimes_i\rho_i^\eta\Vert(\sigma^\eta)^{\otimes n})\to1-\prod_i(1-e_i)$, while the right-hand side of~\eqref{eq:subadd_HS} is equal to $1-\prod_i(1-e_i)$ for every $\eta$. For tensor powers $\rho^{\otimes n}$ and $\sigma^{\otimes n}$, \cref{Prop:n_HS_subADD} reads $E_{\gamma^n}(\rho^{\otimes n}\Vert\sigma^{\otimes n})\le1-(1-E_\gamma(\rho\Vert\sigma))^n$, and equality is attained by the tensor powers of the pairs in \cref{prop:optimality,ex:noncommuting-equality}. For faithful pairs, however, the bound can be far from tight for large $n$: if $D(\rho\Vert\sigma)<\ln\gamma<\Dmax(\rho\Vert\sigma)$, where $D$ is the quantum relative entropy, then $E_\gamma(\rho\Vert\sigma)>0$, so that the right-hand side tends to one, whereas $E_{\gamma^n}(\rho^{\otimes n}\Vert\sigma^{\otimes n})\to0$ as $n\to\infty$, since $\Dup{\delta}(\rho^{\otimes n}\Vert\sigma^{\otimes n})=nD(\rho\Vert\sigma)+O(\sqrt n)$ for every $\delta\in(0,1)$~\cite{datta2014second,tomamichel2013hierarchy,li2014second}. In other words, the bound captures the worst case, in which every leakage is catastrophic, and not the concentration that occurs for many copies of faithful states.
\end{remark}

Finally, the parameter $\gamma$ has to be multiplied.

\begin{example}[Failure at a Common Parameter $\gamma$]
\label{ex:same-error}
Let
\begin{equation}
 \rho=\operatorname{diag}(1/2,1/2),\qquad
 \sigma=\operatorname{diag}(9/10,1/10),\qquad\gamma=4.
\end{equation}
Then, we have that
\begin{equation}
    \rho-\gamma \sigma= \operatorname{diag}(-31/10, 1/10), \qquad \rho^{\otimes 2}-\gamma \sigma^{\otimes 2}= \operatorname{diag}(-299/100,-11/100,-11/100,21/100).
\end{equation}
This leads to
\begin{equation}
    2 E_\gamma(\rho\Vert \sigma)= \frac{2}{10} < \frac{21}{100}= E_\gamma\!\left( \rho^{\otimes 2} \middle \Vert \sigma^{\otimes 2} \right),
\end{equation}
which provides a counterexample for the sub-additivity of hockey-stick divergence at fixed $\gamma$.
\end{example}

\section{Applications}
\label{sec:applications}

We apply the preceding tensor-product inequalities to quantum differential privacy, hypothesis testing, and refined Chernoff asymptotics.

\subsection{Applications to Quantum Differential Privacy}
\label{sec:QDP}

Differential privacy was introduced to protect the data of individuals in a database while still allowing useful statistical analysis~\cite{dwork2006calibrating,dwork2014algorithmic}.
Quantum differential privacy (QDP) is a generalization of the classical framework, which guarantees that the outputs of a channel for any pair of input quantum states declared as neighbors are approximately indistinguishable under every measurement~\cite{QDP_computation17}. QDP has been related to gentle measurements and applied to learning quantum states~\cite{aaronson2019gentle}, studied in the local model~\cite{hirche2023quantum,angrisani_localModel25,guan2024optimal,nuradha2026privacyutilitytradeoffsQI}, and it has been generalized to quantum pufferfish privacy~\cite{nuradha_QPP,measuredHS25}. Next, we define QDP as follows:

\begin{definition}[\cite{QDP_computation17,hirche2023quantum}]\label{def:QDP}
    Fix $\varepsilon \geq 0$ and $\delta \in [0,1]$.
    Let $\mathcal{A}$ be a quantum channel. We say $\mathcal{A}$ is \emph{$(\varepsilon, \delta)$-quantum differentially private}, or $(\varepsilon,\delta)$-QDP, if
\begin{equation} \Tr\left[M \mathcal{A}(\rho)\right] \leq e^\varepsilon \Tr\left[M \mathcal{A}(\sigma)\right] + \delta
\label{eq:QDP-def}
\end{equation}
for all pairs of neighboring states $\rho \sim \sigma$ and measurement operators $M$ satisfying $0 \leq M \leq I.$
\end{definition}

Moreover, $(\varepsilon,\delta)$-QDP is equivalent to the following criterion on the hockey-stick divergence~\cite{hirche2023quantum}:
\begin{equation} \label{eq:QDP_HS_Equivalence}
    \sup_{\rho \sim \sigma} E_{e^\varepsilon}\!\left( \mathcal{A}(\rho) \Vert \mathcal{A}(\sigma)\right) \leq \delta.
\end{equation}

\subsubsection{Previously Known Composition Results}
\label{sec:QDP-prior}

The degradation of the privacy level under the composition of privacy mechanisms has been studied in~\cite{hirche2023quantum,nuradha_QPP,alabi2026quantum,guan2024optimal,bhalerao2026high,nuradha2025nonLinearSDPI}. One of the composition setups considered therein is parallel composition: the privacy level provided by $\mathcal{A}_1 \otimes \mathcal{A}_2$ over product neighboring inputs, while the adversary may use all measurements on the joint output. Here neighboring inputs are defined by $\rho_1 \otimes \rho_2 \sim \sigma_1 \otimes \sigma_2$ if and only if $\rho_i \sim \sigma_i$ for $i\in\{1,2\}$.

Let $\mathcal{A}_i$ be $(\varepsilon_i,\delta_i)$-QDP for $i\in\{1,2\}$ with $\delta_i \in[0,1)$. For general measurements, the following parallel composition results were known: $\mathcal{A}_1 \otimes \mathcal{A}_2$ satisfies
\begin{enumerate}
    \item  $(\varepsilon',\delta')$-QDP~\cite[Theorem~2]{nuradha_QPP},  where  \begin{align}
         \varepsilon'& \coloneqq \varepsilon_1 + \varepsilon_2 + \ln\!\left( \frac{1}{(1-\delta_1)(1-\delta_2)}\right), \\
         \delta'& \coloneqq \sqrt{\delta_1(2-\delta_1)} + \sqrt{\delta_2(2-\delta_2)};   \label{eq:qpp-parallel-composition}
    \end{align}
    \item $(\varepsilon_1 + \varepsilon_2, \delta)$-QDP~\cite[Corollary~III.3]{hirche2023quantum} (see also~\cite[Theorem~2]{nuradha_QPP}), where  \begin{equation} \label{eq:parallelcomposition-similar-QDP}
     \delta \coloneqq \min\{ \delta_1 +e^{\varepsilon_1} \delta_2, \delta_2 + e^{\varepsilon_2} \delta_1 \}.
 \end{equation}
\end{enumerate}
The second result follows from the sub-additivity bound~\eqref{eq:intro-HRF}, applied with either ordering of the two systems and with the roles of $\gamma_1$ and $\gamma_2$ interchanged. A parallel composition rule with parameters $(\varepsilon_1+\varepsilon_2,\delta_1+\delta_2)$ for general measurements was also claimed in~\cite[Theorem~5]{QDP_computation17}. However, as discussed in~\cite[Section~III]{hirche2023quantum}, the composition results of~\cite{QDP_computation17} rely on an erroneous assumption in~\cite[Lemma~1]{QDP_computation17}, so that~\cite[Corollary~III.3]{hirche2023quantum} was the first established parallel composition property for QDP.

For restricted classes of measurements, improved composition results are known. If the adversary is restricted to product measurements, or more generally to a class of separable measurements, then $\mathcal{A}_1 \otimes \mathcal{A}_2$ satisfies $(\varepsilon_1 +\varepsilon_2, \delta_1 + \delta_2)$-QDP~\cite[Theorem~1 and the discussion after it]{nuradha_QPP}. This result is formulated in the more general framework of quantum pufferfish privacy, and it requires each $\mathcal A_i$ to be private only with respect to the chosen class of measurements. Composition theorems of the classical type, including advanced composition, were obtained for local measurements followed by classical postprocessing in~\cite{alabi2026quantum}.
{For pure QDP, i.e., $\delta_i=0$, an advanced composition theorem was proved for tensor-product channels, product neighboring inputs, and global measurements in~\cite[Theorem~33]{alabi2026quantum}.}
The same work showed that composition can fail completely for joint channels that are not tensor products~\cite[Theorem~15]{alabi2026quantum}, and it noted that it was not known whether $\delta_1+\delta_2$ suffices for tensor-product channels and global measurements~\cite[Section~5.2]{alabi2026quantum}. \Cref{thm:parallel_composability_QDP} below shows that even $\delta_1+\delta_2-\delta_1\delta_2$ suffices. 

A different setting is quantum local differential privacy (QLDP), in which all pairs of input states are neighbors~\cite{hirche2023quantum}, so that the inputs of different parties may be entangled. 
{Composition in this setting was studied in~\cite{guan2024optimal}} and in the high privacy regime in~\cite{bhalerao2026high}, and sequential composition of QLDP channels was studied in~\cite{nuradha2025nonLinearSDPI}. In this work, we consider only product neighboring inputs.

\subsubsection{Parallel Composition}

Utilizing the sub-additivity of the hockey-stick divergence in \cref{Prop:n_HS_subADD}, we now prove a parallel composition result for QDP with general measurements. In particular, the following result establishes the claim of~\cite[Theorem~5]{QDP_computation17}.

\begin{theorem}[Parallel Composability of QDP] \label{thm:parallel_composability_QDP}
    Let $\mathcal{A}_i$ be $(\varepsilon_i,\delta_i)$-QDP in the sense of \cref{def:QDP}, with $\varepsilon_i \geq 0$ and $\delta_i \in [0,1]$ for all $i \in\{1,\ldots,n\}$. Then, $\mathcal{A}_1 \otimes \cdots \otimes \mathcal{A}_n$ is $(\varepsilon^*, \delta^*)$-QDP for product neighboring inputs (i.e., $\rho_1 \otimes \cdots \otimes \rho_n \sim \sigma_1 \otimes \cdots \otimes \sigma_n$ if $\rho_i \sim \sigma_i$ for all $i$), where
    \begin{align}
        \varepsilon^* &\coloneqq \sum_{i=1}^n \varepsilon_i ,\\
        \delta^* &\coloneqq 1- \prod_{i=1}^n (1-\delta_i) \leq \sum_{i=1}^n \delta_i.
    \end{align}
\end{theorem}
\begin{proof}
    The proof follows directly by using~\cref{Prop:n_HS_subADD} and~\eqref{eq:QDP_HS_Equivalence}.
\end{proof}

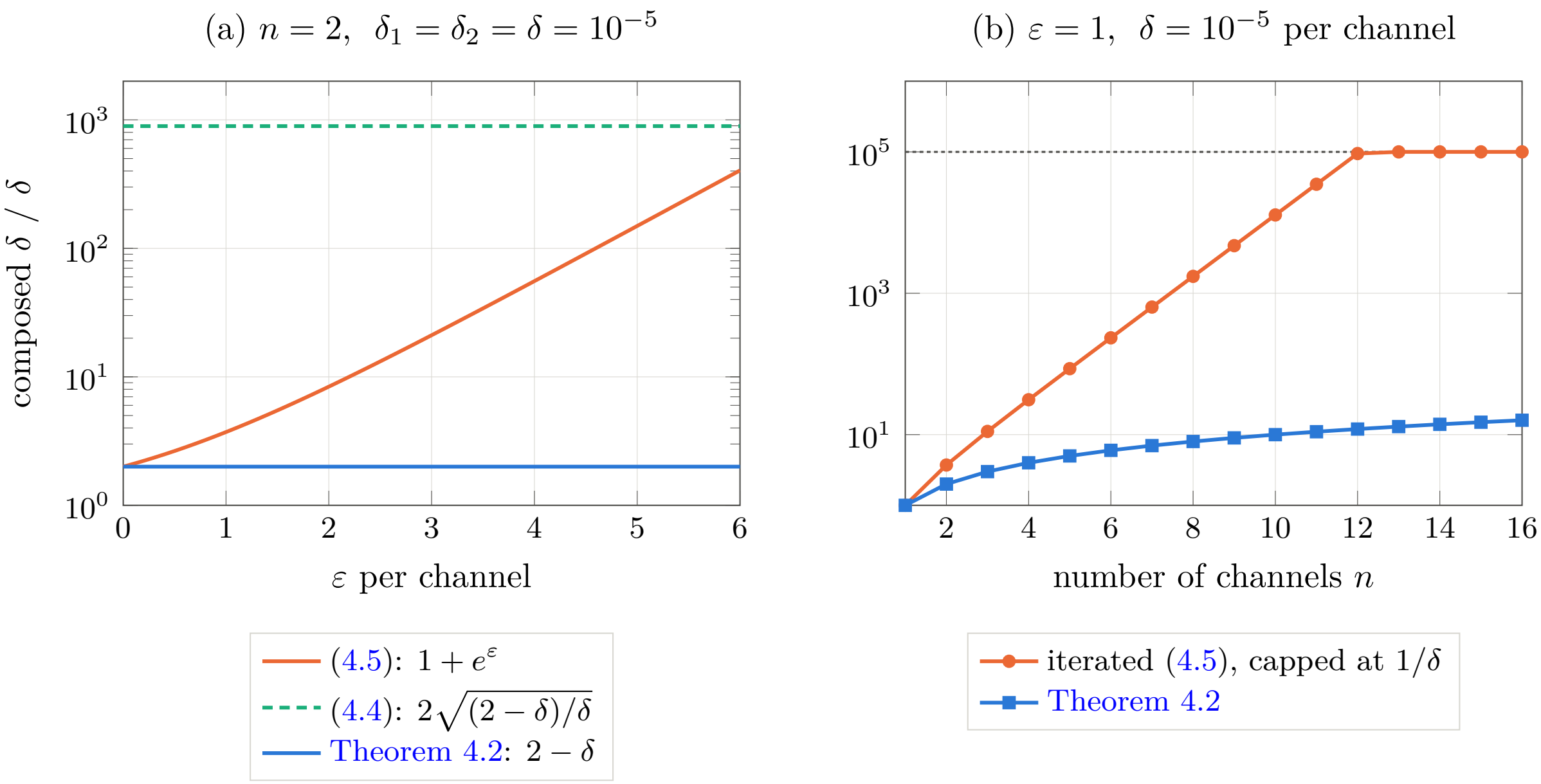
\begin{figure}[t]
\centering
\begin{tikzpicture}
\begin{groupplot}[group style={group size=2 by 1, horizontal sep=17mm}, vizaxis, width=0.48\textwidth, height=0.36\textwidth, ymode=log]
\nextgroupplot[xlabel={$\varepsilon$ per channel}, ylabel={composed $\delta$ / $\delta$}, xmin=0, xmax=6, ymin=1, ymax=2000, title={(a) $n=2$, \ $\delta_1=\delta_2=\delta=10^{-5}$}, legend style={at={(0.5,-0.3)}, anchor=north, legend columns=1}]
\addplot[vizorange, domain=0:6, samples=60] {1+exp(x)};
\addlegendentry{\eqref{eq:parallelcomposition-similar-QDP}: $1+e^{\varepsilon}$}
\addplot[vizaqua, densely dashed, domain=0:6, samples=2] {2*sqrt((2-0.00001)/0.00001)};
\addlegendentry{\eqref{eq:qpp-parallel-composition}: $2\sqrt{(2-\delta)/\delta}$}
\addplot[vizblue, domain=0:6, samples=2] {2-0.00001};
\addlegendentry{\cref{thm:parallel_composability_QDP}: $2-\delta$}
\nextgroupplot[xlabel={number of channels $n$}, xmin=1, xmax=16, ymin=1, ymax=1e6, title={(b) $\varepsilon=1$, \ $\delta=10^{-5}$ per channel}, legend style={at={(0.5,-0.3)}, anchor=north, legend columns=1}]
\addplot[vizgray, densely dotted, line width=0.6pt, domain=1:16, samples=2, forget plot] {1e5};
\addplot[vizorange, mark=*, mark size=1.4pt, domain=1:16, samples=16] {min((exp(x)-1)/(exp(1)-1),1e5)};
\addlegendentry{iterated \eqref{eq:parallelcomposition-similar-QDP}{, capped at $1/\delta$}}
\addplot[vizblue, mark=square*, mark size=1.4pt] coordinates {(1,1.00000) (2,1.99999) (3,2.99997) (4,3.99994) (5,4.99990) (6,5.99985) (7,6.99979) (8,7.99972) (9,8.99964) (10,9.99955) (11,10.99945) (12,11.99934) (13,12.99922) (14,13.99909) (15,14.99895) (16,15.99880)};
\addlegendentry{\cref{thm:parallel_composability_QDP}}
\end{groupplot}
\end{tikzpicture}
\caption{The $\delta$-parameter of the parallel composition of QDP channels with product neighbors, normalized by the per-channel $\delta$. (a) Two channels: the previous bound~\eqref{eq:parallelcomposition-similar-QDP} grows like $e^\varepsilon$ in the low privacy regime, and the bound~\eqref{eq:qpp-parallel-composition} gives $\delta'$ of order $\sqrt{\delta}$, so that $\delta'/\delta$ grows like $1/\sqrt\delta$ (and it comes with the larger $\varepsilon'=2\varepsilon+2\ln\frac1{1-\delta}$), whereas \cref{thm:parallel_composability_QDP} gives $2\delta-\delta^2$ for all $\varepsilon$. (b) $n$ channels at $\varepsilon=1$: iterating~\eqref{eq:parallelcomposition-similar-QDP} gives exponential growth $\delta(e^{n\varepsilon}-1)/(e^\varepsilon-1)$, {which exceeds one for $n\ge13$, so that the composed $\delta$-parameter is then only bounded by its trivial value $1$ (dotted line at $1/\delta=10^5$),} while \cref{thm:parallel_composability_QDP} gives $1-(1-\delta)^n\le n\delta$, which is optimal by \cref{rem:QDP-tight}.}
\label{fig:delta-composition}
\end{figure}

\cref{thm:parallel_composability_QDP} shows that the privacy degradation is at most linear when the neighboring inputs and the private channels are product, whereas the previous results allowed a much stronger degradation, especially in the low privacy regime where $\varepsilon_i \geq 1$. \Cref{fig:delta-composition} compares the $\delta$-parameters. Iterating~\eqref{eq:parallelcomposition-similar-QDP} over $n$ identical $(\varepsilon,\delta)$-QDP channels gives $\delta\sum_{k=0}^{n-1}e^{k\varepsilon}=\delta(e^{n\varepsilon}-1)/(e^\varepsilon-1)$, which grows exponentially in $n$, whereas $\delta^*=1-(1-\delta)^n\le n\delta$.

\begin{remark}[Optimality, Already for Classical Mechanisms]
\label{rem:QDP-tight}
The parameter $\delta^*$ in \cref{thm:parallel_composability_QDP} cannot be improved at $\varepsilon^*=\sum_i\varepsilon_i$, even for classical mechanisms. Consider the binary-input mechanism $\mathcal A$ that maps the two neighboring inputs to the diagonal states
\[
P\coloneqq\Big(\delta,\ \tfrac{(1-\delta)e^\varepsilon}{1+e^\varepsilon},\ \tfrac{1-\delta}{1+e^\varepsilon},\ 0\Big),
\qquad
Q\coloneqq\Big(0,\ \tfrac{1-\delta}{1+e^\varepsilon},\ \tfrac{(1-\delta)e^\varepsilon}{1+e^\varepsilon},\ \delta\Big),
\]
which is $(\varepsilon,\delta)$-QDP. For $n$ copies, the outcome strings that contain the first symbol have probability $1-(1-\delta)^n$ under $P^{\otimes n}$ and probability zero under $Q^{\otimes n}$, while all other strings have likelihood ratio at most $e^{n\varepsilon}$. Hence $E_{e^{n\varepsilon}}(P^{\otimes n}\Vert Q^{\otimes n})=1-(1-\delta)^n$. For different parameters $(\varepsilon_i,\delta_i)$, the tensor product of the corresponding mechanisms gives $E_{\exp(\sum_i\varepsilon_i)}(\bigotimes_iP_i\Vert\bigotimes_iQ_i)=1-\prod_i(1-\delta_i)$ by the same argument. This is the extreme point of the optimal composition theorem for classical differential privacy~\cite{kairouz2017composition}, so \cref{thm:parallel_composability_QDP} shows that collective quantum measurements do not degrade this point. These mechanisms are of the form in \cref{prop:optimality}.
\end{remark}

\subsection{Applications to Asymmetric Hypothesis Testing}
\label{sec:HT}

We derive a sharp tensor-product bound for the hypothesis-testing relative entropy normalized by the acceptance probability and explain its operational meaning.

\subsubsection{Dual Representation of the Type-II Error}

For a state $\rho$, an operator $\sigma\geq0$, and $\ee\in[0,1)$, define
\begin{align}
 \bb_\ee(\rho\|\sigma)
 &\coloneqq\min\left\{\Tr[\sigma T]:0\leq T\leq\id,\ \Tr[\rho T]\geq1-\ee\right\},\label{eq:beta-def}\\
 \Dtest\ee(\rho\|\sigma)&\coloneqq-\log_2\bb_\ee(\rho\|\sigma),\label{eq:DH-def}\\
 \Dhat\ee(\rho\|\sigma)
 &\coloneqq-\log_2\frac{\bb_\ee(\rho\|\sigma)}{1-\ee}
 =\Dtest\ee(\rho\|\sigma)+\log_2(1-\ee).\label{eq:DH-normalized}
\end{align}
For normalized $\sigma$, $\bb_\ee$ is the minimum type-II error probability, where the outcome associated with $T$ is the decision for $\rho$ (acceptance); for a general positive $\sigma$, it is a weighted testing cost. The normalization in~\eqref{eq:DH-normalized} has the following operational meaning.

\begin{lemma}[Normalized Type-II Error as a Likelihood Ratio]
\label{lem:ht-operational}
For a state $\rho$, an operator $\sigma\ge0$, and $\ee\in[0,1)$, the quantity $\bb_\ee$ in~\eqref{eq:beta-def} satisfies
\begin{equation}
\frac{\bb_\ee(\rho\|\sigma)}{1-\ee}=\min\left\{\frac{\Tr[\sigma T]}{\Tr[\rho T]}:0\le T\le\id,\ \Tr[\rho T]\ge1-\ee\right\}.
\label{eq:ht-likelihood}
\end{equation}
In particular, $\ee\mapsto\Dhat\ee(\rho\|\sigma)$ is nondecreasing on $[0,1)$.
\end{lemma}

\begin{proof}
We first show that the normalized type-II error $\ee\mapsto\bb_\ee(\rho\|\sigma)/(1-\ee)$ is nonincreasing, and then we use this property to prove~\eqref{eq:ht-likelihood}.

Fix $0\le\ee'\le\ee<1$, and let $T'$ be an optimal test in~\eqref{eq:beta-def} at the error $\ee'$, so that $\Tr[\sigma T']=\bb_{\ee'}(\rho\|\sigma)$ and $\Tr[\rho T']\ge1-\ee'$. Set $T\coloneqq\frac{1-\ee}{1-\ee'}\,T'.$
Since $0<(1-\ee)/(1-\ee')\le1$, we have $0\le T\le\id$. Furthermore, with $\Tr[\rho T']\ge1-\ee'$, we have 
\begin{align}
\Tr[\rho T]&=\frac{1-\ee}{1-\ee'}\Tr[\rho T'] \geq 1-\ee,
\end{align}
so that $T$ is feasible in~\eqref{eq:beta-def} at the error $\ee$. Then we have that
\begin{align}
\bb_\ee(\rho\|\sigma)&\le\Tr[\sigma T]=\frac{1-\ee}{1-\ee'}\Tr[\sigma T']
=\frac{1-\ee}{1-\ee'}\bb_{\ee'}(\rho\|\sigma),
\end{align}
where the inequality follows because $T$ is feasible at the error $\ee$, and the last equality because $T'$ is optimal at the error $\ee'$. Dividing both sides by $1-\ee$, we arrive at
\begin{equation}
\frac{\bb_\ee(\rho\|\sigma)}{1-\ee}\le\frac{\bb_{\ee'}(\rho\|\sigma)}{1-\ee'}.
\label{eq:beta-monotone}
\end{equation}
By~\eqref{eq:DH-normalized}, this inequality is equivalent to $\Dhat{\ee'}(\rho\|\sigma)\le\Dhat{\ee}(\rho\|\sigma)$, which proves the last claim of the lemma.

We now prove~\eqref{eq:ht-likelihood}. Let $T^\star$ be an optimal test in~\eqref{eq:beta-def} at the error $\ee$. Since $T^\star$ is feasible in the minimization in~\eqref{eq:ht-likelihood}, we have that
\begin{align}
\min\left\{\frac{\Tr[\sigma T]}{\Tr[\rho T]}:0\le T\le\id,\ \Tr[\rho T]\ge1-\ee\right\}
&\le\frac{\Tr[\sigma T^\star]}{\Tr[\rho T^\star]}
\le\frac{\bb_\ee(\rho\|\sigma)}{1-\ee},
\end{align}
where the last inequality follows from $\Tr[\sigma T^\star]=\bb_\ee(\rho\|\sigma)$ and $\Tr[\rho T^\star]\ge1-\ee$. Conversely, let $T$ be feasible in~\eqref{eq:ht-likelihood}, and set $\ee'\coloneqq1-\Tr[\rho T]$, so that $0\le\ee'\le\ee$. Then
\begin{align}
\frac{\Tr[\sigma T]}{\Tr[\rho T]}&=\frac{\Tr[\sigma T]}{1-\ee'}
\ge\frac{\bb_{\ee'}(\rho\|\sigma)}{1-\ee'}
\ge\frac{\bb_\ee(\rho\|\sigma)}{1-\ee},
\end{align}
where 
the first inequality follows because $T$ is feasible in~\eqref{eq:beta-def} at the error $\ee'$, and the last inequality from~\eqref{eq:beta-monotone}. Minimizing the left-hand side over all feasible $T$ and combining the two inequalities concludes the proof of~\eqref{eq:ht-likelihood}.
\end{proof}

For a test $T$, the ratio $\Tr[\rho T]/\Tr[\sigma T]$ is the likelihood ratio of the event that the test accepts. By Bayes' rule, observing this event multiplies the prior odds in favor of $\rho$ by this ratio. Thus, by \cref{lem:ht-operational}, $\Dhat\ee(\rho\|\sigma)$ is the logarithm of the largest likelihood ratio of an acceptance event, among all tests that accept $\rho$ with probability at least $1-\ee$. The normalization removes the nonzero value $-\log_2(1-\ee)$ that the unnormalized divergence~\eqref{eq:DH-def} assigns to two identical states. The equality~\eqref{eq:lambda-dual} in the following lemma was established by Buscemi and Gour~\cite[Eq.~(2)]{buscemi2017quantum}.

\begin{lemma}[One-Dimensional Testing Dual]
\label{lem:beta-dual}
For $0<\ee<1$, the quantity $\bb_\ee$ in~\eqref{eq:beta-def} satisfies
\begin{equation}
 \bb_\ee(\rho\|\sigma)
 =\max_{\lambda\geq0}\left\{\lambda(1-\ee)-\Tr[(\lambda\rho-\sigma)_+]\right\}.
 \label{eq:lambda-dual}
\end{equation}
If $\bb_\ee(\rho\|\sigma)>0$, there is a finite $t>0$ such that the noncommutative minimum in~\eqref{eq:m-def} satisfies
\begin{equation}
 \ov(\rho,t\sigma)=\ee+t\bb_\ee(\rho\|\sigma).
 \label{eq:supporting-t}
\end{equation}
For every $t>0$ and every $\ee\in[0,1)$, whether or not equality is attained,
\begin{equation}
 \bb_\ee(\rho\|\sigma)\geq
 \frac{\ov(\rho,t\sigma)-\ee}{t}.
 \label{eq:beta-lower-t}
\end{equation}
\end{lemma}

\begin{proof}
We first prove a lower bound on $\bb_\ee(\rho\|\sigma)$ for every $\lambda\ge0$, then show that it is attained for $0<\ee<1$, and finally express it through the noncommutative minimum.

\underline{Lower bound}: Fix $\ee\in[0,1)$ and $\lambda\ge0$, and let $T$ be feasible in~\eqref{eq:beta-def}, i.e., $0\le T\le\id$ and $\Tr[\rho T]\ge1-\ee$. Consider that
\begin{align}
\Tr[\sigma T]&\ge\Tr[\sigma T]-\lambda\left(\Tr[\rho T]-(1-\ee)\right)\\
&=\lambda(1-\ee)-\Tr[(\lambda\rho-\sigma)T]\\
&\ge\lambda(1-\ee)-\Tr[(\lambda\rho-\sigma)_+],
\end{align}
where the first inequality follows from $\lambda\ge0$ and $\Tr[\rho T]\ge1-\ee$, and the last inequality from $\Tr[XT]\le\Tr[X_+T]\le\Tr[X_+]$, which holds for every Hermitian operator $X=X_+-X_-$ and every $0\le T\le\id$. Minimizing the left-hand side over all feasible $T$, we arrive at
\begin{equation}
\bb_\ee(\rho\|\sigma)\ge f(\lambda)\qquad\text{for all }\lambda\ge0,
\qquad\text{where}\qquad
f(\lambda)\coloneqq\lambda(1-\ee)-\Tr[(\lambda\rho-\sigma)_+].
\label{eq:weak-duality}
\end{equation}

\underline{Attainment}: Let $0<\ee<1$. The minimization in~\eqref{eq:beta-def} is a semidefinite program, and the maximization of $f(\lambda)$ over $\lambda\ge0$ is its dual. The test $T=(1-\ee/2)\id$ satisfies $0<T<\id$ and $\Tr[\rho T]=1-\ee/2>1-\ee$, i.e., it is strictly feasible. Hence strong duality holds by Slater's condition, so that $\bb_\ee(\rho\|\sigma)=\sup_{\lambda\ge0}f(\lambda)$. The function $f$ is continuous, and, since $\Tr[X_+]\ge\Tr[X]$ for every Hermitian $X$, $
f(\lambda)\le\lambda(1-\ee)-\Tr[\lambda\rho-\sigma]=\Tr[\sigma]-\lambda\ee, $
which tends to $-\infty$ as $\lambda\to\infty$. Thus the supremum is attained at some finite $\lambda^\star\ge0$, which proves~\eqref{eq:lambda-dual}.

\underline{Relation to the noncommutative minimum}: For $\lambda>0$, set $t\coloneqq1/\lambda$. Then we have that
\begin{align}
f(\lambda)&=\frac1t\left(1-\ee-\Tr[(\rho-t\sigma)_+]\right)=\frac{\ov(\rho,t\sigma)-\ee}{t},
\label{eq:dual-to-m}
\end{align}
where the first equality follows from $(\lambda\rho-\sigma)_+=\lambda(\rho-t\sigma)_+$, and the second equality from~\eqref{eq:m-positive-part} with $A=\rho$ and $B=t\sigma$, together with $\Tr[\rho]=1$. Combining~\eqref{eq:weak-duality} and~\eqref{eq:dual-to-m} proves~\eqref{eq:beta-lower-t} for every $t>0$ and every $\ee\in[0,1)$.

Finally, let $0<\ee<1$ and $\bb_\ee(\rho\|\sigma)>0$. Since
$f(0)=-\Tr[(-\sigma)_+]=0<\bb_\ee(\rho\|\sigma)=f(\lambda^\star),$
the maximizer satisfies $\lambda^\star>0$. Setting $t\coloneqq1/\lambda^\star$ in~\eqref{eq:dual-to-m}, we arrive at $\bb_\ee(\rho\|\sigma)=(\ov(\rho,t\sigma)-\ee)/t$, which is equivalent to~\eqref{eq:supporting-t}.
\end{proof}

Geometrically,~\eqref{eq:beta-lower-t} and~\eqref{eq:supporting-t} state that the line $t\mapsto\ee+t\,\bb_\ee(\rho\|\sigma)$ lies above the function $t\mapsto\ov(\rho,t\sigma)$ and touches it at some $t>0$. In the proof of the following theorem, we combine these touching points for the individual pairs by using \cref{thm:overlap}.

\subsubsection{A Sharp Normalized Sub-Additivity Theorem}
The next theorem is the main hypothesis-testing consequence of \cref{thm:overlap}.
\begin{theorem}[Normalized Hypothesis-Testing Sub-Additivity]
\label{thm:ht}
Let $\rho_i$ be states, $\sigma_i\geq0$, and $\ee_i\in[0,1)$ for $i\in\{1,\ldots,n\}$, and set $\bar\ee\coloneqq\prod_{i=1}^n\ee_i$. Then the quantities in~\eqref{eq:beta-def}--\eqref{eq:DH-normalized} satisfy
\begin{equation}
 \bb_{\bar\ee}(\rhot\|\sigt)
 \geq(1-\bar\ee)\prod_{i=1}^n\frac{\bb_{\ee_i}(\rho_i\|\sigma_i)}{1-\ee_i}.
 \label{eq:beta-main}
\end{equation}
Equivalently,
\begin{equation}
 \Dhat{\bar\ee}(\rhot\|\sigt)
 \leq\sum_{i=1}^n\Dhat{\ee_i}(\rho_i\|\sigma_i).
 \label{eq:ht-normalized-main}
\end{equation}
In the unnormalized convention,
\begin{align}
 \Dtest{\bar\ee}(\rhot\|\sigt)
 &\leq\sum_{i=1}^n\Dtest{\ee_i}(\rho_i\|\sigma_i)
 +\log_2\frac{\prod_i(1-\ee_i)}{1-\bar\ee}\label{eq:ht-correction}\\
 &\leq\sum_{i=1}^n\Dtest{\ee_i}(\rho_i\|\sigma_i).
 \label{eq:ht-weak}
\end{align}
\end{theorem}

\begin{proof}
We first prove~\eqref{eq:beta-main}, and then we derive~\eqref{eq:ht-normalized-main}--\eqref{eq:ht-weak} from it. Throughout the proof, write $b_i\coloneqq\bb_{\ee_i}(\rho_i\|\sigma_i)$ and $[n]\coloneqq\{1,\ldots,n\}$. We use the expansion
\begin{equation}
\prod_{i=1}^n(x_i+y_i)=\sum_{S\subseteq[n]}\left(\prod_{i\in S}y_i\right)\left(\prod_{j\notin S}x_j\right),
\label{eq:product-expansion}
\end{equation}
which holds for all real numbers $x_i$ and $y_i$, and in which the term with $S=\varnothing$ is equal to $\prod_jx_j$.

\underline{Case 1: $b_i=0$ for some $i$}. Then the right-hand side of~\eqref{eq:beta-main} is equal to zero, and~\eqref{eq:beta-main} holds because $\bb_{\bar\ee}(\rhot\|\sigt)\ge0$.

\underline{Case 2: $\ee_i\in(0,1)$ and $b_i>0$ for all $i$}. By \cref{lem:beta-dual}, for every $i\in[n]$, there exists $t_i>0$ such that
\begin{equation}
 \ov(\rho_i,t_i\sigma_i)=\ee_i+t_ib_i.
 \label{eq:local-dual}
\end{equation}
Set $\Gamma\coloneqq\prod_{i=1}^nt_i$. Then we have that
\begin{align}
 \bb_{\bar\ee}(\rhot\|\sigt)
 &\geq\frac{\ov(\rhot,\Gamma\,\sigt)-\bar\ee}{\Gamma}\\
 &\geq\frac{\prod_{i=1}^n\ov(\rho_i,t_i\sigma_i)-\prod_{i=1}^n\ee_i}{\prod_{i=1}^nt_i}\\
 &=\frac{\prod_{i=1}^n(\ee_i+t_ib_i)-\prod_{i=1}^n\ee_i}{\prod_{i=1}^nt_i}\\
 &=\sum_{\varnothing\neq S\subseteq[n]}
 \left(\prod_{i\in S}b_i\right)
 \left(\prod_{j\notin S}\frac{\ee_j}{t_j}\right),
 \label{eq:beta-expansion}
\end{align}
where the first inequality follows from~\eqref{eq:beta-lower-t} applied to the pair $(\rhot,\sigt)$ with $t=\Gamma$ and the error $\bar\ee$, the second inequality from $\Gamma\,\sigt=\bigotimes_{i}(t_i\sigma_i)$ and \cref{thm:overlap} applied $n-1$ times, the first equality from~\eqref{eq:local-dual}, and the last equality from~\eqref{eq:product-expansion} with $x_i=\ee_i$ and $y_i=t_ib_i$, in which the term with $S=\varnothing$ cancels $\prod_i\ee_i$. 

It remains to remove the dependence on $t_1,\ldots,t_n$. By~\eqref{eq:m-bounds}, $\ov(\rho_i,t_i\sigma_i)\le\Tr[\rho_i]=1$, so that~\eqref{eq:local-dual} gives $t_ib_i\le1-\ee_i$. Since $b_i>0$, this is equivalent to
\begin{equation}
 \frac{\ee_i}{t_i}\geq\frac{\ee_i b_i}{1-\ee_i}.
 \label{eq:t-bound}
\end{equation}
Since every term in~\eqref{eq:beta-expansion} is nonnegative, we arrive at
\begin{align}
 \bb_{\bar\ee}(\rhot\|\sigt)
 &\geq\sum_{\varnothing\neq S\subseteq[n]}
 \left(\prod_{i\in S}b_i\right)
 \left(\prod_{j\notin S}\frac{\ee_jb_j}{1-\ee_j}\right)\\
 &=\left(\prod_{i=1}^n\frac{b_i}{1-\ee_i}\right)
 \sum_{\varnothing\neq S\subseteq[n]}
 \left(\prod_{i\in S}(1-\ee_i)\right)
 \left(\prod_{j\notin S}\ee_j\right)\\
 &=\left(1-\prod_{i=1}^n\ee_i\right)\prod_{i=1}^n\frac{b_i}{1-\ee_i},
 \label{eq:beta-finish}
\end{align}
where the inequality follows from~\eqref{eq:t-bound}, the first equality from writing $b_i=(1-\ee_i)\cdot b_i/(1-\ee_i)$ for $i\in S$, and the last equality from~\eqref{eq:product-expansion} with $x_i=\ee_i$ and $y_i=1-\ee_i$, which gives
\begin{equation}
\sum_{\varnothing\neq S\subseteq[n]}\left(\prod_{i\in S}(1-\ee_i)\right)\left(\prod_{j\notin S}\ee_j\right)
=\prod_{i=1}^n\left[(1-\ee_i)+\ee_i\right]-\prod_{i=1}^n\ee_i
=1-\prod_{i=1}^n\ee_i.
\label{eq:binomial-identity}
\end{equation}
This proves~\eqref{eq:beta-main} in Case~2.

\underline{Case 3: $\ee_i=0$ for some $i$}. Then $\bar\ee=0$. We first show that, for a state $\rho$ and an operator $\sigma\ge0$,
\begin{equation}
 \bb_0(\rho\|\sigma)=\Tr[P\sigma],
 \label{eq:beta-zero}
\end{equation}
where $P$ is the projection onto $\supp(\rho)$. Let $T$ be feasible in~\eqref{eq:beta-def} at the error $0$, i.e., $0\le T\le\id$ and $\Tr[\rho(\id-T)]=0$. Since $\Tr[\rho(\id-T)]=\lVert(\id-T)^{1/2}\rho^{1/2}\rVert_2^2$, we have $(\id-T)\rho=0$, so that $(\id-T)P=0$ by~\eqref{eq:range-support}. Thus $TP=PT=P$, and hence $T=P+(\id-P)T(\id-P)\ge P.$
This gives $\Tr[\sigma T]\ge\Tr[\sigma P]$, with equality for the feasible test $T=P$, which proves~\eqref{eq:beta-zero}. Since $\supp(\rhot)=\bigotimes_i\supp(\rho_i)$, the projection onto $\supp(\rhot)$ is $\bigotimes_iP_i$, where $P_i$ is the projection onto $\supp(\rho_i)$. Then we have that
\begin{align}
\bb_0(\rhot\|\sigt)&=\Tr\!\left[\left(\textstyle\bigotimes_{i}P_i\right)\left(\textstyle\bigotimes_{i}\sigma_i\right)\right]=\prod_{i=1}^n\bb_0(\rho_i\|\sigma_i)\ge\prod_{i=1}^n\frac{b_i}{1-\ee_i},
\end{align}
where the first two equalities follow from~\eqref{eq:beta-zero}, and the inequality from~\eqref{eq:beta-monotone} with $\ee'=0\le\ee_i$. Since $1-\bar\ee=1$, this proves~\eqref{eq:beta-main} in Case~3.

\underline{Logarithmic forms}: We have that
\begin{align}
\Dhat{\bar\ee}(\rhot\|\sigt)&=-\log_2\frac{\bb_{\bar\ee}(\rhot\|\sigt)}{1-\bar\ee}\\
&\le-\log_2\prod_{i=1}^n\frac{\bb_{\ee_i}(\rho_i\|\sigma_i)}{1-\ee_i}\\
&=\sum_{i=1}^n\Dhat{\ee_i}(\rho_i\|\sigma_i),
\end{align}
where the equalities follow from~\eqref{eq:DH-normalized}, and the inequality from~\eqref{eq:beta-main} and the monotonicity of the logarithm. This proves~\eqref{eq:ht-normalized-main}. Using~\eqref{eq:DH-normalized} again, we find that
\begin{align}
\Dtest{\bar\ee}(\rhot\|\sigt)&=\Dhat{\bar\ee}(\rhot\|\sigt)-\log_2(1-\bar\ee)\\
&\le\sum_{i=1}^n\left[\Dtest{\ee_i}(\rho_i\|\sigma_i)+\log_2(1-\ee_i)\right]-\log_2(1-\bar\ee)\\
&=\sum_{i=1}^n\Dtest{\ee_i}(\rho_i\|\sigma_i)+\log_2\frac{\prod_i(1-\ee_i)}{1-\bar\ee},
\end{align}
where the inequality follows from~\eqref{eq:ht-normalized-main}. This proves~\eqref{eq:ht-correction}. Finally, the term with $S=[n]$ in~\eqref{eq:binomial-identity} is equal to $\prod_i(1-\ee_i)$, and all the terms are nonnegative, so that $\prod_i(1-\ee_i)\le1-\prod_i\ee_i$. Hence the last term in~\eqref{eq:ht-correction} is nonpositive, which proves~\eqref{eq:ht-weak}.
\end{proof}

\begin{remark}[Sharpness of the Normalization]
\label{rem:ht-sharp}
For $q_i\in[0,1]$, consider the states
\begin{equation}
 \rho_i=\proj{0},\qquad
 \sigma_i=q_i\proj{0}+(1-q_i)\proj{1}.
 \label{eq:sharp-family}
\end{equation}
For a test $T$ that is feasible in~\eqref{eq:beta-def}, we have $\langle0|T|0\rangle=\Tr[\rho_iT]\ge1-\ee_i$, so that $\Tr[\sigma_iT]=q_i\langle0|T|0\rangle+(1-q_i)\langle1|T|1\rangle\ge q_i(1-\ee_i),$ with equality for $T=(1-\ee_i)\proj0$. Hence $\bb_{\ee_i}(\rho_i\|\sigma_i)=(1-\ee_i)q_i$. The product pair has the same structure, since $\rhot$ is the projection onto $|0\cdots0\rangle$ and $\sigt$ is diagonal in the computational basis with the entry $\prod_iq_i$ at $|0\cdots0\rangle$. The same argument gives $\bb_{\bar\ee}(\rhot\|\sigt)=(1-\bar\ee)\prod_iq_i$. Thus \eqref{eq:beta-main} and \eqref{eq:ht-correction} are equalities. In particular, the coefficient $1-\bar\ee$ in \eqref{eq:beta-main} cannot be increased uniformly. When $q_i=1$ for every $i$, all normalized divergences vanish, whereas the unnormalized divergence is $-\log_2(1-\ee_i)$.
\end{remark}

The change in the error parameter is not optional. The following classical example rules out the statement with the same error on both sides.
\begin{example}[Failure at a Common Error Parameter]
\label{ex:same-error_HT}
Let
\begin{equation}
 \rho=\operatorname{diag}(1/2,1/2),\qquad
 \sigma=\operatorname{diag}(9/10,1/10),\qquad\ee=1/5.
\end{equation}
For commuting states, an optimal test in~\eqref{eq:beta-def} accepts the outcomes in decreasing order of the likelihood ratio $\rho(x)/\sigma(x)$ until the acceptance probability reaches $1-\ee$ (Neyman--Pearson lemma). For one copy, the likelihood ratios of the first and second outcomes are $5/9$ and $5$, respectively. Hence the optimal test accepts the second outcome with probability one and the first outcome with probability $3/5$, which gives the acceptance probability $\frac12+\frac35\cdot\frac12=\frac45$ and
\begin{equation}
 \bb_{1/5}(\rho\|\sigma)=\frac1{10}+\frac35\cdot\frac9{10}=\frac{16}{25}.
\end{equation}
For two copies, the likelihood ratios are $25$ for the outcome $(2,2)$, $25/9$ for the outcomes $(1,2)$ and $(2,1)$, and $25/81$ for the outcome $(1,1)$. Hence the optimal test accepts the first three outcomes with probability one and the outcome $(1,1)$ with probability $1/5$, which gives the acceptance probability $\frac34+\frac15\cdot\frac14=\frac45$ and
\begin{equation}
 \bb_{1/5}(\rho^{\otimes2}\|\sigma^{\otimes2})
 =\left(\frac1{100}+\frac9{100}+\frac9{100}\right)+\frac15\cdot\frac{81}{100}
 =\frac{44}{125}<\left(\frac{16}{25}\right)^2.
\end{equation}
Therefore $\Dtest{1/5}(\rho^{\otimes2}\|\sigma^{\otimes2})>2\Dtest{1/5}(\rho\|\sigma)$.
\end{example}

\subsubsection{Operational Interpretation: Global versus Product Tests}
\label{sec:ht-operational}

We now explain the meaning of \cref{thm:ht} for tests on $n$ systems. Let all $\sigma_i$ be states, and consider the product pair $\rhot$ and $\sigt$. For a test $T$ on the $n$ systems, let
\begin{equation}
L(T)\coloneqq\frac{\Tr\!\left[(\rhot)\,T\right]}{\Tr\!\left[(\sigt)\,T\right]}
\end{equation}
be the likelihood ratio of its acceptance event. By \cref{lem:ht-operational}, every test $T$ with type-I error at most $\ee$ satisfies $\log_2L(T)\le\Dhat{\ee}(\rhot\|\sigt)$, and equality is attained by some such test.

Let $T_i$ be an optimal test in~\eqref{eq:beta-def} for the pair $(\rho_i,\sigma_i)$ at the error $\ee_i$, so that $\Tr[\rho_iT_i]\ge1-\ee_i$ and $\Tr[\sigma_iT_i]=\bb_{\ee_i}(\rho_i\|\sigma_i)$. We compare global tests with two product tests.
\begin{itemize}
\item The AND rule $T_\wedge\coloneqq T_1\otimes\cdots\otimes T_n$ accepts if all local tests accept. Its acceptance probability is $\prod_i\Tr[\rho_iT_i]\ge\prod_i(1-\ee_i)$, so that its type-I error is at most $1-\prod_i(1-\ee_i)$. Moreover,
\begin{equation}
L(T_\wedge)=\prod_{i=1}^n\frac{\Tr[\rho_iT_i]}{\bb_{\ee_i}(\rho_i\|\sigma_i)}\ge\prod_{i=1}^n\frac{1-\ee_i}{\bb_{\ee_i}(\rho_i\|\sigma_i)},
\end{equation}
i.e., $\log_2L(T_\wedge)\ge\sum_i\Dhat{\ee_i}(\rho_i\|\sigma_i)$ by~\eqref{eq:DH-normalized}.
\item The OR rule $T_\vee\coloneqq\id-\bigotimes_i(\id-T_i)$ accepts if at least one local test accepts. Its type-I error is $\prod_i\Tr[\rho_i(\id-T_i)]\le\prod_i\ee_i$. However, $\Tr[(\rhot)T_\vee]\le1$ and
\begin{equation}
\Tr[(\sigt)T_\vee]=1-\prod_{i=1}^n\left(1-\bb_{\ee_i}(\rho_i\|\sigma_i)\right)\ge\max_i\bb_{\ee_i}(\rho_i\|\sigma_i),
\end{equation}
so that $\log_2L(T_\vee)\le\min_i\Dtest{\ee_i}(\rho_i\|\sigma_i)$ by~\eqref{eq:DH-def}.
\end{itemize}
Thus the OR rule has a small type-I error, and the AND rule has a large likelihood ratio. \Cref{thm:ht} shows that a global test, possibly based on an entangled measurement, cannot do better than combining these two properties. Indeed, every global test $T$ with type-I error at most $\prod_i\ee_i$ satisfies $\log_2L(T)\le\sum_i\Dhat{\ee_i}(\rho_i\|\sigma_i)$, which is the value achieved by the AND rule. In terms of the normalized divergence, we obtain
\begin{equation}
\Dhat{\bar\ee}(\rhot\|\sigt)\le\sum_{i=1}^n\Dhat{\ee_i}(\rho_i\|\sigma_i)\le\Dhat{1-\prod_i(1-\ee_i)}(\rhot\|\sigt).
\label{eq:Dhat-sandwich}
\end{equation}
The first inequality is~\eqref{eq:ht-normalized-main}. The second inequality follows from \cref{lem:ht-operational}, because $T_\wedge$ is feasible in~\eqref{eq:ht-likelihood} for the product pair at the error $1-\prod_i(1-\ee_i)$ and satisfies $\log_2L(T_\wedge)\ge\sum_i\Dhat{\ee_i}(\rho_i\|\sigma_i)$. Both inequalities are equalities for the pairs in~\eqref{eq:sharp-family}, since their normalized divergences do not depend on the error parameter.

\Cref{thm:ht} is also an equality in two further cases. First, it is an equality at $\ee_1=\cdots=\ee_n=0$, since the proof of Case~3 shows that $\bb_0(\rhot\|\sigt)=\prod_i\bb_0(\rho_i\|\sigma_i)$. Second, let all $\sigma_i$ be faithful, and let $\ee_i$ be sufficiently close to one. Since $\rho_i\le e^{\Dmax(\rho_i\Vert\sigma_i)}\sigma_i$, every test $T$ satisfies $\Tr[\rho_iT]\le e^{\Dmax(\rho_i\Vert\sigma_i)}\Tr[\sigma_iT]$, so that \cref{lem:ht-operational} gives $\Dhat{\ee}(\rho_i\|\sigma_i)\le\Dmax(\rho_i\Vert\sigma_i)/\ln2$ for every $\ee$. Conversely, let $v_i$ be an eigenvector of $\sigma_i^{-1/2}\rho_i\sigma_i^{-1/2}$ with the largest eigenvalue $e^{\Dmax(\rho_i\Vert\sigma_i)}$, and let $T_i$ be the projection onto the vector $w_i\coloneqq\sigma_i^{-1/2}v_i$. Then
\begin{equation}
\frac{\Tr[\rho_iT_i]}{\Tr[\sigma_iT_i]}=\frac{\langle w_i|\rho_i|w_i\rangle}{\langle w_i|\sigma_i|w_i\rangle}=\frac{\langle v_i|\sigma_i^{-1/2}\rho_i\sigma_i^{-1/2}|v_i\rangle}{\langle v_i|v_i\rangle}=e^{\Dmax(\rho_i\Vert\sigma_i)},
\end{equation}
and $T_i$ is feasible whenever $\ee_i\ge1-\Tr[\rho_iT_i]$. For such $\ee_i$, \cref{lem:ht-operational} gives $\Dhat{\ee_i}(\rho_i\|\sigma_i)=\Dmax(\rho_i\Vert\sigma_i)/\ln2$. The same argument applies to the product pair with the vector $\bigotimes_iv_i$, which is an eigenvector of $\bigotimes_i\sigma_i^{-1/2}\rho_i\sigma_i^{-1/2}$ with the eigenvalue $e^{\sum_i\Dmax(\rho_i\Vert\sigma_i)}$. The corresponding projection $\bigotimes_iT_i$ is feasible at the error $\bar\ee=\prod_i\ee_i$ as soon as $\prod_i\ee_i\ge1-\prod_i\Tr[\rho_iT_i]$, which holds if all $\ee_i$ are sufficiently close to one.  Thus both sides of~\eqref{eq:ht-normalized-main} are equal to $\sum_i\Dmax(\rho_i\Vert\sigma_i)/\ln2$ if all $\ee_i$ are sufficiently close to one.

For $n$ copies of a single pair, i.e., $\rho_i=\rho$, $\sigma_i=\sigma$, and $\ee_i=\ee$, \cref{thm:ht} gives a converse that holds for every $n$. Let $T$ be a test on $n$ copies with type-I error $1-\Tr[\rho^{\otimes n}T]\le\ee^n$. Then
\begin{align}
\Tr\!\left[\sigma^{\otimes n}T\right]&\ge\bb_{\ee^n}\!\left(\rho^{\otimes n}\middle\|\sigma^{\otimes n}\right)\\
&\ge\left(1-\ee^n\right)\left(\frac{\bb_\ee(\rho\|\sigma)}{1-\ee}\right)^n\\
&=\left(1-\ee^n\right)2^{-n\Dhat{\ee}(\rho\|\sigma)},
\label{eq:ht-many-copies}
\end{align}
where the first inequality follows from~\eqref{eq:beta-def}, the second inequality from~\eqref{eq:beta-main}, and the equality from~\eqref{eq:DH-normalized}. In particular, let $B(r)$ be the Hoeffding exponent, i.e., the optimal exponent of the type-II error when the type-I error is required to decay as $e^{-nr}$~\cite{hayashi2007error,nagaoka2006converse,audenaert2008asymptotic}. Applying~\eqref{eq:ht-many-copies} with $\ee=e^{-r}$, taking $-\frac1n\ln$ of both sides, and letting $n\to\infty$ gives $B(r)\le\Dhat{e^{-r}}(\rho\|\sigma)\ln2$. This bound is not tight in general, since $B(r)$ is given by the Petz R\'enyi divergences. However,~\eqref{eq:ht-many-copies} holds for every finite $n$ and involves only a single copy.

\subsection{Refined Quantum Chernoff Bounds and Sample Complexity}
\label{sec:chernoff}

In this subsection, we apply the tools developed in \cref{sec:supermult} to symmetric binary quantum hypothesis testing. Here, $n$ copies of a quantum system are prepared either in the state $\rho^{\otimes n}$, with prior probability $p\in(0,1)$, or in the state $\sigma^{\otimes n}$, with prior probability $1-p$. The goal is to identify the state by performing a measurement $\{T,\id-T\}$, where the outcome associated with $\id-T$ is the guess $\rho^{\otimes n}$. The optimal error probability is
\begin{equation}
P_{e,n}^*(p)\coloneqq\min_{0\le T\le\id}\left\{p\Tr\!\left[\rho^{\otimes n}T\right]+(1-p)\Tr\!\left[\sigma^{\otimes n}(\id-T)\right]\right\}.
\label{eq:ch-Pe}
\end{equation}
By~\eqref{eq:m-variational}, it is given by the noncommutative minimum,
\begin{equation}
P_{e,n}^*(p)=m\!\left(p\rho^{\otimes n},(1-p)\sigma^{\otimes n}\right).
\label{eq:ch-prior-profile}
\end{equation}
Since $m$ is symmetric and $m(cA,cB)=c\,m(A,B)$ for $c>0$, we have $P_{e,n}^*(p)=p\,m(\rho^{\otimes n},\frac{1-p}p\sigma^{\otimes n})=(1-p)\,m(\sigma^{\otimes n},\frac p{1-p}\rho^{\otimes n})$. Applying~\eqref{eq:m-hsd} to the form whose weight is at least one gives
\begin{equation}
P_{e,n}^*(p)=
\begin{cases}
p\left[1-E_{\frac{1-p}p}\!\left(\rho^{\otimes n}\middle\Vert\sigma^{\otimes n}\right)\right], & p\le\frac12,\\[4pt]
(1-p)\left[1-E_{\frac{p}{1-p}}\!\left(\sigma^{\otimes n}\middle\Vert\rho^{\otimes n}\right)\right], & p\ge\frac12.
\end{cases}
\label{eq:ch-prior-HS}
\end{equation}
The quantum Chernoff bound~\cite{audenaert2007chernoff,nussbaum2009chernoff} states that $P_{e,n}^*(p)$ decays exponentially fast in $n$ with the exponent
\begin{equation}
\xi\coloneqq-\min_{0\le\alpha\le1}\Phi(\alpha),
\qquad
\Phi(\alpha)\coloneqq\ln\Tr\!\left[\rho^\alpha\sigma^{1-\alpha}\right].
\label{eq:ch-Phi}
\end{equation}
In this subsection, we go beyond the exponent: we determine the prefactor of $e^{-n\xi}$ in $P_{e,n}^*(p)$ and, under the quantitative nonlattice condition (QN) in \cref{def:ch-QN} below, its expansion to every fixed order in $1/n$. We also determine the sample complexity.

The main idea is as follows. Since $m(p\rho^{\otimes n},(1-p)\sigma^{\otimes n})$ is difficult to evaluate directly for noncommuting states, we compare it with the harmonic overlap $\Har$ in~\eqref{eq:ch-Har-def}. The harmonic overlap is a lower bound on $m$ that, for noncommuting faithful states, differs from it by a term exponentially smaller than $P_{e,n}^*(p)$ (\cref{thm:ch-three-bounds,cor:ch-gap}); this is where \cref{lem:decomposition_general_to_rank} enters. Moreover, it is equal to a classical quantity of the Nussbaum--Szko{\l}a distributions (\cref{prop:ch-classical-rep}) and has an explicit integral representation in terms of $\Phi$ (\cref{prop:ch-integral-rep}), which reduce the refined asymptotics to a classical calculation.

\subsubsection{Setting and Main Result}
\label{sec:ch-main}

We introduce the classical spectral distributions and regularity assumptions needed to state the refined Chernoff expansion.

Let $\rho=\sum_ir_i|u_i\rangle\!\langle u_i|$ and $\sigma=\sum_js_j|v_j\rangle\!\langle v_j|$ be spectral decompositions in complete orthonormal eigenbases, including zero eigenvalues. The Nussbaum--Szko{\l}a distributions~\cite{nussbaum2009chernoff} are
\begin{equation}
P_{\NS}(i,j)\coloneqq r_i|\langle u_i,v_j\rangle|^2,
\qquad
Q_{\NS}(i,j)\coloneqq s_j|\langle u_i,v_j\rangle|^2.
\label{eq:ch-NS}
\end{equation}
Thus $\sum_{i,j}P_{\NS}(i,j)=\sum_{i,j}Q_{\NS}(i,j)=1$. For $0<\operatorname{Re}\beta<1$, write $Q_\beta^{\NS}$ for the Chernoff quantity of these distributions, extended to complex orders:
\begin{equation}
\begin{aligned}
Q_\beta^{\NS}\coloneqq Q_\beta(P_{\NS},Q_{\NS})
&\coloneqq\sum_{i,j}P_{\NS}(i,j)^\beta Q_{\NS}(i,j)^{1-\beta}\\
&=\sum_{i,j}r_i^{\beta}s_j^{1-\beta}|\langle u_i,v_j\rangle|^2.
\end{aligned}
\label{eq:ch-F}
\end{equation}
Here $0^\beta=0$ for $\operatorname{Re}\beta>0$. 
If $\beta=\alpha$ is real, with $\alpha\in(0,1)$, then $Q_{\alpha}^{\NS}=\Tr[\rho^\alpha\sigma^{1-\alpha}]=e^{\Phi(\alpha)}$. In what follows, $\alpha$ denotes a real order, and complex orders appear only in the form $\beta=\alpha+iu$ with $u\in\mathbb R$.
For $\alpha\in(0,1)$ with $Q_{\alpha}^{\NS}>0$, we define the tilted probability distribution and the log-likelihood ratio on the common positive support $P_{\NS}(i,j)Q_{\NS}(i,j)>0$ by
\begin{equation}
\Pr_{\alpha}[(i,j)]\coloneqq\frac{P_{\NS}(i,j)^\alpha Q_{\NS}(i,j)^{1-\alpha}}{Q_{\alpha}^{\NS}},
\qquad
Z(i,j)\coloneqq\ln\frac{P_{\NS}(i,j)}{Q_{\NS}(i,j)}=\ln\frac{r_i}{s_j}.
\label{eq:ch-tilt}
\end{equation}
Let 
\begin{equation}
    S_n\coloneqq Z_1+\cdots+Z_n, \label{eq:S_n_with_alpha}
\end{equation} where $Z_1,\ldots,Z_n$ are independent copies of $Z$ under $\Pr_\alpha$. Then $\mathbb E_\alpha[e^{uS_n}]=(Q_{\alpha+u}^{\NS}/Q_{\alpha}^{\NS})^n$ for real $u$ with $0<\alpha+u<1$, so that
\begin{equation}
\mathbb E_\alpha[Z]=\Phi'(\alpha),
\qquad
\operatorname{Var}_\alpha[Z]=\Phi''(\alpha),
\label{eq:ch-tilt-moments}
\end{equation}
and $\mathbb E_\alpha[e^{iuZ}]=Q_{\alpha+iu}^{\NS}/Q_{\alpha}^{\NS}$ is the characteristic function of $Z$.

We now introduce the regularity assumptions under which we state the refined asymptotics.

\begin{definition}[Nonlattice and Regular Pairs]
\label{def:ch-regular}
We say that the pair $(\rho,\sigma)$ is nonlattice if the values $\ln(r_i/s_j)$, over the pairs $(i,j)$ with $P_{\NS}(i,j)Q_{\NS}(i,j)>0$, are not contained in a set of the form $z_0+d\mathbb Z$ with $z_0\in\mathbb R$ and $d>0$; equivalently, $|Q_{\alpha+iu}^{\NS}|<Q_{\alpha}^{\NS}$ for every $u\neq0$ and every $\alpha\in(0,1)$. We say that the pair $(\rho,\sigma)$ is regular if $\Phi$ in~\eqref{eq:ch-Phi} has a minimizer $\alpha_*\in(0,1)$ on $[0,1]$ with
\begin{equation}
\Phi'(\alpha_*)=0,
\qquad
V_*\coloneqq\Phi''(\alpha_*)>0,
\label{eq:ch-regular}
\end{equation}
and the pair is nonlattice.
\end{definition}

\begin{definition}[Quantitative Nonlattice Condition]
\label{def:ch-QN}
We say that a nonlattice pair $(\rho,\sigma)$ satisfies the quantitative nonlattice condition (QN) if, for some $\alpha\in(0,1)$ and every $\eta>0$, there exists $U_\eta>0$ such that
\begin{equation}
1-\frac{|Q_{\alpha+iu}^{\NS}|}{Q_\alpha^{\NS}}\ge e^{-\eta|u|}
\qquad\text{for all }|u|\ge U_\eta.
\label{eq:ch-QN}
\end{equation}
Thus the inverse of the gap from modulus one grows subexponentially. The stronger weak Cram\'er condition of~\cite{angst2017weak} imposes a polynomial lower bound on this gap and therefore implies~\eqref{eq:ch-QN}; see also~\cite[Definition~2.1]{song2018uniform}.
\end{definition}

The condition (QN) does not depend on the choice of $\alpha$, since all the tilted laws in~\eqref{eq:ch-tilt} have the same finite support with weights that are comparable up to constant factors. The nonlattice condition alone does not exclude that the almost-periodic function $u\mapsto|Q_{\alpha+iu}^{\NS}|/Q_{\alpha}^{\NS}$ comes arbitrarily close to one as $|u|\to\infty$, and~\eqref{eq:ch-QN} restricts how fast this can happen.

Nonlattice assumptions are standard in precise large deviation asymptotics~\cite{bahadur1960deviations,blackwell1959probability,petrov1975sums,chaganty1993strong} and in refinements of error exponents with exact prefactors~\cite{altug2014refinement,altug2014sphere,moulin2017log,vazquezvilar2018saddlepoint,cheng2019sphere}. In quantum hypothesis testing, Bahadur--Rao estimates for the Nussbaum--Szko{\l}a distributions give lower bounds of order $e^{-n\xi}/\sqrt n$ with unspecified constants~\cite[Proposition~4.3 and Corollary~4.5]{audenaert2012finite}, and the same assumptions as in \cref{def:ch-regular} are used in~\cite[Theorem~5.2]{cheng2026multiple}; $V_*>0$ is the analogue of the positive-variance condition in second-order asymptotics~\cite{tomamichel2013hierarchy,li2014second}. 
Edgeworth expansions of discrete nonlattice laws usually require a polynomial lower bound on the gap in~\eqref{eq:ch-QN}~\cite{angst2017weak,song2018uniform,fernando2021higher}; here the weaker condition~\eqref{eq:ch-QN} suffices, because the Fourier transform of the harmonic kernel decays exponentially (Appendix~\ref{sec:ch-auxiliary}). For qubit pairs, (QN) is equivalent to the small-divisor condition (R) in~\cite[p.~343]{bounemoura2019optimality} (see the proof of \cref{prop:ch-regular-pairs}).

To state the main result, define, for $\alpha\in(0,1)$ with $\Phi''(\alpha)>0$,
\begin{equation}
\tau(\alpha)\coloneqq-\pi\cot(\pi\alpha)-\frac{\Phi'''(\alpha)}{2\Phi''(\alpha)}.
\label{eq:ch-tau}
\end{equation}

\begin{theorem}[Refined Quantum Chernoff Bound]
\label{thm:ch-main}
Let $\rho$ and $\sigma$ be states with $[\rho,\sigma]\neq0$ such that the pair $(\rho,\sigma)$ is regular in the sense of \cref{def:ch-regular}, with $\alpha_*$ and $V_*$ as in~\eqref{eq:ch-regular}, and let $p\in(0,1)$. Then the optimal error probability in~\eqref{eq:ch-Pe} satisfies
\begin{equation}
P_{e,n}^*(p)=A_{\mathrm q}\,\frac{e^{-n\xi}}{\sqrt n}\,[1+o(1)],
\qquad
A_{\mathrm q}\coloneqq\frac{\pi\,p^{\alpha_*}(1-p)^{1-\alpha_*}}{\sin(\pi\alpha_*)\sqrt{2\pi V_*}},
\label{eq:ch-prefactor}
\end{equation}
where $\xi$ is the Chernoff exponent in~\eqref{eq:ch-Phi}. If, in addition, the pair satisfies the quantitative nonlattice condition of \cref{def:ch-QN}, then, for every fixed integer $M\ge1$,
\begin{equation}
-\ln P_{e,n}^*(p)=n\xi+\frac12\ln n-\ln A_{\mathrm q}+\sum_{j=1}^{M}\frac{C_j(p)}{n^j}+O(n^{-M-1}),
\label{eq:ch-higher}
\end{equation}
where every $C_j(p)$ is given by the finite coefficient rule~\eqref{eq:app-coefficient} in Appendix~\ref{sec:ch-coefficients}. For example, with $\Phi^{(k)}\coloneqq\Phi^{(k)}(\alpha_*)$ and $\delta_p\coloneqq\ln\frac{1-p}{p}-\tau(\alpha_*)$ for $\tau$ in~\eqref{eq:ch-tau},
\begin{equation}
C_1(p)\coloneqq\frac{\delta_p^2}{2V_*}-\frac{\Phi^{(4)}}{8V_*^2}+\frac{(\Phi''')^2}{12V_*^3}+\frac{\pi^2}{2V_*\sin^2(\pi\alpha_*)},
\label{eq:ch-C1}
\end{equation}
and $C_2(p)$ and $C_3(p)$ are given in~\eqref{eq:ch-C2} and~\eqref{eq:ch-C3}. Each $C_j(p)$ depends only on $p$, $\alpha_*$, and the derivatives of $\Phi$ at $\alpha_*$ up to order $2j+2$. The remainder constant may depend on $M$, $p$, and the states; no convergence of the infinite series is asserted.
\end{theorem}

\begin{remark}[Comparison with Commuting States and Previous Bounds]
\label{rem:ch-compare}
For commuting states with the same function $\Phi$, the classical precise asymptotics~\cite{bahadur1960deviations} give the prefactor
\begin{equation}
A_{\mathrm{cl}}\coloneqq\frac{p^{\alpha_*}(1-p)^{1-\alpha_*}}{\alpha_*(1-\alpha_*)\sqrt{2\pi V_*}}
\label{eq:ch-Acl}
\end{equation}
(see \cref{ex:ch-commuting}). Thus the prefactor in~\eqref{eq:ch-prefactor} is
\begin{equation}
A_{\mathrm q}=\kappa_{\alpha_*}A_{\mathrm{cl}},
\qquad
\kappa_\alpha\coloneqq\frac{\pi\alpha(1-\alpha)}{\sin(\pi\alpha)}\in\left[\frac\pi4,1\right).
\label{eq:ch-kappa}
\end{equation}
The factor $\kappa_\alpha$ is the ratio of the integrals of the kernels $e^{(1-\alpha)z}/(1+e^z)$ and $\min\{e^{(1-\alpha)z},e^{-\alpha z}\}$ that govern the noncommuting and the commuting cases, respectively (see~\eqref{eq:ch-kernel-identity} and \cref{ex:ch-commuting}). Under the assumptions of \cref{thm:ch-main}, Cheng and Liu~\cite[Theorem~5.2 and Remark~5.3]{cheng2026multiple} showed that $\sqrt n\,e^{n\xi}P_{e,n}^*(p)$ lies asymptotically in $[A_{\mathrm{cl}}/2,2A_{\mathrm{cl}}]$, which determines $-\ln P_{e,n}^*(p)$ up to an additive term of size at most $\ln2$. \Cref{thm:ch-main} identifies the prefactor exactly.
\end{remark}

\Cref{tab:ch-regimes} summarizes the assumptions and the different leading behaviors.

\begin{table}[tbp]
\centering
\small
\setlength{\tabcolsep}{4pt}
\renewcommand{\arraystretch}{1.18}
\begin{tabular}{@{}>{\raggedright\arraybackslash}p{0.18\textwidth}>{\raggedright\arraybackslash}p{0.25\textwidth}>{\raggedright\arraybackslash}p{0.29\textwidth}>{\raggedright\arraybackslash}p{0.22\textwidth}@{}}
\toprule
Regime & Assumptions & Optimal error $P_{e,n}^*(p)$ & Prefactor and refinements\\
\midrule
Regular, noncommuting & $[\rho,\sigma]\ne0$; interior $\alpha_*$, $V_*>0$; nonlattice & $A_{\mathrm q}\,n^{-1/2}e^{-n\xi}[1+o(1)]$ & Fixed prefactor; every fixed order in $1/n$ under (QN). \Cref{thm:ch-main}.\\[5pt]
Regular, commuting & $[\rho,\sigma]=0$; interior $\alpha_*$, $V_*>0$; nonlattice & $A_{\mathrm{cl}}\,n^{-1/2}e^{-n\xi}[1+o(1)]$ & Classical fixed prefactor. \Cref{ex:ch-commuting}.\\[5pt]
Noncommuting, lattice & Interior $\alpha_*$, $V_*>0$; $Z\in z_0+d\mathbb Z$ & $B_n(p)\,n^{-1/2}e^{-n\xi}[1+o(1)]$ & $B_n(p)$ may oscillate with $n$. \Cref{ex:ch-lattice}.\\[5pt]
Pure states & $r=|\langle\psi,\varphi\rangle|^2\in(0,1)$; $V_*=0$ & $p(1-p)r^n[1+O(r^n)]$ & No $\frac12\ln n$ term; exact formula in~\eqref{eq:ch-pure}.\\

\bottomrule
\end{tabular}
\caption{Asymptotic regimes for a fixed prior $p\in(0,1)$. An interior minimizer means $\alpha_*\in(0,1)$; (QN) is the condition in \cref{def:ch-QN}. The constants $A_{\mathrm q}$ and $A_{\mathrm{cl}}$ are defined in~\eqref{eq:ch-prefactor} and~\eqref{eq:ch-Acl}, and $B_n(p)$ is the lattice prefactor in~\eqref{eq:ch-lattice-Pe}.}
\label{tab:ch-regimes}
\end{table}

As a consequence of \cref{thm:ch-main}, we obtain the sample complexity of symmetric binary quantum hypothesis testing~\cite{cheng2024invitation}, defined for $\delta\in(0,1)$ as
\begin{equation}
n^*(p,\delta)\coloneqq\min\left\{n\in\mathbb N:P_{e,n}^*(p)\le\delta\right\}.
\label{eq:ch-sample-def}
\end{equation}
Refer to \cite[Theorem II.7]{cheng2024invitation} and \cite[Theorem~3]{NuradhaQueryComplexity_2025} for existing bounds on the sample complexity for this setting. 

\begin{corollary}[Sample Complexity up to One Sample]
\label{cor:ch-sample}
Under the assumptions of the first part of \cref{thm:ch-main}, let $x_\delta$ be the unique positive solution of
\begin{equation}
x\,\xi+\frac12\ln x-\ln A_{\mathrm q}=\ln\frac1\delta,
\label{eq:ch-sample-root}
\end{equation}
with $A_{\mathrm q}$ in~\eqref{eq:ch-prefactor}, which exists for every $\delta\in(0,1)$. Then the sample complexity in~\eqref{eq:ch-sample-def} satisfies
\begin{equation}
-o(1)\le n^*(p,\delta)-x_\delta\le1+o(1)\qquad\text{as }\delta\to0,
\label{eq:ch-sample}
\end{equation}
and
\begin{equation}
x_\delta=\frac1\xi\left[\ln\frac1\delta-\frac12\ln\!\left(\frac1\xi\ln\frac1\delta\right)+\ln A_{\mathrm q}\right]+o(1).
\label{eq:ch-sample-expansion}
\end{equation}
\end{corollary}

\begin{proof}
We first show that the function $g(x)\coloneqq x\xi+\frac12\ln x-\ln A_{\mathrm q}$ on $x>0$ has a Lipschitz inverse, and then we compare $-\ln P_{e,n}^*(p)$ with $g(n)$. The function $g$ is continuous with $g'(x)=\xi+\frac1{2x}\ge\xi>0$, and it tends to $-\infty$ as $x\to0$ and to $\infty$ as $x\to\infty$. Hence $g$ is a bijection from $(0,\infty)$ onto $\mathbb R$, so that $x_\delta=g^{-1}(\ln\frac1\delta)$ exists and is unique for every $\delta\in(0,1)$. Moreover, $g^{-1}$ is increasing with derivative $1/g'(g^{-1}(\cdot))\le1/\xi$, so that, by the mean value theorem,
\begin{equation}
\left|g^{-1}(L+\eta)-g^{-1}(L)\right|\le\frac{|\eta|}\xi\qquad\text{for all }L,\eta\in\mathbb R.
\label{eq:ch-ginv}
\end{equation}
Taking the logarithm of~\eqref{eq:ch-prefactor} gives $-\ln P_{e,n}^*(p)=g(n)+\eta_n$ with $\eta_n\to0$. Fix $\eta>0$ and choose $N$ such that $|\eta_n|\le\eta$ for all $n\ge N$.
Since $[\rho,\sigma]\neq0$, $\Tr[\rho\sigma]>0$, so that $P^*_{e,n}(p)>0$ for every finite $n$. Let $\delta_0\coloneqq\min_{n<N}P^*_{e,n}(p)>0$. Since we consider the limit $\delta\rightarrow0$, we may restrict to $0<\delta<\delta_0$. For every $n<N$, we then have $P^*_{e,n}(p)\geq\delta_0>\delta$. Hence $n^*\coloneqq n^*(p,\delta)\geq N$.

\underline{Lower bound}: Since $n^*\ge N$ and $P_{e,n^*}^*(p)\le\delta$ by the definition of $n^*$, we have $\ln\frac1\delta\le-\ln P_{e,n^*}^*(p)=g(n^*)+\eta_{n^*}\le g(n^*)+\eta$. Applying the increasing function $g^{-1}$ and then~\eqref{eq:ch-ginv} gives $n^*\ge g^{-1}(\ln\frac1\delta-\eta)\ge x_\delta-\eta/\xi$.

\underline{Upper bound}: Let $n_1\coloneqq\lceil g^{-1}(\ln\frac1\delta+\eta)\rceil$. Since $g^{-1}(\ln\frac1\delta+\eta)\to\infty$ as $\delta\to0$, we have $n_1\ge N$ for all sufficiently small $\delta$. Then $-\ln P_{e,n_1}^*(p)=g(n_1)+\eta_{n_1}\ge g(n_1)-\eta\ge\ln\frac1\delta$, where the last inequality follows because $g$ is increasing and $n_1\ge g^{-1}(\ln\frac1\delta+\eta)$. Hence $P_{e,n_1}^*(p)\le\delta$, so that $n^*\le n_1$ by the definition of $n^*$ as a minimum, and~\eqref{eq:ch-ginv} gives $n_1\le g^{-1}(\ln\frac1\delta+\eta)+1\le x_\delta+\eta/\xi+1$.

Combining the two bounds, for every $\eta>0$, there exists $\delta_0>0$ such that $-\eta/\xi\le n^*(p,\delta)-x_\delta\le1+\eta/\xi$ for all $\delta<\delta_0$. This proves~\eqref{eq:ch-sample}.

\underline{Expansion of $x_\delta$}: Write $L\coloneqq\ln\frac1\delta$. Since $g(x_\delta)=L$ and $g$ is increasing, $x_\delta\to\infty$ as $L\to\infty$. Moreover, $g(x_\delta)=L$ is equivalent to
\begin{equation}
x_\delta=\frac1\xi\left[L-\frac12\ln x_\delta+\ln A_{\mathrm q}\right].
\label{eq:ch-xdelta-fixed}
\end{equation}
Since $\ln x_\delta\ge0$ for large $L$,~\eqref{eq:ch-xdelta-fixed} gives $x_\delta\le(L+\ln A_{\mathrm q})/\xi$, so that $\ln x_\delta=O(\ln L)$ and, again by~\eqref{eq:ch-xdelta-fixed}, $x_\delta=\frac L\xi\bigl(1+O(\frac{\ln L}L)\bigr)$. Hence $\ln x_\delta=\ln\frac L\xi+O(\frac{\ln L}L)$, and substituting this into~\eqref{eq:ch-xdelta-fixed} proves~\eqref{eq:ch-sample-expansion}, with an $o(1)$ term of order $\ln L/L$.
\end{proof}
In comparison, for sufficiently small $\delta$, the bound $n^*(p,\delta)\le\lceil\ln(p^{\alpha_*}(1-p)^{1-\alpha_*}/\delta)/\xi\rceil$ obtained from~\eqref{eq:ch-audenaert} exceeds the true value by $\frac1{2\xi}\ln(\frac1\xi\ln\frac1\delta)+O(1)$ samples. Furthermore, by \cref{rem:ch-compare}, a pair of noncommuting states needs about $\ln(1/\kappa_{\alpha_*})/\xi$ fewer samples than a pair of commuting states with the same function $\Phi$ (see \cref{ex:ch-petz-curve}). \Cref{tab:sample} compares these predictions with numerical values and with the constant-factor bounds of~\cite[Theorem~II.7]{cheng2024invitation}.

\subsubsection{Which Pairs Are Regular}
\label{sec:ch-regular}

The following result shows that the assumptions of \cref{thm:ch-main} hold for generic pairs of faithful states, and in particular after adding generic noise.

\begin{proposition}[Regular Pairs]
\label{prop:ch-regular-pairs}
The following hold.
\begin{enumerate}
\item Let $\rho\neq\sigma$ be faithful states. Then the function $\Phi$ in~\eqref{eq:ch-Phi} is strictly convex on $[0,1]$ with $\Phi(0)=\Phi(1)=0$, it has a unique minimizer $\alpha_*\in(0,1)$, and $\Phi''(\alpha_*)>0$. Hence $(\rho,\sigma)$ is regular in the sense of \cref{def:ch-regular} if and only if it is nonlattice.
\item Let $\rho$ and $\sigma$ be faithful qubit states whose Bloch vectors have lengths $a,b\in(0,1)$ and are not collinear. Then $(\rho,\sigma)$ is regular if and only if $\theta\coloneqq\operatorname{artanh}(a)/\operatorname{artanh}(b)$ is irrational. If, moreover, there are constants $c,\mu>0$ such that $|k\theta-l|\ge c\,k^{-\mu}$ for all integers $k\ge1$ and $l$, then $(\rho,\sigma)$ satisfies (QN) in \cref{def:ch-QN}; more precisely, for each $\alpha\in(0,1)$, the gap in~\eqref{eq:ch-QN} is bounded below by a positive constant times $|u|^{-2\mu}$ for all sufficiently large $|u|$. This holds for almost every $\theta$ with respect to the Lebesgue measure~\cite{khintchine1924}, and, by Baker's theorem on linear forms in logarithms~\cite{baker1975transcendental}, whenever $\theta$ is irrational and $a$ and $b$ are algebraic numbers.
\item Let $\rho$ and $\sigma$ be states on $\mathbb C^d$ with $[\rho,\sigma]\neq0$, and for $\eps,\eps'\in(0,1)$ let $\rho_\eps\coloneqq(1-\eps)\rho+\eps\pi$ and $\sigma_{\eps'}\coloneqq(1-\eps')\sigma+\eps'\pi$, where $\pi$ is the maximally mixed state. Then, for every fixed $\eps'$, the pair $(\rho_\eps,\sigma_{\eps'})$ is regular and noncommuting for all $\eps\in(0,1)$ except at most countably many, and it satisfies (QN) for almost every $\eps\in(0,1)$ with respect to the Lebesgue measure.
\end{enumerate}
\end{proposition}

\begin{proof}
    The proof is given in Appendix~\ref{sec:ch-prop-regular-pairs}.
\end{proof}

\begin{remark}[Examples]
\label{rem:ch-regular-examples}
\Cref{prop:ch-regular-pairs} applies to several natural classes of states. 

\emph{Noisy pure states:} For pure qubit states $\psi$ and $\varphi$ with $|\langle\psi,\varphi\rangle|\in(0,1)$ and $\eps,\eps'\in(0,1)$, the states $(1-\eps)\psi+\eps\pi$ and $(1-\eps')\varphi+\eps'\pi$ have Bloch vectors of lengths $1-\eps$ and $1-\eps'$. If $\eps$ and $\eps'$ are rational, the pair satisfies all assumptions of \cref{thm:ch-main}, including (QN), unless $\operatorname{artanh}(1-\eps)/\operatorname{artanh}(1-\eps')$ is rational. \Cref{ex:ch-numerics} is of this form. 

\emph{Thermal states:} For qubit Gibbs states $\rho\propto e^{-\beta_1H_1}$ and $\sigma\propto e^{-\beta_2H_2}$ with $[H_1,H_2]\neq0$ and energy gaps $\Delta_1,\Delta_2>0$, we have $\theta=\beta_1\Delta_1/(\beta_2\Delta_2)$, so that the pair is regular unless the ratio of the inverse temperatures is a rational multiple of $\Delta_2/\Delta_1$. 

\emph{Equal noise need not help:} For faithful noncommuting qubit states with $\sigma=U\rho U^\dagger$ for a unitary $U$, the values of $Z$ are $0$ and $\pm\ln(r_1/r_2)$, where $r_1>r_2$ are the eigenvalues of $\rho$, so the pair is lattice, also after adding the same amount of depolarizing noise to both states. Then \cref{thm:ch-main} does not apply, and the prefactor is given by the lattice formula in \cref{ex:ch-lattice}. Adding different amounts of noise, as in the third item of \cref{prop:ch-regular-pairs}, generically restores the nonlattice condition.
\end{remark}

\subsubsection{Main Tools: Harmonic Overlap and its Representations}
\label{sec:ch-tools}

We now introduce the tools used in the proofs. For $A,B\ge0$ and $\alpha\in(0,1)$, define
\begin{align}
Q_\alpha(A,B)&\coloneqq\Tr\!\left[A^\alpha B^{1-\alpha}\right],
\label{eq:ch-Qalpha}\\
Q_{\alpha,2}(A,B)&\coloneqq\Tr\!\left[\left(B^{(1-\alpha)/4}A^{\alpha/2}B^{(1-\alpha)/4}\right)^2\right].
\label{eq:ch-Qalpha2}
\end{align}
The first is the Petz R\'enyi quasi-entropy and the second is the $\alpha$-$z$ R\'enyi quasi-entropy with $z=2$. Both are multiplicative under tensor products, homogeneous of degree one under the common scaling $(A,B)\mapsto(cA,cB)$, and satisfy $Q_\alpha(A,tB)=t^{1-\alpha}Q_\alpha(A,B)$ and $Q_{\alpha,2}(A,tB)=t^{1-\alpha}Q_{\alpha,2}(A,B)$.

\begin{definition}[Harmonic Overlap]
\label{def:ch-Har}
For $A,B\ge0$, define
\begin{equation}
\Har(A,B)\coloneqq\min_{X}\left\{\left\lVert A^{1/2}(\id-X)\right\rVert_2^2+\left\lVert XB^{1/2}\right\rVert_2^2\right\},
\label{eq:ch-Har-def}
\end{equation}
where the minimum is over all linear operators $X$.
\end{definition}

\begin{lemma}[Properties of the Harmonic Overlap]
\label{lem:ch-harmonic}
Let $A=\sum_ia_i|u_i\rangle\!\langle u_i|$ and $B=\sum_jb_j|v_j\rangle\!\langle v_j|$ be spectral decompositions with orthonormal bases $\{u_i\}_i$ and $\{v_j\}_j$. Then the harmonic overlap in~\eqref{eq:ch-Har-def} satisfies
\begin{equation}
\Har(A,B)=\sum_{i,j}\frac{a_ib_j}{a_i+b_j}\left|\langle u_i,v_j\rangle\right|^2,
\label{eq:ch-Har-spectral}
\end{equation}
where terms with $a_i=b_j=0$ are set to zero. Moreover, $\Har(A,B)\le m(A,B)$ for $m(\cdot, \cdot)$ in~\eqref{eq:m-def}, and $\Har(\cdot, \cdot)$ is super-additive: $\Har(\sum_jA_j,\sum_jB_j)\ge\sum_j\Har(A_j,B_j)$.
\end{lemma}

\begin{proof}
    The proof is given in Appendix~\ref{sec:ch-auxiliary}.
\end{proof}

In particular, for commuting operators $A=\sum_xa_x\proj x$ and $B=\sum_xb_x\proj x$, we have that $\Har(A,B)=\sum_xa_xb_x/(a_x+b_x)$, the harmonic analogue of~\eqref{eq:m-commuting}.

\begin{theorem}[Comparisons with the Noncommutative Minimum]
\label{thm:ch-three-bounds}
For all $A,B\ge0$ and $\alpha\in(0,1)$, the quantities in~\eqref{eq:m-def}, \eqref{eq:ch-Qalpha}, \eqref{eq:ch-Qalpha2}, and~\eqref{eq:ch-Har-def} satisfy
\begin{align}
\Har(A,B)&\le m(A,B)\le2\Har(A,B),
\label{eq:ch-Har-m}\\
m(A,B)&\le Q_\alpha(A,B),
\label{eq:ch-audenaert}\\
0\le m(A,B)-\Har(A,B)&\le\frac12Q_{\alpha,2}(A,B).
\label{eq:ch-remainder}
\end{align}
Furthermore, the constant $1/2$ in~\eqref{eq:ch-remainder} cannot be improved for any $\alpha\in(0,1)$.
\end{theorem}

\begin{proof}
\underline{Rank-one operators}: If $A$ or $B$ is zero, all the quantities vanish. Otherwise, let $A=a|u\rangle\!\langle u|$ and $B=b|v\rangle\!\langle v|$ with $\lVert u\rVert=\lVert v\rVert=1$ and $a,b>0$, and set $r\coloneqq|\langle u,v\rangle|^2$ and $x\coloneqq m(A,B)$. By~\eqref{eq:xquad} in Appendix~\ref{Sec:Lemma_rank_one} and~\eqref{eq:ch-Har-spectral},
\begin{equation}
x(a+b-x)=abr,
\qquad
\Har(A,B)=\frac{abr}{a+b},
\qquad
Q_\alpha(A,B)=a^\alpha b^{1-\alpha}r,
\qquad
Q_{\alpha,2}(A,B)=a^\alpha b^{1-\alpha}r^2.
\label{eq:ch-rank-one-identities}
\end{equation}
Since $0\le x\le\min\{a,b\}$, we have $\max\{a,b\}\le a+b-x\le a+b$, so that the first identity in~\eqref{eq:ch-rank-one-identities} gives
\begin{equation}
\Har(A,B)=\frac{abr}{a+b}\le \frac{abr}{a+b-x}=x\le\frac{abr}{\max\{a,b\}}=r\min\{a,b\}.
\label{eq:ch-rank-one-x}
\end{equation}
This proves $\Har(A,B)\le m(A,B)$. Furthermore,
\begin{align}
x&=\frac{abr}{a+b-x}\le\frac{abr}{(a+b)/2}=2\Har(A,B),\\
x&\le r\min\{a,b\}\le a^\alpha b^{1-\alpha}r=Q_\alpha(A,B),\\
x-\Har(A,B)&=x-\frac{x(a+b-x)}{a+b}=\frac{x^2}{a+b}\le\frac{r^2\min\{a,b\}^2}{2\min\{a,b\}}\le\frac12a^\alpha b^{1-\alpha}r^2=\frac12Q_{\alpha,2}(A,B),
\end{align}
where the first line follows from $x\le\min\{a,b\}\le(a+b)/2$, the second line from~\eqref{eq:ch-rank-one-x} and $\min\{a,b\}\le a^\alpha b^{1-\alpha}$, and the third line from~\eqref{eq:ch-rank-one-identities}, \eqref{eq:ch-rank-one-x}, $a+b\ge2\min\{a,b\}$, and $\min\{a,b\}\le a^\alpha b^{1-\alpha}$.

\underline{General operators}: Let $(A_j,B_j)$ be the pairs given by \cref{lem:decomposition_general_to_rank}, so that $m(A,B)=\sum_jm(A_j,B_j)$. Recall that $\Har$ is super-additive by \cref{lem:ch-harmonic}, and that $Q_\alpha$ and $Q_{\alpha,2}$ are super-additive by \cref{lem:ch-trace-concavity} in Appendix~\ref{sec:ch-auxiliary}. Then
\begin{align}
m(A,B)-\Har(A,B)&\le\sum_j\left[m(A_j,B_j)-\Har(A_j,B_j)\right]
\le\frac12\sum_jQ_{\alpha,2}(A_j,B_j)
\le\frac12Q_{\alpha,2}(A,B),\\
m(A,B)&=\sum_jm(A_j,B_j)\le2\sum_j\Har(A_j,B_j)\le2\Har(A,B),\\
m(A,B)&=\sum_jm(A_j,B_j)\le\sum_jQ_\alpha(A_j,B_j)\le Q_\alpha(A,B),
\end{align}
where the first inequality in the first line follows from $m(A,B)=\sum_jm(A_j,B_j)$ and the super-additivity of $\Har$, the middle inequality in each line from the rank-one case applied to each pair $(A_j,B_j)$, and the last inequality in each line from the super-additivity of $Q_{\alpha,2}$, $\Har$, and $Q_\alpha$, respectively. The lower bound in~\eqref{eq:ch-Har-m} is in \cref{lem:ch-harmonic}. For the optimality of the constant $1/2$, take $A=B=|0\rangle\!\langle0|$, for which $m(A,B)=1$, $\Har(A,B)=1/2$, and $Q_{\alpha,2}(A,B)=1$.
\end{proof}

The inequality~\eqref{eq:ch-audenaert} is the Audenaert inequality~\cite{audenaert2007chernoff}, and the comparison~\eqref{eq:ch-Har-m} with the harmonic overlap was established in~\cite[Theorem~4.2]{cheng2026multiple}. Here, the exact paired rank-one decomposition gives both of them, together with the remainder bound~\eqref{eq:ch-remainder}, which, for noncommuting faithful states, is exponentially smaller than the optimal error as shown next.

\begin{corollary}[Exponentially Small Remainder for Noncommuting States]
\label{cor:ch-gap}
For states $\rho,\sigma$ and $\alpha\in(0,1)$, the quantities in~\eqref{eq:ch-Qalpha2} and~\eqref{eq:ch-Har-def} satisfy
\begin{equation}
\Tr\!\left[\rho^\alpha\sigma^{1-\alpha}\right]-Q_{\alpha,2}(\rho,\sigma)=\frac12\left\lVert\left[\rho^{\alpha/2},\sigma^{(1-\alpha)/2}\right]\right\rVert_2^2.
\label{eq:ch-commutator}
\end{equation}
In particular, $Q_{\alpha,2}(\rho,\sigma)<\Tr[\rho^\alpha\sigma^{1-\alpha}]$ if $[\rho,\sigma]\neq0$. Furthermore, for all $t>0$ and $n\in\mathbb N$,
\begin{equation}
0\le t^{\alpha-1}\left[m\!\left(\rho^{\otimes n},t\sigma^{\otimes n}\right)-\Har\!\left(\rho^{\otimes n},t\sigma^{\otimes n}\right)\right]\le\frac12\left[Q_{\alpha,2}(\rho,\sigma)\right]^n.
\label{eq:ch-uniform-remainder}
\end{equation}
\end{corollary}

\begin{proof}
Let $X\coloneqq\rho^{\alpha/2}$ and $Y\coloneqq\sigma^{(1-\alpha)/2}$. Then
\begin{align}
\left\lVert XY-YX\right\rVert_2^2&=\Tr[(YX-XY)(XY-YX)]
=2\Tr[X^2Y^2]-2\Tr[(XY)^2]\\
&=2\Tr[\rho^\alpha\sigma^{1-\alpha}]-2\Tr\!\left[(Y^{1/2}XY^{1/2})^2\right]
=2\Tr[\rho^\alpha\sigma^{1-\alpha}]-2Q_{\alpha,2}(\rho,\sigma),
\end{align}
where the first equality follows because $X$ and $Y$ are Hermitian, the second and third equalities from the cyclicity of the trace, and the last equality from~\eqref{eq:ch-Qalpha2}. This proves~\eqref{eq:ch-commutator}. 

If $[\rho,\sigma]\neq0$, then $[\rho^{\alpha/2},\sigma^{(1-\alpha)/2}]\neq0$, because $\rho$ and $\sigma$ are functions of $\rho^{\alpha/2}$ and $\sigma^{(1-\alpha)/2}$, respectively. Finally,~\eqref{eq:ch-remainder} with $A=\rho^{\otimes n}$ and $B=t\sigma^{\otimes n}$ bounds the difference $m\!\left(\rho^{\otimes n},t\sigma^{\otimes n}\right)-\Har\!\left(\rho^{\otimes n},t\sigma^{\otimes n}\right)$ by $\frac12Q_{\alpha,2}(\rho^{\otimes n},t\sigma^{\otimes n})=\frac12t^{1-\alpha}[Q_{\alpha,2}(\rho,\sigma)]^n$, where the equality follows from the multiplicativity and the scaling property of $Q_{\alpha,2}$. Multiplying by $t^{\alpha-1}$ proves~\eqref{eq:ch-uniform-remainder}.
\end{proof}

{For a regular pair, we define}
\begin{equation}
\Delta_*\coloneqq\ln\frac{\Tr[\rho^{\alpha_*}\sigma^{1-\alpha_*}]}{Q_{\alpha_*,2}(\rho,\sigma)},
\label{eq:ch-Delta}
\end{equation}
which is positive for noncommuting states by \cref{cor:ch-gap}. {In the proof of \cref{thm:ch-main}, $\Delta_*$ is a lower bound on the exponential decay rate of the remainder relative to the optimal error (see~\eqref{eq:ch-rn-ratio}).}
{Since~\eqref{eq:ch-uniform-remainder} holds for every $\alpha\in(0,1)$, the same argument also gives the lower bound $\ln\bigl(\Tr[\rho^{\alpha_*}\sigma^{1-\alpha_*}]/\inf_{\alpha\in(0,1)}Q_{\alpha,2}(\rho,\sigma)\bigr)\ge\Delta_*$.}
By~\eqref{eq:ch-commutator}, $\Delta_*$ is governed by the commutator $[\rho^{\alpha_*/2},\sigma^{(1-\alpha_*)/2}]$, which can be small even when $[\rho,\sigma]$ is not, for instance when both states are close to the maximally mixed state.
{In that case, the remainder estimate~\eqref{eq:ch-rn-ratio} becomes small only for large $n$, and, for a given $n$, the finite-$n$ bounds in~\eqref{eq:ch-finite-bounds} control the remainder quantitatively.} 

\begin{proposition}[Classical Representation of the Harmonic Overlap]
\label{prop:ch-classical-rep}
Regard the Nussbaum--Szko{\l}a distributions in~\eqref{eq:ch-NS} as diagonal operators. Then, for all $n\in\mathbb N$ and $t>0$, the harmonic overlap in~\eqref{eq:ch-Har-def} satisfies
\begin{equation}
\Har\!\left(\rho^{\otimes n},t\sigma^{\otimes n}\right)=\Har\!\left(P_{\NS}^{\otimes n},tQ_{\NS}^{\otimes n}\right)=\sum_{x}\frac{P_{\NS}^{\otimes n}(x)\,tQ_{\NS}^{\otimes n}(x)}{P_{\NS}^{\otimes n}(x)+tQ_{\NS}^{\otimes n}(x)}.
\label{eq:ch-classical-rep}
\end{equation}
\end{proposition}

\begin{proof}
For $n=1$, we have that
\begin{equation}
\Har(\rho,t\sigma)=\sum_{i,j}\frac{r_its_j}{r_i+ts_j}|\langle u_i,v_j\rangle|^2=\sum_{i,j}\frac{P_{\NS}(i,j)\,tQ_{\NS}(i,j)}{P_{\NS}(i,j)+tQ_{\NS}(i,j)},
\end{equation}
where the first equality follows from~\eqref{eq:ch-Har-spectral}, and the second from multiplying the numerator and the denominator by $|\langle u_i,v_j\rangle|^2$ and using~\eqref{eq:ch-NS}; terms with $r_i=0$ or $s_j=0$ vanish on both sides. For general $n$, the eigenvalues of $\rho^{\otimes n}$ and $\sigma^{\otimes n}$ are the products $r_{i_1}\cdots r_{i_n}$ and $s_{j_1}\cdots s_{j_n}$ with the eigenvectors $u_{i_1}\otimes\cdots\otimes u_{i_n}$ and $v_{j_1}\otimes\cdots\otimes v_{j_n}$, and $|\langle u_{i_1}\otimes\cdots\otimes u_{i_n},v_{j_1}\otimes\cdots\otimes v_{j_n}\rangle|^2=\prod_k|\langle u_{i_k},v_{j_k}\rangle|^2$. Hence the same computation applies with $P_{\NS}^{\otimes n}$ and $Q_{\NS}^{\otimes n}$ in place of $P_{\NS}$ and $Q_{\NS}$.
\end{proof}

\medskip
In the following, we use the shorthand $P^{(n)}\coloneqq P_{\NS}^{\otimes n}$ and $Q^{(n)}\coloneqq Q_{\NS}^{\otimes n}$, and we denote their outcomes by $\mathbf k$. To state the integral representation, we use the functions
\begin{equation}
g_\alpha(z)\coloneqq\frac{e^{(1-\alpha)z}}{1+e^z},
\qquad
\Lambda(\beta)\coloneqq\ln\frac{\pi}{\sin(\pi\beta)},
\label{eq:ch-kernel}
\end{equation}
where $z\in\mathbb R$ and $0<\operatorname{Re}\beta<1$, and we use the branch of the logarithm with $\Lambda$ real on $(0,1)$. Note that $\Lambda'(\alpha)=-\pi\cot(\pi\alpha)$ and $\Lambda''(\alpha)=\pi^2/\sin^2(\pi\alpha)$, so that $\tau(\alpha)=\Lambda'(\alpha)-\Phi'''(\alpha)/(2\Phi''(\alpha))$ in~\eqref{eq:ch-tau}.

\begin{proposition}[Integral Representations of the Harmonic Overlap]
\label{prop:ch-integral-rep}
For all $\alpha\in(0,1)$, $n\in\mathbb N$, and $x\in\mathbb R$, with $Q_\beta^{\NS}$, $S_n$, $g_\alpha$, and $\Lambda$ as in~\eqref{eq:ch-F}, \eqref{eq:ch-tilt}, \eqref{eq:S_n_with_alpha}, and~\eqref{eq:ch-kernel} (we assume $\Tr[\rho\sigma]>0$, so that the tilted law in~\eqref{eq:ch-tilt} is defined),
\begin{align}
e^{(\alpha-1)x}\,\Har\!\left(\rho^{\otimes n},e^x\sigma^{\otimes n}\right)
&=\bigl(Q_{\alpha}^{\NS}\bigr)^{n}\,\mathbb E_\alpha\!\left[g_\alpha(S_n-x)\right]
\label{eq:ch-kernel-identity}\\
&=\frac1{2\pi}\int_{\mathbb R}\exp\!\left(\Lambda(\alpha+iu)+n\ln Q_{\alpha+iu}^{\NS}-iux\right)du,
\label{eq:ch-MB}
\end{align}
where $e^{n\ln Q_{\alpha+iu}^{\NS}}$ stands for $\bigl(Q_{\alpha+iu}^{\NS}\bigr)^{n}$.
\end{proposition}

\begin{proof}
For $P,Q>0$ with $P/Q=e^z$, we have that
\begin{equation}
e^{(\alpha-1)x}\frac{P\,e^xQ}{P+e^xQ}=P^\alpha Q^{1-\alpha}\,\frac{e^{(1-\alpha)(z-x)}}{1+e^{z-x}}=P^\alpha Q^{1-\alpha}\,g_\alpha(z-x).
\label{eq:ch-term-identity}
\end{equation}
Let $S_n(\mathbf k)\coloneqq\ln(P^{(n)}(\mathbf k)/Q^{(n)}(\mathbf k))$ on the common support of $P^{(n)}$ and $Q^{(n)}$. Then
\begin{align}
e^{(\alpha-1)x}\Har\!\left(\rho^{\otimes n},e^x\sigma^{\otimes n}\right)
&=\sum_{\mathbf k}e^{(\alpha-1)x}\frac{P^{(n)}(\mathbf k)\,e^xQ^{(n)}(\mathbf k)}{P^{(n)}(\mathbf k)+e^xQ^{(n)}(\mathbf k)}\\
&=\sum_{\mathbf k}P^{(n)}(\mathbf k)^\alpha Q^{(n)}(\mathbf k)^{1-\alpha}\,g_\alpha(S_n(\mathbf k)-x)
=\bigl(Q_{\alpha}^{\NS}\bigr)^{n}\,\mathbb E_\alpha\!\left[g_\alpha(S_n-x)\right],
\end{align}
where the first equality follows from \cref{prop:ch-classical-rep} with $t=e^x$, the second from~\eqref{eq:ch-term-identity}, since terms outside the common support vanish, and the last from~\eqref{eq:ch-tilt}. This proves~\eqref{eq:ch-kernel-identity}.

For~\eqref{eq:ch-MB}, we compute the Fourier transform of $g_\alpha$. For $0<\operatorname{Re}w<1$, $\int_{\mathbb R}e^{wz}(1+e^z)^{-1}dz=\pi/\sin(\pi w)$. Taking $w=1-\alpha-iu$ and using $\sin(\pi(1-w))=\sin(\pi w)$ gives
\begin{equation}
\widehat g_\alpha(u)\coloneqq\int_{\mathbb R}g_\alpha(z)e^{-iuz}\,dz=\frac{\pi}{\sin(\pi(\alpha+iu))}=e^{\Lambda(\alpha+iu)}.
\label{eq:ch-g-fourier}
\end{equation}
Since $|\sin(\pi(\alpha+iu))|\ge\sinh(\pi|u|)$, the function $\widehat g_\alpha$ is integrable, and Fourier inversion gives $g_\alpha(z)=(2\pi)^{-1}\int\widehat g_\alpha(u)e^{iuz}du$. Then
\begin{align}
\bigl(Q_{\alpha}^{\NS}\bigr)^{n}\mathbb E_\alpha\!\left[g_\alpha(S_n-x)\right]
&=\frac1{2\pi}\int_{\mathbb R}\widehat g_\alpha(u)\bigl(Q_{\alpha}^{\NS}\bigr)^{n}\mathbb E_\alpha\!\left[e^{iuS_n}\right]e^{-iux}\,du\\
&=\frac1{2\pi}\int_{\mathbb R}e^{\Lambda(\alpha+iu)}\bigl(Q_{\alpha+iu}^{\NS}\bigr)^{n}e^{-iux}\,du,
\end{align}
where the first equality follows from Fourier inversion and Fubini's theorem, and the second from~\eqref{eq:ch-g-fourier} and $\bigl(Q_{\alpha}^{\NS}\bigr)^{n}\mathbb E_\alpha[e^{iuS_n}]=\bigl(Q_{\alpha+iu}^{\NS}\bigr)^{n}$. Together with~\eqref{eq:ch-kernel-identity}, this proves~\eqref{eq:ch-MB}.
\end{proof}

By~\eqref{eq:ch-kernel-identity}, the local limit theorem gives the leading order of the harmonic overlap. Let
\begin{equation}
    \phi_v(y)\coloneqq \frac{e^{-y^2/(2v)}}{\sqrt{2\pi v} }\label{eq:Gaussian_density}
\end{equation}
denote the Gaussian density with variance $v>0$.

\begin{lemma}[Leading Order of the Harmonic Overlap]
\label{lem:ch-LLT-harmonic}
Let $\alpha\in(0,1)$ with $\Phi''(\alpha)>0$, and suppose that the pair $(\rho,\sigma)$ is nonlattice. Then, uniformly in $x\in\mathbb R$,
\begin{equation}
\frac{e^{(\alpha-1)x}\Har(\rho^{\otimes n},e^x\sigma^{\otimes n})}{\bigl(Q_{\alpha}^{\NS}\bigr)^{n}}=\frac{\pi}{\sin(\pi\alpha)}\,\phi_{n\Phi''(\alpha)}\!\left(x-n\Phi'(\alpha)\right)+o(n^{-1/2}).
\label{eq:ch-LLT-harmonic}
\end{equation}
\end{lemma}

\begin{proof}
By~\eqref{eq:ch-kernel-identity}, the left-hand side of~\eqref{eq:ch-LLT-harmonic} is equal to $\mathbb E_\alpha[g_\alpha(S_n-x)]$. We apply \cref{lem:ch-LLT} in Appendix~\ref{sec:ch-auxiliary} to $Z$ under $\Pr_\alpha$, which is nonlattice and has the mean $\Phi'(\alpha)$ and the variance $\Phi''(\alpha)>0$ by~\eqref{eq:ch-tilt-moments}, and to $h=g_\alpha$, which is continuous and satisfies $g_\alpha(z)\le e^{-\min\{\alpha,1-\alpha\}|z|}$ and $\int g_\alpha=\widehat g_\alpha(0)=\pi/\sin(\pi\alpha)$ by~\eqref{eq:ch-g-fourier}. This gives~\eqref{eq:ch-LLT-harmonic}.
\end{proof}

Under (QN), Laplace's method applied to~\eqref{eq:ch-MB} gives the following local expansion of the harmonic overlap. Its proof is given in Appendix~\ref{sec:ch-auxiliary}, and the computation of its coefficients in Appendix~\ref{sec:ch-coefficients}.

\begin{proposition}[Local Expansion of the Harmonic Overlap]
\label{prop:ch-local-expansion}
Let $\alpha\in(0,1)$ with $\Phi''(\alpha)>0$, and suppose that the pair $(\rho,\sigma)$ satisfies (QN) in \cref{def:ch-QN}. Then, for every fixed integer $M\ge1$ and every $Y>0$, uniformly for $|y|\le Y$,
\begin{equation}
\begin{aligned}
\ln\frac{e^{(\alpha-1)x}\Har(\rho^{\otimes n},e^x\sigma^{\otimes n})}{\bigl(Q_{\alpha}^{\NS}\bigr)^{n}}\bigg|_{x=n\Phi'(\alpha)+y}
&=\Lambda(\alpha)-\frac12\ln\!\left(2\pi n\Phi''(\alpha)\right)\\
&\quad+\sum_{j=1}^{M}\frac{c_j(y)}{n^j}+O(n^{-M-1}),
\end{aligned}
\label{eq:ch-local-expansion}
\end{equation}
where $c_j(y)$ is given by~\eqref{eq:app-coefficient} in Appendix~\ref{sec:ch-coefficients}; in particular, $c_1$ and $c_2$ are the polynomials in~\eqref{eq:ch-c1} and~\eqref{eq:ch-c2}. Each $c_j(y)$ is a polynomial in $y$ and involves the derivatives of $\Phi$ up to order $2j+2$ and of $\Lambda$ up to order $2j$. \end{proposition}

These tools also recover the quantum Chernoff exponent~\cite{audenaert2007chernoff,nussbaum2009chernoff}: \eqref{eq:ch-audenaert} gives $P_{e,n}^*(p)\le p^\alpha(1-p)^{1-\alpha}(\Tr[\rho^\alpha\sigma^{1-\alpha}])^n$, and, by \cref{lem:ch-harmonic,prop:ch-classical-rep} and $ab/(a+b)\ge\frac12\min\{a,b\}$, $P_{e,n}^*(p)$ is at least one half of the optimal classical error for $P^{(n)}$ and $Q^{(n)}$, whose exponent is $\xi$~\cite{chernoff1952measure}.

\subsubsection{Comparison with Numerical Computations}
\label{sec:ch-numerics}

The tools above also give rigorous upper and lower bounds for every finite $n$. By~\eqref{eq:ch-prior-profile}, \eqref{eq:ch-Har-m}, and~\eqref{eq:ch-uniform-remainder} at $t=(1-p)/p$, for every $\alpha\in(0,1)$,
\begin{equation}
p\,\Har\!\left(\rho^{\otimes n},\tfrac{1-p}p\sigma^{\otimes n}\right)\le P_{e,n}^*(p)\le p\,\Har\!\left(\rho^{\otimes n},\tfrac{1-p}p\sigma^{\otimes n}\right)+\frac12p^\alpha(1-p)^{1-\alpha}\left[Q_{\alpha,2}(\rho,\sigma)\right]^n,
\label{eq:ch-finite-bounds}
\end{equation}
where the harmonic overlap is computed from the Nussbaum--Szko{\l}a distributions by \cref{prop:ch-classical-rep}. We now compare these predictions with numerical evaluations.

\begin{figure}[t]
\centering
\begin{tikzpicture}
\begin{groupplot}[group style={group size=2 by 1, horizontal sep=19mm}, vizaxis, width=0.48\textwidth, height=0.40\textwidth, xmin=0, xmax=205]
\nextgroupplot[xlabel={number of copies $n$}, ylabel={$\sqrt n\,e^{n\xi}P_{e,n}^*(1/2)$}, xmin=100, xmax=200, xtick={100,125,150,175,200}, ymin=0.68, ymax=1.10, ytick={0.7,0.8,0.9,1,1.1}, title={(a) Rescaled error, $p=1/2$}, legend style={at={(0.5,-0.25)}, anchor=north, legend columns=1}]
\addplot[draw=none,fill=vizaqua!20,forget plot] coordinates {(100,1.060992219887) (101,1.05612253534) (102,1.051305160879) (103,1.046540223902) (104,1.04182781458) (105,1.037167987387) (106,1.032560762615) (107,1.028006127873) (108,1.023504039574) (109,1.019054424417) (110,1.01465718087) (111,1.010312180644) (112,1.006019270163) (113,1.001778272037) (114,9.975889865162e-1) (115,9.934511929405e-1) (116,9.893646511689e-1) (117,9.853291029861e-1) (118,9.813442734813e-1) (119,9.774098723909e-1) (120,9.735255954006e-1) (121,9.696911253979e-1) (122,9.659061336705e-1) (123,9.62170281044e-1) (124,9.584832189526e-1) (125,9.548445904398e-1) (126,9.512540310828e-1) (127,9.4771116984e-1) (128,9.442156298173e-1) (129,9.407670289538e-1) (130,9.373649806263e-1) (131,9.340090941745e-1) (132,9.306989753478e-1) (133,9.274342266794e-1) (134,9.242144477907e-1) (135,9.210392356308e-1) (136,9.179081846601e-1) (137,9.14820886981e-1) (138,9.117769324277e-1) (139,9.087759086183e-1) (140,9.058174009818e-1) (141,9.029009927651e-1) (142,9.000262650301e-1) (143,8.971927966473e-1) (144,8.944001642956e-1) (145,8.916479424731e-1) (146,8.88935703527e-1) (147,8.862630177072e-1) (148,8.836294532486e-1) (149,8.810345764866e-1) (150,8.784779520066e-1) (151,8.759591428314e-1) (152,8.734777106455e-1) (153,8.710332160561e-1) (154,8.6862521889e-1) (155,8.662532785225e-1) (156,8.639169542361e-1) (157,8.616158056038e-1) (158,8.593493928927e-1) (159,8.571172774818e-1) (160,8.549190222878e-1) (161,8.527541921941e-1) (162,8.506223544737e-1) (163,8.485230792025e-1) (164,8.464559396544e-1) (165,8.444205126735e-1) (166,8.424163790171e-1) (167,8.404431236648e-1) (168,8.385003360885e-1) (169,8.365876104808e-1) (170,8.347045459383e-1) (171,8.328507465977e-1) (172,8.310258217238e-1) (173,8.292293857504e-1) (174,8.274610582728e-1) (175,8.257204639962e-1) (176,8.240072326409e-1) (177,8.223209988087e-1) (178,8.206614018145e-1) (179,8.190280854881e-1) (180,8.174206979517e-1) (181,8.158388913782e-1) (182,8.142823217382e-1) (183,8.127506485394e-1) (184,8.112435345672e-1) (185,8.09760645631e-1) (186,8.083016503225e-1) (187,8.068662197921e-1) (188,8.054540275475e-1) (189,8.040647492795e-1) (190,8.026980627189e-1) (191,8.013536475264e-1) (192,8.000311852197e-1) (193,7.987303591365e-1) (194,7.97450854436e-1) (195,7.961923581377e-1) (196,7.949545591959e-1) (197,7.937371486081e-1) (198,7.925398195546e-1) (199,7.913622675655e-1) (200,7.90204190711e-1) (200,7.178969569494e-1) (199,7.178195534094e-1) (198,7.177414593374e-1) (197,7.176626740659e-1) (196,7.175831957337e-1) (195,7.175030210767e-1) (194,7.174221452442e-1) (193,7.173405616427e-1) (192,7.172582618121e-1) (191,7.171752353384e-1) (190,7.170914698018e-1) (189,7.170069507648e-1) (188,7.169216617993e-1) (187,7.168355845524e-1) (186,7.167486988497e-1) (185,7.166609828347e-1) (184,7.165724131401e-1) (183,7.164829650888e-1) (182,7.16392612919e-1) (181,7.163013300299e-1) (180,7.162090892411e-1) (179,7.161158630603e-1) (178,7.16021623955e-1) (177,7.159263446187e-1) (176,7.158299982291e-1) (175,7.157325586891e-1) (174,7.156340008477e-1) (173,7.155343006935e-1) (172,7.154334355162e-1) (171,7.153313840342e-1) (170,7.152281264813e-1) (169,7.151236446532e-1) (168,7.150179219107e-1) (167,7.149109431385e-1) (166,7.148026946607e-1) (165,7.146931641139e-1) (164,7.145823402799e-1) (163,7.144702128812e-1) (162,7.143567723425e-1) (161,7.142420095247e-1) (160,7.141259154342e-1) (159,7.140084809151e-1) (158,7.138896963307e-1) (157,7.137695512396e-1) (156,7.136480340741e-1) (155,7.135251318281e-1) (154,7.134008297601e-1) (153,7.13275111118e-1) (152,7.131479568923e-1) (151,7.130193456018e-1) (150,7.128892531176e-1) (149,7.127576525284e-1) (148,7.12624514051e-1) (147,7.124898049864e-1) (146,7.123534897243e-1) (145,7.122155297945e-1) (144,7.120758839644e-1) (143,7.11934508381e-1) (142,7.117913567534e-1) (141,7.116463805712e-1) (140,7.114995293554e-1) (139,7.11350750934e-1) (138,7.111999917375e-1) (137,7.110471971062e-1) (136,7.10892311602e-1) (135,7.107352793181e-1) (134,7.105760441775e-1) (133,7.104145502148e-1) (132,7.102507418318e-1) (131,7.100845640223e-1) (130,7.099159625583e-1) (129,7.097448841332e-1) (128,7.095712764574e-1) (127,7.093950883024e-1) (126,7.092162694908e-1) (125,7.090347708315e-1) (124,7.088505439991e-1) (123,7.086635413594e-1) (122,7.08473715742e-1) (121,7.082810201643e-1) (120,7.08085407511e-1) (119,7.078868301734e-1) (118,7.076852396571e-1) (117,7.074805861621e-1) (116,7.072728181455e-1) (115,7.07061881873e-1) (114,7.068477209685e-1) (113,7.066302759702e-1) (112,7.064094839004e-1) (111,7.061852778593e-1) (110,7.059575866475e-1) (109,7.057263344275e-1) (108,7.054914404273e-1) (107,7.052528186943e-1) (106,7.050103779018e-1) (105,7.047640212126e-1) (104,7.045136462012e-1) (103,7.04259144836e-1) (102,7.0400040352e-1) (101,7.037373031902e-1) (100,7.034697194705e-1)} -- cycle;
\addplot[vizaqua!75!black,line width=0.8pt,forget plot] coordinates {(100,7.034697194705e-1) (101,7.037373031902e-1) (102,7.0400040352e-1) (103,7.04259144836e-1) (104,7.045136462012e-1) (105,7.047640212126e-1) (106,7.050103779018e-1) (107,7.052528186943e-1) (108,7.054914404273e-1) (109,7.057263344275e-1) (110,7.059575866475e-1) (111,7.061852778593e-1) (112,7.064094839004e-1) (113,7.066302759702e-1) (114,7.068477209685e-1) (115,7.07061881873e-1) (116,7.072728181455e-1) (117,7.074805861621e-1) (118,7.076852396571e-1) (119,7.078868301734e-1) (120,7.08085407511e-1) (121,7.082810201643e-1) (122,7.08473715742e-1) (123,7.086635413594e-1) (124,7.088505439991e-1) (125,7.090347708315e-1) (126,7.092162694908e-1) (127,7.093950883024e-1) (128,7.095712764574e-1) (129,7.097448841332e-1) (130,7.099159625583e-1) (131,7.100845640223e-1) (132,7.102507418318e-1) (133,7.104145502148e-1) (134,7.105760441775e-1) (135,7.107352793181e-1) (136,7.10892311602e-1) (137,7.110471971062e-1) (138,7.111999917375e-1) (139,7.11350750934e-1) (140,7.114995293554e-1) (141,7.116463805712e-1) (142,7.117913567534e-1) (143,7.11934508381e-1) (144,7.120758839644e-1) (145,7.122155297945e-1) (146,7.123534897243e-1) (147,7.124898049864e-1) (148,7.12624514051e-1) (149,7.127576525284e-1) (150,7.128892531176e-1) (151,7.130193456018e-1) (152,7.131479568923e-1) (153,7.13275111118e-1) (154,7.134008297601e-1) (155,7.135251318281e-1) (156,7.136480340741e-1) (157,7.137695512396e-1) (158,7.138896963307e-1) (159,7.140084809151e-1) (160,7.141259154342e-1) (161,7.142420095247e-1) (162,7.143567723425e-1) (163,7.144702128812e-1) (164,7.145823402799e-1) (165,7.146931641139e-1) (166,7.148026946607e-1) (167,7.149109431385e-1) (168,7.150179219107e-1) (169,7.151236446532e-1) (170,7.152281264813e-1) (171,7.153313840342e-1) (172,7.154334355162e-1) (173,7.155343006935e-1) (174,7.156340008477e-1) (175,7.157325586891e-1) (176,7.158299982291e-1) (177,7.159263446187e-1) (178,7.16021623955e-1) (179,7.161158630603e-1) (180,7.162090892411e-1) (181,7.163013300299e-1) (182,7.16392612919e-1) (183,7.164829650888e-1) (184,7.165724131401e-1) (185,7.166609828347e-1) (186,7.167486988497e-1) (187,7.168355845524e-1) (188,7.169216617993e-1) (189,7.170069507648e-1) (190,7.170914698018e-1) (191,7.171752353384e-1) (192,7.172582618121e-1) (193,7.173405616427e-1) (194,7.174221452442e-1) (195,7.175030210767e-1) (196,7.175831957337e-1) (197,7.176626740659e-1) (198,7.177414593374e-1) (199,7.178195534094e-1) (200,7.178969569494e-1)};
\addplot[vizaqua!75!black,line width=0.8pt,forget plot] coordinates {(100,1.060992219887) (101,1.05612253534) (102,1.051305160879) (103,1.046540223902) (104,1.04182781458) (105,1.037167987387) (106,1.032560762615) (107,1.028006127873) (108,1.023504039574) (109,1.019054424417) (110,1.01465718087) (111,1.010312180644) (112,1.006019270163) (113,1.001778272037) (114,9.975889865162e-1) (115,9.934511929405e-1) (116,9.893646511689e-1) (117,9.853291029861e-1) (118,9.813442734813e-1) (119,9.774098723909e-1) (120,9.735255954006e-1) (121,9.696911253979e-1) (122,9.659061336705e-1) (123,9.62170281044e-1) (124,9.584832189526e-1) (125,9.548445904398e-1) (126,9.512540310828e-1) (127,9.4771116984e-1) (128,9.442156298173e-1) (129,9.407670289538e-1) (130,9.373649806263e-1) (131,9.340090941745e-1) (132,9.306989753478e-1) (133,9.274342266794e-1) (134,9.242144477907e-1) (135,9.210392356308e-1) (136,9.179081846601e-1) (137,9.14820886981e-1) (138,9.117769324277e-1) (139,9.087759086183e-1) (140,9.058174009818e-1) (141,9.029009927651e-1) (142,9.000262650301e-1) (143,8.971927966473e-1) (144,8.944001642956e-1) (145,8.916479424731e-1) (146,8.88935703527e-1) (147,8.862630177072e-1) (148,8.836294532486e-1) (149,8.810345764866e-1) (150,8.784779520066e-1) (151,8.759591428314e-1) (152,8.734777106455e-1) (153,8.710332160561e-1) (154,8.6862521889e-1) (155,8.662532785225e-1) (156,8.639169542361e-1) (157,8.616158056038e-1) (158,8.593493928927e-1) (159,8.571172774818e-1) (160,8.549190222878e-1) (161,8.527541921941e-1) (162,8.506223544737e-1) (163,8.485230792025e-1) (164,8.464559396544e-1) (165,8.444205126735e-1) (166,8.424163790171e-1) (167,8.404431236648e-1) (168,8.385003360885e-1) (169,8.365876104808e-1) (170,8.347045459383e-1) (171,8.328507465977e-1) (172,8.310258217238e-1) (173,8.292293857504e-1) (174,8.274610582728e-1) (175,8.257204639962e-1) (176,8.240072326409e-1) (177,8.223209988087e-1) (178,8.206614018145e-1) (179,8.190280854881e-1) (180,8.174206979517e-1) (181,8.158388913782e-1) (182,8.142823217382e-1) (183,8.127506485394e-1) (184,8.112435345672e-1) (185,8.09760645631e-1) (186,8.083016503225e-1) (187,8.068662197921e-1) (188,8.054540275475e-1) (189,8.040647492795e-1) (190,8.026980627189e-1) (191,8.013536475264e-1) (192,8.000311852197e-1) (193,7.987303591365e-1) (194,7.97450854436e-1) (195,7.961923581377e-1) (196,7.949545591959e-1) (197,7.937371486081e-1) (198,7.925398195546e-1) (199,7.913622675655e-1) (200,7.90204190711e-1)};
\addplot[only marks,mark=*,mark size=1.3pt,color=vizgray] coordinates {(1,3.561721123121e-1) (5,5.616560838735e-1) (10,6.348486551759e-1) (20,6.774563054846e-1) (30,6.870647919347e-1) (40,6.908772226695e-1) (50,6.94017496325e-1) (60,6.970458519902e-1) (70,6.998625159907e-1) (80,7.02396229301e-1) (90,7.046335322981e-1) (100,7.066042168557e-1) (110,7.083590589675e-1) (120,7.09934797714e-1) (130,7.113463604862e-1) (140,7.126101998943e-1) (150,7.137547771769e-1) (160,7.148024969109e-1) (170,7.15758404576e-1) (180,7.166257183173e-1) (190,7.174195739509e-1) (200,7.1815587817e-1)};
\addlegendentry{numerical $\sqrt n\,e^{n\xi}P_{e,n}^*(1/2)$}
\addplot[vizblue,domain=100:200,samples=2] {0.73422908402453};
\addlegendentry{$A_{\mathrm q}$, \cref{thm:ch-main}}
\addplot[vizorange,densely dashed,domain=100:200,samples=2] {0.93297346902172};
\addlegendentry{$A_{\mathrm{cl}}$, classical prefactor}
\addlegendimage{area legend,fill=vizaqua!20,draw=vizaqua!75!black}
\addlegendentry{finite-$n$ bounds in~\eqref{eq:ch-finite-bounds}}
\nextgroupplot[xlabel={number of copies $n$}, ylabel={relative error}, ymode=log, ymin=1e-04, ymax=1, xmin=0, xmax=200, xtick={0,50,100,150,200}, ytick={1,1e-1,1e-2,1e-3,1e-4}, title={(b) Relative error with respect to $P_{e,n}^*(1/2)$}, legend style={at={(0.5,-0.25)}, anchor=north, legend columns=1}]
\addplot[vizblue] coordinates {(10,1.565419222966e-1) (11,1.421921715653e-1) (12,1.30365014974e-1) (13,1.205375272651e-1) (14,1.123189545277e-1) (15,1.054104889763e-1) (16,9.957877992886e-2) (17,9.463842086912e-2) (18,9.04399761758e-2) (19,8.686157991557e-2) (20,8.380286386043e-2) (21,8.118048892861e-2) (22,7.892479753096e-2) (23,7.697726178574e-2) (24,7.528850880801e-2) (25,7.381676977739e-2) (26,7.252664423545e-2) (27,7.138810219052e-2) (28,7.037566834465e-2) (29,6.946774760694e-2) (30,6.864606168663e-2) (31,6.789517419373e-2) (32,6.720208718307e-2) (33,6.65558960989e-2) (34,6.594749304636e-2) (35,6.536931051837e-2) (36,6.481509935167e-2) (37,6.427973592433e-2) (38,6.375905454616e-2) (39,6.324970170977e-2) (40,6.274900942246e-2) (41,6.225488526813e-2) (42,6.176571718683e-2) (43,6.128029122927e-2) (44,6.079772076196e-2) (45,6.031738577843e-2) (46,5.983888112205e-2) (47,5.936197255392e-2) (48,5.888655970998e-2) (49,5.841264508849e-2) (50,5.794030829542e-2) (51,5.746968485367e-2) (52,5.700094895221e-2) (53,5.653429957647e-2) (54,5.606994952071e-2) (55,5.560811683778e-2) (56,5.514901833242e-2) (57,5.46928647507e-2) (58,5.423985736074e-2) (59,5.379018565942e-2) (60,5.334402597504e-2) (61,5.290154076871e-2) (62,5.24628784663e-2) (63,5.202817367954e-2) (64,5.159754769815e-2) (65,5.117110915609e-2) (66,5.074895479392e-2) (67,5.033117025515e-2) (68,4.99178308694e-2) (69,4.950900238744e-2) (70,4.910474164369e-2) (71,4.870509713162e-2) (72,4.831010948445e-2) (73,4.791981186057e-2) (74,4.753423023786e-2) (75,4.715338362564e-2) (76,4.67772842056e-2) (77,4.640593741566e-2) (78,4.603934199184e-2) (79,4.567748998397e-2) (80,4.532036676114e-2) (81,4.496795102196e-2) (82,4.462021482415e-2) (83,4.427712364602e-2) (84,4.393863649112e-2) (85,4.360470604502e-2) (86,4.327527889119e-2) (87,4.295029579071e-2) (88,4.262969202813e-2) (89,4.231339782361e-2) (90,4.200133880924e-2) (91,4.169343656546e-2) (92,4.138960921131e-2) (93,4.1089772041e-2) (94,4.079383819741e-2) (95,4.050171937219e-2) (96,4.021332652098e-2) (97,3.99285705821e-2) (98,3.964736318611e-2) (99,3.936961734421e-2) (100,3.909524810329e-2) (101,3.882417315642e-2) (102,3.855631339787e-2) (103,3.829159341337e-2) (104,3.802994189718e-2) (105,3.777129198925e-2) (106,3.751558152712e-2) (107,3.726275320932e-2) (108,3.701275466828e-2) (109,3.676553845313e-2) (110,3.65210619241e-2) (111,3.627928706236e-2) (112,3.604018020055e-2) (113,3.5803711681e-2) (114,3.556985544983e-2) (115,3.533858859649e-2) (116,3.51098908493e-2) (117,3.488374403815e-2) (118,3.466013153625e-2) (119,3.443903769305e-2) (120,3.422044727032e-2) (121,3.400434489341e-2) (122,3.379071452901e-2) (123,3.357953900007e-2) (124,3.337079954781e-2) (125,3.316447544914e-2) (126,3.29605436972e-2) (127,3.275897875058e-2) (128,3.255975235578e-2) (129,3.236283344552e-2) (130,3.216818811403e-2) (131,3.197577966862e-2) (132,3.178556875536e-2) (133,3.159751355502e-2) (134,3.14115700439e-2) (135,3.122769231307e-2) (136,3.104583293799e-2) (137,3.086594338982e-2) (138,3.068797447873e-2) (139,3.051187681892e-2) (140,3.033760130487e-2) (141,3.016509958796e-2) (142,2.999432454296e-2) (143,2.9825230714e-2) (144,2.965777473035e-2) (145,2.949191568298e-2) (146,2.932761545393e-2) (147,2.916483899148e-2) (148,2.900355452555e-2) (149,2.884373371893e-2) (150,2.868535175151e-2) (151,2.85283873362e-2) (152,2.837282266652e-2) (153,2.821864329765e-2) (154,2.806583796389e-2) (155,2.791439833696e-2) (156,2.776431873089e-2) (157,2.761559576019e-2) (158,2.746822795909e-2) (159,2.732221537032e-2) (160,2.717755911262e-2) (161,2.703426093636e-2) (162,2.689232277699e-2) (163,2.675174631584e-2) (164,2.66125325578e-2) (165,2.647468143453e-2) (166,2.633819144167e-2) (167,2.620305931746e-2) (168,2.606927976906e-2) (169,2.593684525222e-2) (170,2.58057458082e-2) (171,2.567596896103e-2) (172,2.554749967639e-2) (173,2.542032038257e-2) (174,2.529441105204e-2) (175,2.516974934144e-2) (176,2.504631078593e-2) (177,2.492406904327e-2) (178,2.480299618163e-2) (179,2.46830630042e-2) (180,2.456423940325e-2) (181,2.444649473534e-2) (182,2.432979820928e-2) (183,2.421411927805e-2) (184,2.40994280261e-2) (185,2.398569554343e-2) (186,2.38728942783e-2) (187,2.376099836099e-2) (188,2.364998389183e-2) (189,2.353982918722e-2) (190,2.343051497886e-2) (191,2.332202456197e-2) (192,2.321434388987e-2) (193,2.310746161305e-2) (194,2.300136906264e-2) (195,2.289606017885e-2) (196,2.279153138649e-2) (197,2.268778142081e-2) (198,2.258481110778e-2) (199,2.248262310404e-2) (200,2.238122160257e-2)};
\addlegendentry{leading term $A_{\mathrm q}e^{-n\xi}/\sqrt n$}
\addplot[vizblue, densely dashed] coordinates {(10,2.808862907416e-1) (11,2.584580676739e-1) (12,2.392379068067e-1) (13,2.225275431636e-1) (14,2.078149987818e-1) (15,1.947185280209e-1) (16,1.829493723618e-1) (17,1.722861730706e-1) (18,1.625571041995e-1) (19,1.536271870484e-1) (20,1.453891769423e-1) (21,1.377569188734e-1) (22,1.306604412304e-1) (23,1.240422933907e-1) (24,1.178547839101e-1) (25,1.120578799616e-1) (26,1.066175983366e-1) (27,1.015047656471e-1) (28,9.6694058152e-2) (29,9.216325510189e-2) (30,8.789265621756e-2) (31,8.386462602254e-2) (32,8.006323664627e-2) (33,7.64739873135e-2) (34,7.308358367678e-2) (35,6.987976389194e-2) (36,6.685116119846e-2) (37,6.398719497341e-2) (38,6.127798394323e-2) (39,5.871427658123e-2) (40,5.628739477699e-2) (41,5.398918770201e-2) (42,5.181199346124e-2) (43,4.974860665042e-2) (44,4.779225036202e-2) (45,4.593655151951e-2) (46,4.417551868828e-2) (47,4.25035217247e-2) (48,4.09152727933e-2) (49,3.940580841394e-2) (50,3.797047230409e-2) (51,3.660489885984e-2) (52,3.530499717927e-2) (53,3.406693557574e-2) (54,3.288712656041e-2) (55,3.17622122951e-2) (56,3.068905053056e-2) (57,2.966470105317e-2) (58,2.868641266653e-2) (59,2.77516107342e-2) (60,2.685788530763e-2) (61,2.600297985891e-2) (62,2.518478063291e-2) (63,2.440130662746e-2) (64,2.365070020441e-2) (65,2.293121832817e-2) (66,2.224122442289e-2) (67,2.1579180834e-2) (68,2.094364187502e-2) (69,2.033324743641e-2) (70,1.974671712948e-2) (71,1.918284493541e-2) (72,1.864049432685e-2) (73,1.81185938281e-2) (74,1.76161329781e-2) (75,1.713215866052e-2) (76,1.666577176431e-2) (77,1.621612413928e-2) (78,1.578241581156e-2) (79,1.53638924255e-2) (80,1.495984287995e-2) (81,1.456959712935e-2) (82,1.419252412192e-2) (83,1.382802985013e-2) (84,1.347555549144e-2) (85,1.313457561983e-2) (86,1.280459647205e-2) (87,1.248515425506e-2) (88,1.217581348456e-2) (89,1.187616534675e-2) (90,1.158582607879e-2) (91,1.130443536558e-2) (92,1.103165475304e-2) (93,1.076716608e-2) (94,1.05106699329e-2) (95,1.026188412883e-2) (96,1.002054223377e-2) (97,9.786392123989e-3) (98,9.559194598892e-3) (99,9.338722054207e-3) (100,9.124757224232e-3) (101,8.91709200166e-3) (102,8.715526342909e-3) (103,8.519867266076e-3) (104,8.329927947598e-3) (105,8.145526922462e-3) (106,7.966487391427e-3) (107,7.792636637191e-3) (108,7.62380554985e-3) (109,7.459828260345e-3) (110,7.300541878947e-3) (111,7.145786334228e-3) (112,6.995404306402e-3) (113,6.849241247518e-3) (114,6.707145479669e-3) (115,6.568968361263e-3) (116,6.434564510481e-3) (117,6.303792074319e-3) (118,6.176513031123e-3) (119,6.052593514281e-3) (120,5.931904144732e-3) (121,5.814320360178e-3) (122,5.699722729377e-3) (123,5.587997240578e-3) (124,5.479035554085e-3) (125,5.37273521003e-3) (126,5.268999783705e-3) (127,5.16773898221e-3) (128,5.06886867769e-3) (129,4.972310874005e-3) (130,4.87799360535e-3) (131,4.785850766916e-3) (132,4.695821879366e-3) (133,4.607851790374e-3) (134,4.52189031799e-3) (135,4.437891841892e-3) (136,4.355814849773e-3) (137,4.275621447119e-3) (138,4.197276839448e-3) (139,4.120748796676e-3) (140,4.046007109672e-3) (141,3.973023049191e-3) (142,3.901768837345e-3) (143,3.832217141437e-3) (144,3.764340599492e-3) (145,3.698111386118e-3) (146,3.633500826395e-3) (147,3.570479064482e-3) (148,3.509014792392e-3) (149,3.449075043097e-3) (150,3.39062505073e-3) (151,3.333628179214e-3) (152,3.278045919183e-3) (153,3.223837951625e-3) (154,3.170962275274e-3) (155,3.119375393448e-3) (156,3.069032554833e-3) (157,3.0198880416e-3) (158,2.971895497317e-3) (159,2.925008286364e-3) (160,2.879179875953e-3) (161,2.834364231517e-3) (162,2.790516216011e-3) (163,2.747591983732e-3) (164,2.705549359491e-3) (165,2.664348194397e-3) (166,2.623950690125e-3) (167,2.584321684352e-3) (168,2.545428890949e-3) (169,2.50724308962e-3) (170,2.46973826083e-3) (171,2.432891663117e-3) (172,2.396683851217e-3) (173,2.361098634704e-3) (174,2.326122978225e-3) (175,2.291746845639e-3) (176,2.257962991624e-3) (177,2.22476670543e-3) (178,2.192155512478e-3) (179,2.160128840375e-3) (180,2.12868765666e-3) (181,2.097834086158e-3) (182,2.067571016181e-3) (183,2.037901698041e-3) (184,2.008829353339e-3) (185,1.980356793336e-3) (186,1.952486059324e-3) (187,1.925218091437e-3) (188,1.898552432627e-3) (189,1.872486973713e-3) (190,1.847017744478e-3) (191,1.82213875472e-3) (192,1.797841888063e-3) (193,1.774116850144e-3) (194,1.750951171607e-3) (195,1.728330265108e-3) (196,1.706237534416e-3) (197,1.684654532524e-3) (198,1.663561164679e-3) (199,1.642935931298e-3) (200,1.622756204891e-3)};
\addlegendentry{with the $1/n$ term}
\addplot[vizblue, dotted, line width=1.4pt] coordinates {(10,2.524508057779e-1) (11,1.72948053722e-1) (12,1.183670329474e-1) (13,7.9609588989e-2) (14,5.139758903784e-2) (15,3.048677296798e-2) (16,1.478767077718e-2) (17,2.901886582169e-3) (18,6.136977962018e-3) (19,1.301514630487e-2) (20,1.823133735801e-2) (21,2.215531674617e-2) (22,2.506601742649e-2) (23,2.717702221507e-2) (24,2.865399970587e-2) (25,2.962688663983e-2) (26,3.019856262894e-2) (27,3.045113508888e-2) (28,3.045056565688e-2) (29,3.02501262114e-2) (30,2.989301609136e-2) (31,2.941436944825e-2) (32,2.884281316328e-2) (33,2.820168931295e-2) (34,2.751002420054e-2) (35,2.678330368118e-2) (36,2.603409877424e-2) (37,2.527257432053e-2) (38,2.450690533461e-2) (39,2.374361979287e-2) (40,2.298788225124e-2) (41,2.224372945836e-2) (42,2.151426671285e-2) (43,2.080183188629e-2) (44,2.010813264207e-2) (45,1.943436130953e-2) (46,1.87812910429e-2) (47,1.814935624489e-2) (48,1.753871972109e-2) (49,1.694932862184e-2) (50,1.638096089807e-2) (51,1.58332637292e-2) (52,1.530578516013e-2) (53,1.479800000159e-2) (54,1.430933089458e-2) (55,1.383916531009e-2) (56,1.33868691453e-2) (57,1.295179748319e-2) (58,1.253330300094e-2) (59,1.213074244279e-2) (60,1.174348151159e-2) (61,1.137089848067e-2) (62,1.10123867813e-2) (63,1.066735678095e-2) (64,1.033523693246e-2) (65,1.001547444375e-2) (66,9.707535591371e-3) (67,9.410905777879e-3) (68,9.125089413594e-3) (69,8.849609685784e-3) (70,8.584008263874e-3) (71,8.327844976643e-3) (72,8.08069748686e-3) (73,7.842160979893e-3) (74,7.611847875552e-3) (75,7.389387566486e-3) (76,7.174426181751e-3) (77,6.966626370599e-3) (78,6.765667098948e-3) (79,6.571243449243e-3) (80,6.383066413493e-3) (81,6.200862668984e-3) (82,6.024374326422e-3) (83,5.85335864108e-3) (84,5.687587678633e-3) (85,5.526847928876e-3) (86,5.370939862192e-3) (87,5.219677425514e-3) (88,5.072887476466e-3) (89,4.930409156342e-3) (90,4.7920932045e-3) (91,4.657801218643e-3) (92,4.527404867111e-3) (93,4.400785060914e-3) (94,4.277831094535e-3) (95,4.158439765634e-3) (96,4.04251448465e-3) (97,3.929964385839e-3) (98,3.820703451605e-3) (99,3.714649661969e-3) (100,3.611724180769e-3) (101,3.511850589633e-3) (102,3.414954179989e-3) (103,3.320961312348e-3) (104,3.229798850868e-3) (105,3.14139367982e-3) (106,3.055672306994e-3) (107,2.972560557472e-3) (108,2.891983359433e-3) (109,2.813864621911e-3) (110,2.738127202689e-3) (111,2.664692962761e-3) (112,2.593482902234e-3) (113,2.524417370958e-3) (114,2.457416345876e-3) (115,2.392399765844e-3) (116,2.329287913729e-3) (117,2.268001834785e-3) (118,2.208463779806e-3) (119,2.150597661237e-3) (120,2.094329510411e-3) (121,2.03958792425e-3) (122,1.986304490243e-3) (123,1.934414179172e-3) (124,1.883855695948e-3) (125,1.834571780002e-3) (126,1.78650944792e-3) (127,1.739620172397e-3) (128,1.693859993085e-3) (129,1.649189556492e-3) (130,1.605574083679e-3) (131,1.562983266159e-3) (132,1.52139109197e-3) (133,1.480775605451e-3) (134,1.441118605666e-3) (135,1.402405289781e-3) (136,1.364623848817e-3) (137,1.327765024243e-3) (138,1.291821634632e-3) (139,1.256788082245e-3) (140,1.222659849715e-3) (141,1.189432997215e-3) (142,1.157103670357e-3) (143,1.125667628795e-3) (144,1.09511980498e-3) (145,1.065453901786e-3) (146,1.036662036815e-3) (147,1.008734440141e-3) (148,9.816592110378e-4) (149,9.554221378918e-4) (150,9.30006584158e-4) (151,9.053934417164e-4) (152,8.815611515552e-4) (153,8.584857902442e-4) (154,8.361412192616e-4) (155,8.144992929081e-4) (156,7.935301193238e-4) (157,7.732023680274e-4) (158,7.534836164611e-4) (159,7.343407272591e-4) (160,7.157402473745e-4) (161,6.976488198254e-4) (162,6.800335986389e-4) (163,6.628626576074e-4) (164,6.461053837041e-4) (165,6.297328464384e-4) (166,6.13718135049e-4) (167,5.980366562216e-4) (168,5.826663859584e-4) (169,5.675880702979e-4) (170,5.527853707562e-4) (171,5.382449516197e-4) (172,5.239565075196e-4) (173,5.099127310489e-4) (174,4.961092214986e-4) (175,4.825443370761e-4) (176,4.692189941847e-4) (177,4.561364184763e-4) (178,4.433018534017e-4) (179,4.307222328682e-4) (180,4.184058253382e-4) (181,4.063618572677e-4) (182,3.946001241654e-4) (183,3.831305977508e-4) (184,3.719630377018e-4) (185,3.611066163084e-4) (186,3.505695639906e-4) (187,3.403588431158e-4) (188,3.304798568643e-4) (189,3.209361990656e-4) (190,3.117294499808e-4) (191,3.028590219563e-4) (192,2.943220577517e-4) (193,2.861133831643e-4) (194,2.782255143758e-4) (195,2.706487192448e-4) (196,2.633711305993e-4) (197,2.563789084635e-4) (198,2.496564471156e-4) (199,2.431866219293e-4) (200,2.369510701342e-4)};
\addlegendentry{with the $1/n$ and $1/n^2$ terms}
\addplot[vizblue!60!red, dashdotted] coordinates {(10,5.926576918781e-1) (11,4.955866757085e-1) (12,4.161624540718e-1) (13,3.525053775841e-1) (14,3.017726184772e-1) (15,2.612308601247e-1) (16,2.285938394972e-1) (17,2.020593162531e-1) (18,1.802444629743e-1) (19,1.62100211904e-1) (20,1.468330546245e-1) (21,1.338417301017e-1) (22,1.226685863636e-1) (23,1.129631283413e-1) (24,1.044549984642e-1) (25,9.693402252034e-2) (26,9.023546574326e-2) (27,8.422910644937e-2) (28,7.881110320267e-2) (29,7.389791004474e-2) (30,6.942169881473e-2) (31,6.532689574915e-2) (32,6.156754632717e-2) (33,5.810529918527e-2) (34,5.490785533831e-2) (35,5.194776904294e-2) (36,4.920151579465e-2) (37,4.664876426014e-2) (38,4.427180461613e-2) (39,4.205509735619e-2) (40,3.998491525286e-2) (41,3.804905762013e-2) (42,3.623662088375e-2) (43,3.453781314861e-2) (44,3.29438032547e-2) (45,3.144659695608e-2) (46,3.003893450384e-2) (47,2.871420518329e-2) (48,2.746637533798e-2) (49,2.628992717556e-2) (50,2.517980624257e-2) (51,2.413137591743e-2) (52,2.314037763006e-2) (53,2.22028957975e-2) (54,2.131532668287e-2) (55,2.047435055574e-2) (56,1.967690666354e-2) (57,1.892017062649e-2) (58,1.820153394751e-2) (59,1.751858538958e-2) (60,1.686909401999e-2) (61,1.62509937564e-2) (62,1.566236927736e-2) (63,1.510144318006e-2) (64,1.45665642841e-2) (65,1.405619699184e-2) (66,1.356891162439e-2) (67,1.310337565953e-2) (68,1.26583458027e-2) (69,1.223266082599e-2) (70,1.182523511373e-2) (71,1.143505285542e-2) (72,1.106116282924e-2) (73,1.07026737216e-2) (74,1.035874993005e-2) (75,1.002860779911e-2) (76,9.711512240844e-3) (77,9.406773694122e-3) (78,9.113745379345e-3) (79,8.831820807865e-3) (80,8.560431508323e-3) (81,8.299044935142e-3) (82,8.047162527561e-3) (83,7.80431789091e-3) (84,7.57007507514e-3) (85,7.344026929046e-3) (86,7.125793511951e-3) (87,6.915020547998e-3) (88,6.711377911381e-3) (89,6.514558133992e-3) (90,6.324274929873e-3) (91,6.140261733595e-3) (92,5.962270252198e-3) (93,5.790069032533e-3) (94,5.62344204781e-3) (95,5.462187308795e-3) (96,5.306115506416e-3) (97,5.155048693546e-3) (98,5.008819014392e-3) (99,4.867267490285e-3) (100,4.730242870677e-3) (101,4.597600557913e-3) (102,4.469201613764e-3) (103,4.344911854937e-3) (104,4.224601043727e-3) (105,4.108142178743e-3) (106,3.995410889256e-3) (107,3.886284935209e-3) (108,3.780643813311e-3) (109,3.678368468008e-3) (110,3.579341104467e-3) (111,3.483445099083e-3) (112,3.390565001513e-3) (113,3.300586620754e-3) (114,3.213397186522e-3) (115,3.128885576063e-3) (116,3.04694259556e-3) (117,2.967461304637e-3) (118,2.890337371916e-3) (119,2.815469449374e-3) (120,2.742759553213e-3) (121,2.672113439227e-3) (122,2.603440961087e-3) (123,2.536656400696e-3) (124,2.471678760669e-3) (125,2.408432010096e-3) (126,2.346845276015e-3) (127,2.286852974423e-3) (128,2.228394876192e-3) (129,2.171416104809e-3) (130,2.115867064521e-3) (131,2.061703299093e-3) (132,2.008885282991e-3) (133,1.957378148364e-3) (134,1.907151352626e-3) (135,1.858178292802e-3) (136,1.810435873957e-3) (137,1.763904040028e-3) (138,1.718565276216e-3) (139,1.674404092653e-3) (140,1.631406499478e-3) (141,1.589559483577e-3) (142,1.548850497186e-3) (143,1.509266968244e-3) (144,1.470795841873e-3) (145,1.433423161658e-3) (146,1.397133698458e-3) (147,1.361910633481e-3) (148,1.32773530108e-3) (149,1.294586995473e-3) (150,1.262442844167e-3) (151,1.231277749421e-3) (152,1.201064397636e-3) (153,1.171773335106e-3) (154,1.14337310715e-3) (155,1.115830456353e-3) (156,1.089110574378e-3) (157,1.063177400769e-3) (158,1.037993961188e-3) (159,1.013522736787e-3) (160,9.897260558392e-4) (161,9.665664983518e-4) (162,9.440073042409e-4) (163,9.220127756509e-4) (164,9.005486642566e-4) (165,8.795825348096e-4) (166,8.590840968132e-4) (167,8.390254969976e-4) (168,8.193815662082e-4) (169,8.00130015392e-4) (170,7.812515765415e-4) (171,7.627300857113e-4) (172,7.445525065279e-4) (173,7.267088939395e-4) (174,7.091922992734e-4) (175,6.919986189507e-4) (176,6.751263904297e-4) (177,6.585765400777e-4) (178,6.423520886874e-4) (179,6.264578212377e-4) (180,6.108999282256e-4) (181,5.956856264588e-4) (182,5.808227675819e-4) (183,5.663194428086e-4) (184,5.521835923439e-4) (185,5.384226278049e-4) (186,5.250430755958e-4) (187,5.120502486646e-4) (188,4.994479533866e-4) (189,4.872382374933e-4) (190,4.754211840178e-4) (191,4.639947551795e-4) (192,4.529546890057e-4) (193,4.422944503126e-4) (194,4.320052364672e-4) (195,4.220760371522e-4) (196,4.124937461858e-4) (197,4.032433223299e-4) (198,3.94307994981e-4) (199,3.856695096961e-4) (200,3.773084076861e-4)};
\addlegendentry{through the $1/n^3$ term}
\addplot[vizorange] coordinates {(10,4.695998194456e-1) (11,4.513658145407e-1) (12,4.363372307462e-1) (13,4.238495950763e-1) (14,4.134063826184e-1) (15,4.046279029704e-1) (16,3.972176410525e-1) (17,3.909400036905e-1) (18,3.856051053662e-1) (19,3.810580929349e-1) (20,3.771714300516e-1) (21,3.738392190422e-1) (22,3.709729470923e-1) (23,3.684982437337e-1) (24,3.663523716098e-1) (25,3.644822557308e-1) (26,3.62842913285e-1) (27,3.613961856296e-1) (28,3.601097016999e-1) (29,3.5895602095e-1) (30,3.579119174403e-1) (31,3.569577763904e-1) (32,3.560770815136e-1) (33,3.552559765608e-1) (34,3.544828882715e-1) (35,3.537482007324e-1) (36,3.53043973229e-1) (37,3.523636952547e-1) (38,3.517020735316e-1) (39,3.510548468104e-1) (40,3.50418624914e-1) (41,3.49790749042e-1) (42,3.491691707748e-1) (43,3.485523475661e-1) (44,3.479391527843e-1) (45,3.47328798597e-1) (46,3.467207701779e-1) (47,3.461147698833e-1) (48,3.455106701822e-1) (49,3.44908474249e-1) (50,3.443082832378e-1) (51,3.437102693555e-1) (52,3.431146539412e-1) (53,3.425216898427e-1) (54,3.419316474544e-1) (55,3.413448038524e-1) (56,3.407614345274e-1) (57,3.40181807271e-1) (58,3.396061778315e-1) (59,3.390347869996e-1) (60,3.38467858833e-1) (61,3.379055997685e-1) (62,3.373481984085e-1) (63,3.36795825802e-1) (64,3.3624863607e-1) (65,3.35706767252e-1) (66,3.351703422747e-1) (67,3.346394699639e-1) (68,3.3411424604e-1) (69,3.335947540512e-1) (70,3.330810662163e-1) (71,3.325732441552e-1) (72,3.320713394998e-1) (73,3.31575394384e-1) (74,3.310854418175e-1) (75,3.306015059551e-1) (76,3.301236022759e-1) (77,3.296517376898e-1) (78,3.291859105907e-1) (79,3.287261108764e-1) (80,3.282723199557e-1) (81,3.278245107612e-1) (82,3.27382647787e-1) (83,3.269466871671e-1) (84,3.265165768078e-1) (85,3.260922565874e-1) (86,3.256736586305e-1) (87,3.25260707663e-1) (88,3.24853321452e-1) (89,3.244514113293e-1) (90,3.240548827969e-1) (91,3.236636362081e-1) (92,3.232775675181e-1) (93,3.228965690927e-1) (94,3.225205305638e-1) (95,3.221493397193e-1) (96,3.21782883412e-1) (97,3.214210484726e-1) (98,3.210637226114e-1) (99,3.207107952932e-1) (100,3.203621585692e-1) (101,3.200177078534e-1) (102,3.196773426274e-1) (103,3.193409670634e-1) (104,3.190084905543e-1) (105,3.186798281421e-1) (106,3.183549008379e-1) (107,3.180336358297e-1) (108,3.177159665747e-1) (109,3.17401832778e-1) (110,3.170911802577e-1) (111,3.167839607041e-1) (112,3.164801313364e-1) (113,3.161796544691e-1) (114,3.158824969961e-1) (115,3.155886298052e-1) (116,3.152980271374e-1) (117,3.150106659044e-1) (118,3.147265249787e-1) (119,3.144455844734e-1) (120,3.141678250255e-1) (121,3.138932270987e-1) (122,3.136217703198e-1) (123,3.133534328621e-1) (124,3.13088190889e-1) (125,3.128260180682e-1) (126,3.125668851652e-1) (127,3.123107597248e-1) (128,3.120576058458e-1) (129,3.118073840515e-1) (130,3.115600512584e-1) (131,3.113155608416e-1) (132,3.110738627947e-1) (133,3.108349039786e-1) (134,3.105986284532e-1) (135,3.103649778826e-1) (136,3.101338920053e-1) (137,3.099053091564e-1) (138,3.096791668311e-1) (139,3.094554022757e-1) (140,3.092339530925e-1) (141,3.090147578458e-1) (142,3.087977566542e-1) (143,3.085828917576e-1) (144,3.083701080456e-1) (145,3.081593535356e-1) (146,3.079505797918e-1) (147,3.077437422746e-1) (148,3.075388006141e-1) (149,3.073357188026e-1) (150,3.071344653018e-1) (151,3.069350130624e-1) (152,3.067373394579e-1) (153,3.065414261337e-1) (154,3.063472587744e-1) (155,3.061548267976e-1) (156,3.059641229781e-1) (157,3.057751430141e-1) (158,3.055878850429e-1) (159,3.05402349119e-1) (160,3.052185366639e-1) (161,3.050364499017e-1) (162,3.048560912918e-1) (163,3.046774629707e-1) (164,3.045005662152e-1) (165,3.043254009384e-1) (166,3.041519652289e-1) (167,3.03980254942e-1) (168,3.038102633521e-1) (169,3.036419808728e-1) (170,3.034753948498e-1) (171,3.033104894301e-1) (172,3.031472455097e-1) (173,3.029856407605e-1) (174,3.028256497339e-1) (175,3.026672440386e-1) (176,3.025103925874e-1) (177,3.023550619077e-1) (178,3.022012165068e-1) (179,3.02048819285e-1) (180,3.018978319847e-1) (181,3.017482156679e-1) (182,3.015999312085e-1) (183,3.014529397911e-1) (184,3.013072034024e-1) (185,3.011626853069e-1) (186,3.01019350495e-1) (187,3.008771660945e-1) (188,3.007361017365e-1) (189,3.005961298678e-1) (190,3.004572260047e-1) (191,3.003193689206e-1) (192,3.001825407666e-1) (193,3.00046727121e-1) (194,2.999119169685e-1) (195,2.99778102609e-1) (196,2.996452794995e-1) (197,2.995134460329e-1) (198,2.993826032588e-1) (199,2.99252754553e-1) (200,2.991239052434e-1)};
\addlegendentry{commuting prefactor $A_{\mathrm{cl}}e^{-n\xi}/\sqrt n$}
\addplot[vizaqua] coordinates {(10,9.927081018034e-2) (11,9.210888029173e-2) (12,8.598188414225e-2) (13,8.068635869651e-2) (14,7.606266068029e-2) (15,7.198444984103e-2) (16,6.83514175992e-2) (17,6.508369621108e-2) (18,6.211755636095e-2) (19,5.940205953988e-2) (20,5.68964621352e-2) (21,5.456817222787e-2) (22,5.239112569794e-2) (23,5.034448752552e-2) (24,4.841160645081e-2) (25,4.657917048832e-2) (26,4.483652370748e-2) (27,4.317511346038e-2) (28,4.158804352317e-2) (29,4.006971369362e-2) (30,3.861553017173e-2) (31,3.722167397123e-2) (32,3.588491692267e-2) (33,3.460247667066e-2) (34,3.33719035443e-2) (35,3.219099338136e-2) (36,3.105772137679e-2) (37,2.997019284549e-2) (38,2.89266074752e-2) (39,2.792523422221e-2) (40,2.696439449078e-2) (41,2.604245165072e-2) (42,2.515780529894e-2) (43,2.430888896902e-2) (44,2.349417024589e-2) (45,2.271215245683e-2) (46,2.196137729009e-2) (47,2.124042784399e-2) (48,2.054793173476e-2) (49,1.988256399544e-2) (50,1.924304958231e-2) (51,1.862816537408e-2) (52,1.803674160201e-2) (53,1.746766269176e-2) (54,1.691986752849e-2) (55,1.639234917986e-2) (56,1.588415412719e-2) (57,1.539438106471e-2) (58,1.492217933227e-2) (59,1.446674704802e-2) (60,1.402732900663e-2) (61,1.360321440512e-2) (62,1.319373445367e-2) (63,1.279825992298e-2) (64,1.241619867387e-2) (65,1.204699320796e-2) (66,1.169011827242e-2) (67,1.134507854522e-2) (68,1.101140642194e-2) (69,1.068865991961e-2) (70,1.037642070857e-2) (71,1.007429227886e-2) (72,9.781898244106e-3) (73,9.498880782727e-3) (74,9.224899213752e-3) (75,8.959628702296e-3) (76,8.702759088303e-3) (77,8.453993830867e-3) (78,8.213049059693e-3) (79,7.979652724794e-3) (80,7.75354383533e-3) (81,7.534471778633e-3) (82,7.322195710684e-3) (83,7.116484009793e-3) (84,6.917113785756e-3) (85,6.723870437426e-3) (86,6.536547252273e-3) (87,6.354945042244e-3) (88,6.178871810877e-3) (89,6.008142447321e-3) (90,5.842578443505e-3) (91,5.682007631276e-3) (92,5.526263936839e-3) (93,5.375187150275e-3) (94,5.228622708304e-3) (95,5.086421488778e-3) (96,4.948439615676e-3) (97,4.814538273566e-3) (98,4.684583530703e-3) (99,4.558446170022e-3) (100,4.436001527429e-3) (101,4.317129336819e-3) (102,4.201713581322e-3) (103,4.089642350312e-3) (104,3.980807701718e-3) (105,3.875105529198e-3) (106,3.772435433772e-3) (107,3.67270059947e-3) (108,3.575807672641e-3) (109,3.481666644516e-3) (110,3.390190736706e-3) (111,3.301296289309e-3) (112,3.21490265137e-3) (113,3.130932073449e-3) (114,3.049309602142e-3) (115,2.969962976408e-3) (116,2.892822525641e-3) (117,2.817821069464e-3) (118,2.744893819265e-3) (119,2.673978281552e-3) (120,2.605014163243e-3) (121,2.537943279035e-3) (122,2.472709461021e-3) (123,2.409258470763e-3) (124,2.347537914015e-3) (125,2.287497158303e-3) (126,2.229087253577e-3) (127,2.172260856107e-3) (128,2.116972155818e-3) (129,2.063176807181e-3) (130,2.010831863804e-3) (131,1.959895716774e-3) (132,1.910328036817e-3) (133,1.862089720255e-3) (134,1.815142838738e-3) (135,1.769450592666e-3) (136,1.72497726818e-3) (137,1.681688197576e-3) (138,1.639549722947e-3) (139,1.598529162855e-3) (140,1.558594781775e-3) (141,1.519715762071e-3) (142,1.481862178222e-3) (143,1.445004973036e-3) (144,1.409115935554e-3) (145,1.374167680388e-3) (146,1.340133628202e-3) (147,1.306987987106e-3) (148,1.274705734709e-3) (149,1.243262600612e-3) (150,1.212635049164e-3) (151,1.182800262296e-3) (152,1.153736122302e-3) (153,1.125421194457e-3) (154,1.097834709382e-3) (155,1.070956545109e-3) (156,1.044767208816e-3) (157,1.019247818234e-3) (158,9.943800827419e-4) (159,9.701462842007e-4) (160,9.465292575866e-4) (161,9.235123714993e-4) (162,9.010795086381e-4) (163,8.792150463434e-4) (164,8.579038373089e-4) (165,8.3713119057e-4) (166,8.16882852876e-4) (167,7.971449905474e-4) (168,7.779041719146e-4) (169,7.591473504254e-4) (170,7.408618484981e-4) (171,7.230353421844e-4) (172,7.056558466932e-4) (173,6.887117028137e-4) (174,6.721915642568e-4) (175,6.560843859261e-4) (176,6.403794131091e-4) (177,6.250661715687e-4) (178,6.101344585034e-4) (179,5.955743343285e-4) (180,5.813761152269e-4) (181,5.675303664036e-4) (182,5.540278959754e-4) (183,5.408597494207e-4) (184,5.280172045109e-4) (185,5.154917666446e-4) (186,5.032751645054e-4) (187,4.913593459669e-4) (188,4.797364741713e-4) (189,4.683989237138e-4) (190,4.573392768697e-4) (191,4.465503198113e-4) (192,4.360250387646e-4) (193,4.257566160702e-4) (194,4.157384261147e-4) (195,4.059640311151e-4) (196,3.964271767406e-4) (197,3.871217875687e-4) (198,3.78041962379e-4) (199,3.691819692959e-4) (200,3.605362407963e-4)};
\addlegendentry{lower bound in~\eqref{eq:ch-finite-bounds}}
\end{groupplot}

\begin{axis}[at={(group c1r1.north east)},anchor=north east,xshift=-6pt,yshift=-15pt,
width=0.17\textwidth,height=0.13\textwidth,scale only axis,
xmin=0,xmax=205,ymin=0.2,ymax=2.05,xtick={0,200},ytick={0.5,1,2},
tick label style={font=\tiny},title={Cheng--Liu window},title style={font=\tiny,yshift=-2pt},
axis background/.style={fill=white},axis line style={vizgray,line width=0.4pt},
tick style={vizgray,line width=0.4pt},grid=none,clip=true]
\path[fill=vizgray!12] (axis cs:0,0.46648673451086) rectangle (axis cs:205,1.8659469380434);
\draw[vizgray,densely dotted,line width=0.4pt] (axis cs:0,0.46648673451086)--(axis cs:205,0.46648673451086);
\draw[vizgray,densely dotted,line width=0.4pt] (axis cs:0,1.8659469380434)--(axis cs:205,1.8659469380434);
\addplot[vizblue,line width=0.65pt,domain=0:205,samples=2] {0.73422908402453};
\addplot[vizorange,densely dashed,line width=0.65pt,domain=0:205,samples=2] {0.93297346902172};
\addplot[only marks,mark=*,mark size=0.55pt,color=vizgray] coordinates {(1,3.561721123121e-1) (5,5.616560838735e-1) (10,6.348486551759e-1) (20,6.774563054846e-1) (30,6.870647919347e-1) (40,6.908772226695e-1) (50,6.94017496325e-1) (60,6.970458519902e-1) (70,6.998625159907e-1) (80,7.02396229301e-1) (90,7.046335322981e-1) (100,7.066042168557e-1) (110,7.083590589675e-1) (120,7.09934797714e-1) (130,7.113463604862e-1) (140,7.126101998943e-1) (150,7.137547771769e-1) (160,7.148024969109e-1) (170,7.15758404576e-1) (180,7.166257183173e-1) (190,7.174195739509e-1) (200,7.1815587817e-1)};
\end{axis}

\end{tikzpicture}
\caption{Optimal error probability for the noisy qubit states of \cref{ex:ch-numerics} with $p=1/2$, evaluated numerically through Schur--Weyl duality at sampled values of $n$ up to $200$. (a) The main view zooms to $100\le n\le200$. The green band shows the rescaled finite-$n$ lower and upper bounds in~\eqref{eq:ch-finite-bounds}, with the upper bound evaluated at $\alpha=0.458035$. Horizontal lines mark $A_{\mathrm q}$ and $A_{\mathrm{cl}}$. The inset shows the full range and Cheng--Liu's asymptotic window $[A_{\mathrm{cl}}/2,2A_{\mathrm{cl}}]$~\cite[Theorem~5.2 and Remark~5.3]{cheng2026multiple}; its gray band is not a finite-$n$ bound. (b) The relative errors $|\widehat P_n/P_{e,n}^*(1/2)-1|$ of the leading term and its corrections through orders $1/n$, $1/n^2$, and $1/n^3$, obtained by exponentiating the corresponding truncations of~\eqref{eq:ch-higher}.}
\label{fig:chernoff-numerics}
\end{figure}
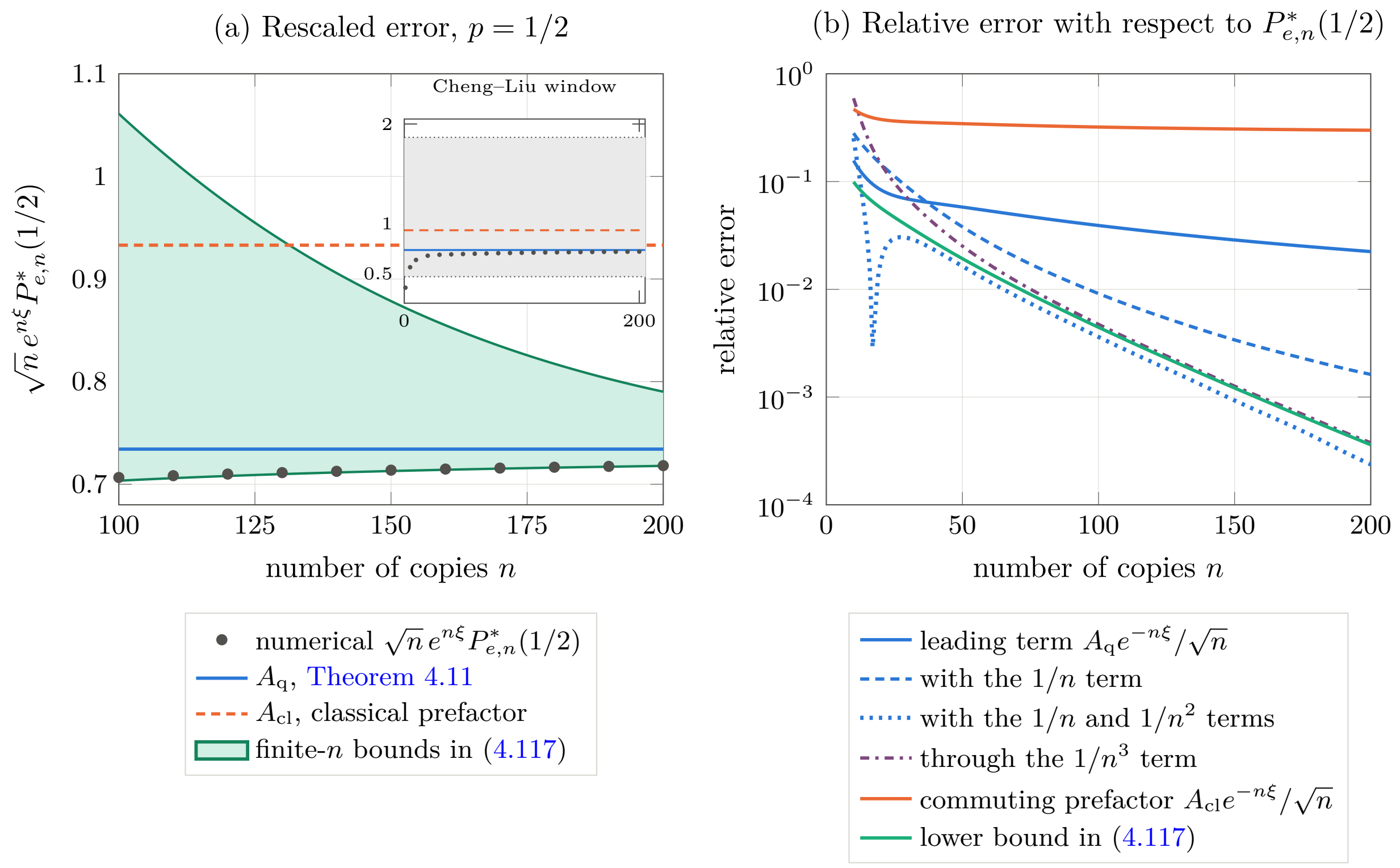

\begin{example}[Noisy Pure Qubit States]
\label{ex:ch-numerics}
Let $\psi$ and $\varphi$ be pure qubit states with $|\langle\psi,\varphi\rangle|^2=\cos^2\frac\pi8\approx0.854$, and let
\[
\rho\coloneqq0.98\,\psi+0.02\,\pi,
\qquad
\sigma\coloneqq0.92\,\varphi+0.08\,\pi.
\]
Equivalently, the Bloch vectors of $\rho$ and $\sigma$ have lengths $a=0.98$ and $b=0.92$ and enclose the angle $\pi/4$. The ratio $\operatorname{artanh}(a)/\operatorname{artanh}(b)=\ln99/\ln24$ is irrational, since $99^k\neq24^l$ for $(k,l)\neq(0,0)$, and $a$ and $b$ are rational. By \cref{prop:ch-regular-pairs}, the pair is regular and satisfies (QN). Numerically, $\alpha_*=0.453650$, $\xi=0.114313$, $V_*=0.744111$, and $\Delta_*=0.019438$. For $p=1/2$, we have $A_{\mathrm q}=0.734229$, $A_{\mathrm{cl}}=0.932973$, $C_1(1/2)=4.75170$, $C_2(1/2)=-55.4838$, and $C_3(1/2)=1123.20367$.

We evaluated $P_{e,n}^*(1/2)$ numerically up to $n=200$ using Schur--Weyl duality~\cite{fulton1991representation}. For qubits,
\[
\begin{gathered}
\rho^{\otimes n}-t\sigma^{\otimes n}\cong\bigoplus_{k=0}^{\lfloor n/2\rfloor}\left(\pi_k(\rho)-t\,\pi_k(\sigma)\right)\otimes I_{m_k},
\\
\pi_k(X)\coloneqq(\det X)^k\operatorname{Sym}^{n-2k}(X),
\qquad
m_k\coloneqq\binom nk-\binom n{k-1},
\end{gathered}
\]
where $\pi_k$ is the irreducible representation of $\mathrm{GL}(2)$ with highest weight $(n-k,k)$ and $m_k$ is its multiplicity. Hence $m(\rho^{\otimes n},t\sigma^{\otimes n})=\sum_km_k\,m\!\left(\pi_k(\rho),t\,\pi_k(\sigma)\right )$, which only requires the eigenvalues of matrices of size at most $n+1$.

\Cref{fig:chernoff-numerics} shows the rescaled error $\sqrt n\,e^{n\xi}P_{e,n}^*(1/2)$ and the relative errors of the approximations. The leading term $A_{\mathrm q}e^{-n\xi}/\sqrt n$ of \cref{thm:ch-main}, obtained by neglecting all terms of lower order, approximates the optimal error with a relative error of $3.9\%$ at $n=100$ and $2.2\%$ at $n=200$, and with the corrections of order $1/n$ and $1/n^2$ in~\eqref{eq:ch-higher}, the relative error decreases to $0.16\%$ and $2.4\times10^{-4}$ at $n=200$, respectively. Including $C_3(p)/n^3$ gives a relative error of $3.8\times10^{-4}$ at $n=200$. A higher truncation order need not be more accurate at each finite $n$, since the remainders can partially cancel. In contrast, the approximation with the commuting prefactor $A_{\mathrm{cl}}$ has a relative error of about $30\%$ at $n=200$, which converges to $1/\kappa_{\alpha_*}-1=0.271$, and the Audenaert bound $\frac12e^{-n\xi}$ is larger than the optimal error by a factor that grows like $\sqrt n$ and reaches $9.8$ at $n=200$. 
The lower bound in~\eqref{eq:ch-finite-bounds} has a relative error that decreases {exponentially}, from $4.4\times10^{-3}$ at $n=100$ to $3.6\times10^{-4}$ at $n=200$, {in agreement with~\eqref{eq:ch-rn-ratio}}. Moreover, the upper bound in~\eqref{eq:ch-finite-bounds} with $\alpha=0.458035$, which minimizes $Q_{\alpha,2}(\rho,\sigma)$, exceeds the optimal error by $10\%$ at $n=200$. 

\Cref{tab:sample} compares numerical estimates of the sample complexity $n^*(1/2,\delta)$ with \cref{cor:ch-sample}. For all the values of $\delta$ in the table, $n^*(1/2,\delta)-x_\delta$ lies in $[-0.44,0.27]$, whereas the bounds of~\cite[Theorem~II.7]{cheng2024invitation} leave an uncertainty of $14$ to $55$ samples.
\end{example}

\begin{table}[t]
\centering
\small
\renewcommand{\arraystretch}{1.15}
\begin{tabular}{@{}lrrrrrr@{}}
\toprule
$\delta$ & $10^{-2}$ & $10^{-3}$ & $10^{-4}$ & $10^{-6}$ & $10^{-8}$ & $10^{-10}$\\
\midrule
numerical $n^*(1/2,\delta)$ & $24$ & $41$ & $60$ & $98$ & $137$ & $176$\\
$x_\delta$ in~\eqref{eq:ch-sample-root} & $23.73$ & $41.44$ & $59.96$ & $98.10$ & $136.92$ & $176.11$\\
lower bound of~\cite[Theorem~II.7]{cheng2024invitation} & $21.0$ & $36.0$ & $51.0$ & $81.0$ & $111.0$ & $141.0$\\
upper bound of~\cite[Theorem~II.7]{cheng2024invitation} & $35$ & $55$ & $75$ & $115$ & $156$ & $196$\\
\bottomrule
\end{tabular}
\caption{Sample complexity of symmetric binary hypothesis testing for the noisy qubit states of \cref{ex:ch-numerics} with $p=1/2$. The numerical estimates are compared with the solution $x_\delta$ of~\eqref{eq:ch-sample-root} in \cref{cor:ch-sample}, which predicts $n^*(1/2,\delta)\in[x_\delta,x_\delta+1]$ up to $o(1)$, and with the lower bound $\ln(p(1-p)/\delta)/(-\ln F(\rho,\sigma))$ (the other term in the maximum of~\cite[Theorem~II.7]{cheng2024invitation} is smaller here) and the upper bound $\lceil\ln(\frac12/\delta)/\xi\rceil$ of~\cite[Theorem~II.7]{cheng2024invitation}, where $F(\rho,\sigma) \coloneqq \lVert\sqrt\rho\sqrt\sigma\rVert_1^2$ is the Uhlmann fidelity.}
\label{tab:sample}
\end{table}

\subsubsection{Application to the Integral R\'enyi Divergence}
\label{sec:ch-integral}

The harmonic-overlap comparison also determines the exact bounded correction to the integral R\'enyi divergence of~\cite{hirche2024quantum}. This application does not require a nonlattice assumption. For states $\rho,\sigma$ and $\alpha\in(0,1)$, recall the Petz R\'enyi divergence
\begin{equation}
D_\alpha(\rho\Vert\sigma)\coloneqq-\frac1{1-\alpha}\ln\Tr\!\left[\rho^\alpha\sigma^{1-\alpha}\right],\qquad\alpha\in(0,1).
\label{eq:ch-petz}
\end{equation}
We also define
\begin{equation}
Q_\alpha^{\mathrm{int}}(\rho\Vert\sigma)\coloneqq\alpha(1-\alpha)\int_0^\infty t^{\alpha-2}m(\rho,t\sigma)\,dt,
\qquad
D_\alpha^{\mathrm{int}}(\rho\Vert\sigma)\coloneqq-\frac1{1-\alpha}\ln Q_\alpha^{\mathrm{int}}(\rho\Vert\sigma).
\label{eq:ch-int}
\end{equation}
With $x=\ln t$, the integral in~\eqref{eq:ch-int} is the area under $e^{(\alpha-1)x}m(\rho,e^x\sigma)$.

Using $m(\rho,t\sigma)=t\,m(\sigma,t^{-1}\rho)$ and~\eqref{eq:m-hsd}, one verifies that\[Q_\alpha^{\mathrm{int}}(\rho\Vert\sigma)=1-\alpha(1-\alpha)\int_1^\infty\left[\gamma^{\alpha-2}E_\gamma(\rho\Vert\sigma)+\gamma^{-\alpha-1}E_\gamma(\sigma\Vert\rho)\right]d\gamma,\]so that $D_\alpha^{\mathrm{int}}$ is the R\'enyi divergence obtained from the integral representation of~\cite{hirche2024quantum}.

\begin{proposition}[Comparison with the Petz R\'enyi Divergence]
\label{prop:ch-integral}
For all states $\rho,\sigma$, $\alpha\in(0,1)$, and $n\in\mathbb N$,
\begin{equation}
\kappa_\alpha\left(\Tr[\rho^\alpha\sigma^{1-\alpha}]\right)^n\le Q_\alpha^{\mathrm{int}}(\rho^{\otimes n}\Vert\sigma^{\otimes n})\le\left(\Tr[\rho^\alpha\sigma^{1-\alpha}]\right)^n,
\label{eq:ch-int-petz}
\end{equation}
where $\kappa_\alpha$ is given in~\eqref{eq:ch-kappa}. If $\Tr[\rho\sigma]>0$, then equivalently
\begin{equation}
0\le D_\alpha^{\mathrm{int}}(\rho^{\otimes n}\Vert\sigma^{\otimes n})-nD_\alpha(\rho\Vert\sigma)\le\frac{\ln(1/\kappa_\alpha)}{1-\alpha}.
\label{eq:ch-int-order}
\end{equation}
\end{proposition}

\begin{proof}
For $a,b>0$ and $\beta\in(0,1)$, the substitution $t=(a/b)s$ and $\int_0^\infty s^{\beta-1}(1+s)^{-1}ds=\pi/\sin(\pi\beta)$ give
\begin{equation}
\int_0^\infty t^{\beta-2}\frac{a\,tb}{a+tb}\,dt=\frac{\pi}{\sin(\pi\beta)}a^\beta b^{1-\beta}.
\label{eq:ch-beta-integral}
\end{equation}
Then we have that
\begin{align}
Q_\alpha^{\mathrm{int}}(\rho^{\otimes n}\Vert\sigma^{\otimes n})
&\ge\alpha(1-\alpha)\int_0^\infty t^{\alpha-2}\Har(\rho^{\otimes n},t\sigma^{\otimes n})\,dt\\
&=\alpha(1-\alpha)\sum_{\mathbf k}\int_0^\infty t^{\alpha-2}\frac{P^{(n)}(\mathbf k)\,tQ^{(n)}(\mathbf k)}{P^{(n)}(\mathbf k)+tQ^{(n)}(\mathbf k)}\,dt
=\frac{\pi\alpha(1-\alpha)}{\sin(\pi\alpha)}\sum_{\mathbf k}P^{(n)}(\mathbf k)^\alpha Q^{(n)}(\mathbf k)^{1-\alpha}\\
&=\kappa_\alpha\left(\Tr[\rho^\alpha\sigma^{1-\alpha}]\right)^n,
\label{eq:ch-int-Har}
\end{align}
where the inequality follows from $\Har (\cdot, \cdot)\le m(\cdot, \cdot)$ in~\eqref{eq:ch-Har-m}, the first equality from \cref{prop:ch-classical-rep}, the second equality from~\eqref{eq:ch-beta-integral}, and the last equality from $\sum_{\mathbf k}P^{(n)}(\mathbf k)^\alpha Q^{(n)}(\mathbf k)^{1-\alpha}=(Q_\alpha^{\NS})^n$ and~\eqref{eq:ch-kappa}. This proves the lower bound in~\eqref{eq:ch-int-petz}, and the upper bound is~\cite[Proposition~3.9]{beigi2025properties}. Finally, if $\Tr[\rho\sigma]>0$, then $\Tr[\rho^\alpha\sigma^{1-\alpha}]>0$, and applying $-\frac1{1-\alpha}\ln$ to~\eqref{eq:ch-int-petz} gives~\eqref{eq:ch-int-order}.
\end{proof}

The constant $\kappa_\alpha$ in~\eqref{eq:ch-int-petz} cannot be improved in general. Indeed, for pure states $\rho=|\psi\rangle\!\langle\psi|$ and $\sigma=|\varphi\rangle\!\langle\varphi|$ with $r\coloneqq|\langle\psi,\varphi\rangle|^2\to0$, we have $m(\rho,t\sigma)=\Har(\rho,t\sigma)(1+O(r))$, so that $Q_\alpha^{\mathrm{int}}(\rho\Vert\sigma)/(\kappa_\alpha\Tr[\rho^\alpha\sigma^{1-\alpha}])\to1$. For noncommuting states and many copies, the lower bound in~\eqref{eq:ch-int-petz} is achieved up to an exponentially small relative error.

\begin{theorem}[Integral R\'enyi Divergence for Noncommuting States]
\label{thm:ch-integral-exact}
Let $\rho,\sigma$ be states with $[\rho,\sigma]\neq0$, and let $\alpha\in(0,1)$. Then there exists $c_\alpha>0$ such that the quantities in~\eqref{eq:ch-int} satisfy
\begin{align}
Q_\alpha^{\mathrm{int}}(\rho^{\otimes n}\Vert\sigma^{\otimes n})&=\kappa_\alpha\left(\Tr[\rho^\alpha\sigma^{1-\alpha}]\right)^n\left[1+O(e^{-nc_\alpha})\right],
\label{eq:ch-int-exact}\\
D_\alpha^{\mathrm{int}}(\rho^{\otimes n}\Vert\sigma^{\otimes n})&=nD_\alpha(\rho\Vert\sigma)+\frac{\ln(1/\kappa_\alpha)}{1-\alpha}+O(e^{-nc_\alpha}),
\label{eq:ch-int-exact-log}
\end{align}
with $\kappa_\alpha$ in~\eqref{eq:ch-kappa} and $D_\alpha$ in~\eqref{eq:ch-petz}. In contrast, for commuting states, $Q_\alpha^{\mathrm{int}}(\rho^{\otimes n}\Vert\sigma^{\otimes n})=\left(\Tr[\rho^\alpha\sigma^{1-\alpha}]\right)^n$ for all $n$.
\end{theorem}
\begin{proof} 
We split $m=\Har+R_n$, where
\begin{equation}
R_n(t)\coloneqq m(\rho^{\otimes n},t\sigma^{\otimes n})-\Har(\rho^{\otimes n},t\sigma^{\otimes n})\ge0
\end{equation}
by~\eqref{eq:ch-Har-m}. By~\eqref{eq:ch-int} and the equalities in~\eqref{eq:ch-int-Har}, the harmonic part contributes exactly $\kappa_\alpha(\Tr[\rho^\alpha\sigma^{1-\alpha}])^n$, so that
\begin{equation}
Q_\alpha^{\mathrm{int}}(\rho^{\otimes n}\Vert\sigma^{\otimes n})-\kappa_\alpha\left(\Tr[\rho^\alpha\sigma^{1-\alpha}]\right)^n=\alpha(1-\alpha)\int_0^\infty t^{\alpha-2}R_n(t)\,dt\ge0.
\label{eq:ch-int-split}
\end{equation}
It remains to show that the right-hand side is exponentially smaller than $(\Tr[\rho^\alpha\sigma^{1-\alpha}])^n$.

\noindent \underline{Pointwise bounds on $R_n$}: For every $\beta\in(0,1)$, \eqref{eq:ch-uniform-remainder} gives $t^{\alpha-2}R_n(t)\le\frac12t^{\alpha-\beta-1}Q_{\beta,2}(\rho,\sigma)^n$. The choice $\beta=\alpha$ gives the bound $\frac12t^{-1}Q_{\alpha,2}(\rho,\sigma)^n$, which is integrable neither at $t=0$ nor at $t=\infty$. We therefore use two orders slightly below and above $\alpha$. Since $[\rho,\sigma]\neq0$, \cref{cor:ch-gap} gives $Q_{\alpha,2}(\rho,\sigma)<\Tr[\rho^\alpha\sigma^{1-\alpha}]$. Since $\beta\mapsto Q_{\beta,2}(\rho,\sigma)$ is continuous, we can choose $0<\eta<\min\{\alpha,1-\alpha\}$ such that
\begin{equation}
q_*\coloneqq\frac{\sqrt{q_-q_+}}{\Tr[\rho^\alpha\sigma^{1-\alpha}]}<1,
\qquad\text{where}\qquad
q_\pm\coloneqq Q_{\alpha\pm\eta,2}(\rho,\sigma).
\end{equation}
Applying the above bound with $\beta=\alpha-\eta$ and $\beta=\alpha+\eta$, we obtain, for all $t>0$,
\begin{equation}
0\le t^{\alpha-2}R_n(t)\le\frac12\min\left\{t^{\eta-1}q_-^n,\,t^{-\eta-1}q_+^n\right\}.
\label{eq:ch-Rn-min}
\end{equation}
Here the first term is integrable near $t=0$, and the second term is integrable near $t=\infty$.

\noindent \underline{Integration}: The two terms in~\eqref{eq:ch-Rn-min} are equal at $t_*\coloneqq(q_+/q_-)^{n/(2\eta)}$, which satisfies $t_*^{\eta}=(q_+/q_-)^{n/2}$. Bounding the minimum by the first term for $t\le t_*$ and by the second term for $t>t_*$, we arrive at
\begin{align}
\alpha(1-\alpha)\int_0^\infty t^{\alpha-2}R_n(t)\,dt
&\le\frac{\alpha(1-\alpha)}2\left(q_-^n\int_0^{t_*}t^{\eta-1}\,dt+q_+^n\int_{t_*}^\infty t^{-\eta-1}\,dt\right)\\
&=\frac{\alpha(1-\alpha)}{2\eta}\left(q_-^n\,t_*^{\eta}+q_+^n\,t_*^{-\eta}\right)\\
&=\frac{\alpha(1-\alpha)}{\eta}\,(q_-q_+)^{n/2}\\
&=\frac{\alpha(1-\alpha)}{\eta}\left(\Tr[\rho^\alpha\sigma^{1-\alpha}]\right)^nq_*^n,
\end{align}
where the inequality follows from~\eqref{eq:ch-Rn-min}, the first equality from direct integration, the second equality from $t_*^{\eta}=(q_+/q_-)^{n/2}$, and the last equality from the definition of $q_*$.

Combining~\eqref{eq:ch-int-split} with the previous display and dividing by $\kappa_\alpha(\Tr[\rho^\alpha\sigma^{1-\alpha}])^n$, we obtain
\begin{equation}
0\le\frac{Q_\alpha^{\mathrm{int}}(\rho^{\otimes n}\Vert\sigma^{\otimes n})}{\kappa_\alpha\left(\Tr[\rho^\alpha\sigma^{1-\alpha}]\right)^n}-1\le\frac{\alpha(1-\alpha)}{\eta\,\kappa_\alpha}\,q_*^n,
\end{equation}
which proves~\eqref{eq:ch-int-exact} with $c_\alpha\coloneqq-\ln q_*>0$. Applying $-\frac1{1-\alpha}\ln$ to~\eqref{eq:ch-int-exact} and using~\eqref{eq:ch-petz} and $\ln(1+O(e^{-nc_\alpha}))=O(e^{-nc_\alpha})$ gives~\eqref{eq:ch-int-exact-log}.

\noindent \underline{Commuting states}: Let $\rho$ and $\sigma$ have joint eigenvalues $(a_x,b_x)$. By~\eqref{eq:m-commuting}, $m(\rho^{\otimes n},t\sigma^{\otimes n})=\sum_{\mathbf x}\min\{a_{\mathbf x},tb_{\mathbf x}\}$, where $a_{\mathbf x}\coloneqq\prod_ka_{x_k}$ and $b_{\mathbf x}\coloneqq\prod_kb_{x_k}$. For $a,b>0$, splitting the integral at $t=a/b$ gives
\begin{equation}
\alpha(1-\alpha)\int_0^\infty t^{\alpha-2}\min\{a,tb\}\,dt
=\alpha(1-\alpha)\left[b\int_0^{a/b}t^{\alpha-1}\,dt+a\int_{a/b}^\infty t^{\alpha-2}\,dt\right]
=a^\alpha b^{1-\alpha},
\end{equation}
and the terms with $a_{\mathbf x}b_{\mathbf x}=0$ vanish. Summing over $\mathbf x$ gives $Q_\alpha^{\mathrm{int}}(\rho^{\otimes n}\Vert\sigma^{\otimes n})=\sum_{\mathbf x}a_{\mathbf x}^\alpha b_{\mathbf x}^{1-\alpha}=(\Tr[\rho^\alpha\sigma^{1-\alpha}])^n$, completing the proof.
\end{proof}

\subsubsection{Proof of the Refined Chernoff Bound}
\label{sec:ch-proofs}

We now prove \cref{thm:ch-main}. Throughout, we write $x_0\coloneqq\ln((1-p)/p)$ for the prior log-odds.

\begin{proof}[Proof of \cref{thm:ch-main}]
We first write the optimal error in terms of the harmonic overlap, up to an exponentially smaller remainder. Then we apply \cref{lem:ch-LLT-harmonic} for the leading term and \cref{prop:ch-local-expansion} for the higher-order terms. Recall that 
since $\alpha_*$ minimizes $\Phi$,
\begin{equation}
\Phi'(\alpha_*)=0,
\qquad
\Phi''(\alpha_*)=V_*,
\qquad
Q_{\alpha_*}^{\NS}=e^{\Phi(\alpha_*)}=e^{-\xi}.
\label{eq:ch-at-alpha-star}
\end{equation}

\noindent \underline{Reduction to the harmonic overlap}: Define
\begin{equation}
H_n\coloneqq e^{(\alpha_*-1)x_0}\Har(\rho^{\otimes n},e^{x_0}\sigma^{\otimes n}),
\qquad
r_n\coloneqq e^{(\alpha_*-1)x_0}\left[m(\rho^{\otimes n},e^{x_0}\sigma^{\otimes n})-\Har(\rho^{\otimes n},e^{x_0}\sigma^{\otimes n})\right],
\end{equation}
where the factor $e^{(\alpha_*-1)x_0}$ is the normalization used in \cref{lem:ch-LLT-harmonic,prop:ch-local-expansion}. Then we have that
\begin{align}
P_{e,n}^*(p)&=p\,m\!\left(\rho^{\otimes n},\tfrac{1-p}p\sigma^{\otimes n}\right)
=p\,e^{(1-\alpha_*)x_0}\cdot e^{(\alpha_*-1)x_0}m(\rho^{\otimes n},e^{x_0}\sigma^{\otimes n})\\
&=p^{\alpha_*}(1-p)^{1-\alpha_*}\left[H_n+r_n\right],
\label{eq:ch-Pe-at-x0}
\end{align}
where the first equality follows from~\eqref{eq:ch-prior-profile} and $m(cA,cB)=c\,m(A,B)$ with $c=p$, the second from $e^{x_0}=(1-p)/p$, and the last from $p\,e^{(1-\alpha_*)x_0}=p\left(\frac{1-p}p\right)^{1-\alpha_*}=p^{\alpha_*}(1-p)^{1-\alpha_*}$ and the definitions of $H_n$ and $r_n$. Moreover, \eqref{eq:ch-uniform-remainder} with $\alpha=\alpha_*$ and $t=e^{x_0}$, together with~\eqref{eq:ch-Delta} and~\eqref{eq:ch-at-alpha-star}, gives
\begin{equation}
0\le r_n\le\frac12\left[Q_{\alpha_*,2}(\rho,\sigma)\right]^n=\frac12e^{-n(\xi+\Delta_*)},
\label{eq:ch-rn}
\end{equation}
where $\Delta_*>0$ by \cref{cor:ch-gap}, since $[\rho,\sigma]\neq0$.

\noindent \underline{Leading term}: We have that
\begin{align}
H_n&=\left(Q_{\alpha_*}^{\NS}\right)^n\left[\frac{\pi}{\sin(\pi\alpha_*)}\,\phi_{nV_*}(x_0)+o(n^{-1/2})\right]\\
&=e^{-n\xi}\left[\frac{\pi}{\sin(\pi\alpha_*)}\frac{e^{-x_0^2/(2nV_*)}}{\sqrt{2\pi nV_*}}+o(n^{-1/2})\right]\\
&=\frac{\pi}{\sin(\pi\alpha_*)}\frac{e^{-n\xi}}{\sqrt{2\pi nV_*}}\,[1+o(1)],
\label{eq:ch-Hn}
\end{align}
where the first equality follows from \cref{lem:ch-LLT-harmonic} at $\alpha=\alpha_*$ and $x=x_0$, together with $\Phi'(\alpha_*)=0$ and $\Phi''(\alpha_*)=V_*$, the second from~\eqref{eq:ch-at-alpha-star} and the definition of $\phi_v$ in~\eqref{eq:Gaussian_density}, and the last from $e^{-x_0^2/(2nV_*)}\to1$, since $x_0$ does not depend on $n$.
By~\eqref{eq:ch-rn} and~\eqref{eq:ch-Hn}, there is a constant $C>0$ such that, for all sufficiently large $n$,
\begin{equation}
0\le\frac{r_n}{H_n}\le C\sqrt n\,e^{-n\Delta_*}.
\label{eq:ch-rn-ratio}
\end{equation}
Hence
\begin{align}
P_{e,n}^*(p)&=p^{\alpha_*}(1-p)^{1-\alpha_*}H_n\left(1+\frac{r_n}{H_n}\right)\\
&=\frac{\pi\,p^{\alpha_*}(1-p)^{1-\alpha_*}}{\sin(\pi\alpha_*)\sqrt{2\pi V_*}}\,\frac{e^{-n\xi}}{\sqrt n}\,[1+o(1)]
=A_{\mathrm q}\frac{e^{-n\xi}}{\sqrt n}\,[1+o(1)],
\end{align}
where the first equality follows from~\eqref{eq:ch-Pe-at-x0}, the second from~\eqref{eq:ch-Hn} and~\eqref{eq:ch-rn-ratio}, and the last from the definition of $A_{\mathrm q}$ in~\eqref{eq:ch-prefactor}. This proves~\eqref{eq:ch-prefactor}.

\noindent \underline{Higher-order terms}: Assume, in addition, that the pair satisfies (QN), and fix $M\ge1$. By~\eqref{eq:ch-at-alpha-star}, \cref{prop:ch-local-expansion} at $\alpha=\alpha_*$ and $x=n\Phi'(\alpha_*)+y$ with $y=x_0$ gives
\begin{equation}
\ln H_n=-n\xi+\Lambda(\alpha_*)-\frac12\ln(2\pi nV_*)+\sum_{j=1}^{M}\frac{c_j(x_0)}{n^j}+O(n^{-M-1}).
\label{eq:ch-lnHn}
\end{equation}
Taking $-\ln$ of~\eqref{eq:ch-Pe-at-x0}, we have that
\begin{align}
-\ln P_{e,n}^*(p)&=-\alpha_*\ln p-(1-\alpha_*)\ln(1-p)-\ln H_n-\ln\!\left(1+\frac{r_n}{H_n}\right)\\
&=n\xi+\frac12\ln n-\alpha_*\ln p-(1-\alpha_*)\ln(1-p)-\Lambda(\alpha_*)+\frac12\ln(2\pi V_*)\notag\\
&\qquad-\sum_{j=1}^{M}\frac{c_j(x_0)}{n^j}+O(n^{-M-1})\\
&=n\xi+\frac12\ln n-\ln A_{\mathrm q}-\sum_{j=1}^{M}\frac{c_j(x_0)}{n^j}+O(n^{-M-1}),
\end{align}
where the second equality follows from~\eqref{eq:ch-lnHn} and $\ln(1+r_n/H_n)=O(\sqrt n\,e^{-n\Delta_*})=O(n^{-M-1})$ by~\eqref{eq:ch-rn-ratio}, and the last equality from $\Lambda(\alpha_*)=\ln\frac{\pi}{\sin(\pi\alpha_*)}$ and the definition of $A_{\mathrm q}$. This proves~\eqref{eq:ch-higher} with $C_j(p)=-c_j(x_0)$. Finally, $x_0-\tau(\alpha_*)=\delta_p$ by the definition of $\delta_p$, so that~\eqref{eq:ch-c1} gives $C_1(p)=-b_1(\alpha_*)+\delta_p^2/(2V_*)$, which is~\eqref{eq:ch-C1} by~\eqref{eq:ch-b1}. In the same way,~\eqref{eq:ch-c2} gives~\eqref{eq:ch-C2}, and Appendix~\ref{sec:ch-coefficients} gives the general rule and $C_3(p)$.
\end{proof}

\subsubsection{Cases Outside the Regular Setting}
\label{sec:ch-outside}

We end this subsection with examples that illustrate the role of the assumptions in \cref{thm:ch-main}.

\begin{example}[Commuting States]
\label{ex:ch-commuting}
If $\rho$ and $\sigma$ commute, then the problem is classical, and by~\eqref{eq:m-commuting}, $e^{(\alpha-1)x}m(\rho^{\otimes n},e^x\sigma^{\otimes n})=\bigl(Q_{\alpha}^{\NS}\bigr)^{n}\,\mathbb E_\alpha[h_\alpha(S_n-x)]$ with the kernel $h_\alpha(z)\coloneqq\min\{e^{(1-\alpha)z},e^{-\alpha z}\}$, whose integral is $1/(\alpha(1-\alpha))$. If the Chernoff minimizer is interior, $V_*>0$, and the pair is nonlattice, \cref{lem:ch-LLT} gives
\begin{equation}
P_{e,n}^*(p)=A_{\mathrm{cl}}\frac{e^{-n\xi}}{\sqrt n}[1+o(1)],
\label{eq:ch-commuting}
\end{equation}
in agreement with the classical precise asymptotics~\cite{bahadur1960deviations}. Therefore, the noncommuting prefactor in \cref{thm:ch-main} differs from the commuting one by the factor $\kappa_{\alpha_*}$ in~\eqref{eq:ch-kappa}.
\end{example}

\begin{example}[The Function $\Phi$ Does Not Determine the Prefactor]
\label{ex:ch-petz-curve}
Let $P,Q$ be probability distributions with full support on a finite set, and let $\psi,\varphi$ be unit vectors with $|\langle\psi,\varphi\rangle|^2=c\in(0,1)$. Consider
\begin{align*}
\rho_{\mathrm q}&\coloneqq\operatorname{diag}(P)\otimes|\psi\rangle\!\langle\psi|,&
\sigma_{\mathrm q}&\coloneqq\operatorname{diag}(Q)\otimes|\varphi\rangle\!\langle\varphi|,\\
\rho_{\mathrm{cl}}&\coloneqq\operatorname{diag}(cP,1-c,0),&
\sigma_{\mathrm{cl}}&\coloneqq\operatorname{diag}(cQ,0,1-c).
\end{align*}
Then, for all $\alpha\in(0,1)$,
\begin{equation}
\Tr\!\left[\rho_{\mathrm q}^\alpha\sigma_{\mathrm q}^{1-\alpha}\right]=\Tr\!\left[\rho_{\mathrm{cl}}^\alpha\sigma_{\mathrm{cl}}^{1-\alpha}\right]=c\sum_xP(x)^\alpha Q(x)^{1-\alpha},
\label{eq:ch-same-petz}
\end{equation}
so that both pairs have the same Chernoff exponent and the same tilted law of $Z$. However, the first pair is noncommuting, and the second pair is commuting. Therefore, if the law of $Z$ is nonlattice, \cref{thm:ch-main} and \cref{ex:ch-commuting} show that their prefactors differ by the factor $\kappa_{\alpha_*}$, and by \cref{cor:ch-sample}, their sample complexities differ by $\ln(1/\kappa_{\alpha_*})/\xi+O(1)$. Thus the Petz R\'enyi divergences determine the Chernoff exponent, but not the prefactor. In contrast, \cref{thm:ch-integral-exact} shows that the integral R\'enyi divergences of the two pairs differ by $\ln(1/\kappa_\alpha)/(1-\alpha)+o(1)$.
\end{example}

\begin{example}[Lattice Spectra]
\label{ex:ch-lattice}
Suppose that the values of $Z$ in~\eqref{eq:ch-tilt} are contained in $z_0+d\mathbb Z$ with maximal span $d>0$, and that $\Phi''>0$. Then the lattice local limit theorem replaces the integral $\int g_\alpha=\pi/\sin(\pi\alpha)$ in the proof of \cref{thm:ch-main} by the periodic function
\begin{equation}
\mathcal G_\alpha(\vartheta)\coloneqq d\sum_{k\in\mathbb Z}g_\alpha(\vartheta+kd),\qquad\vartheta\in\mathbb R.
\label{eq:ch-lattice-G}
\end{equation}
For a noncommuting pair with an interior minimizer $\alpha_*$, this leads to
\begin{equation}
P_{e,n}^*(p)=\frac{p^{\alpha_*}(1-p)^{1-\alpha_*}}{\sqrt{2\pi nV_*}}\,\mathcal G_{\alpha_*}\!\left(nz_0-\ln\frac{1-p}p\right)e^{-n\xi}[1+o(1)].
\label{eq:ch-lattice-Pe}
\end{equation}
Thus, for a fixed prior $p$, the rescaled error $\sqrt{n}\,e^{n\xi}P_{e,n}^*(p)$ can oscillate with $n$, so a single limiting prefactor need not exist.
For example, for qubit states whose Bloch vectors have the same length $a=0.95$ and enclose the angle $\pi/4$ (see \cref{rem:ch-regular-examples}), the values of $Z$ are $0$ and $\pm d$ with $d=\ln39$, so that $z_0=0$, $\alpha_*=1/2$, and, for $p=1/2$, \eqref{eq:ch-lattice-Pe} predicts $-\ln P_{e,n}^*(1/2)-n\xi-\frac12\ln n\to0.25815$, with $\mathcal G_{1/2}(0)=3.1993$ in place of $\pi$. The nonlattice formula~\eqref{eq:ch-prefactor} would instead give $0.27635$.
\end{example}

\begin{example}[Pure States and Boundary Minimizers]
\label{ex:ch-pure}
For pure states $\rho=|\psi\rangle\!\langle\psi|$ and $\sigma=|\varphi\rangle\!\langle\varphi|$ with $r\coloneqq|\langle\psi,\varphi\rangle|^2\in(0,1)$, the variable $Z$ is constant, so $\Phi''\equiv0$, and the formula~\eqref{eq:xformula} gives
\begin{equation}
P_{e,n}^*(p)=\frac{1-\sqrt{1-4p(1-p)r^n}}2=p(1-p)r^n+p^2(1-p)^2r^{2n}+O(r^{3n}).
\label{eq:ch-pure}
\end{equation}
In particular, there is no $\frac12\ln n$ term. For singular pairs, the minimizer of $\Phi$ can also lie at an endpoint of $[0,1]$. For example, for $\rho=|0\rangle\!\langle0|$ and $\sigma=q|0\rangle\!\langle0|+(1-q)|1\rangle\!\langle1|$ with $q\in(0,1)$, we have $P_{e,n}^*(p)=\min\{p,(1-p)q^n\}$, which again has no logarithmic correction.
\end{example}

\begin{example}[Nearly Maximally Mixed States]
\label{ex:ch-faithful}
Let
\begin{equation}
\rho\coloneqq\operatorname{diag}\!\left(\frac e{1+e},\frac1{1+e}\right),
\qquad
\sigma\coloneqq H\operatorname{diag}\!\left(\frac{e^{\sqrt2}}{1+e^{\sqrt2}},\frac1{1+e^{\sqrt2}}\right)H^\dagger,
\qquad
H\coloneqq\frac1{\sqrt2}\begin{pmatrix}1&1\\1&-1\end{pmatrix}.
\label{eq:ch-faithful-example}
\end{equation}
The Bloch vectors have lengths $a=\tanh\frac12$ and $b=\tanh\frac{\sqrt2}2$ and are orthogonal, so that $\theta\coloneqq\operatorname{artanh}(a)/\operatorname{artanh}(b)=1/\sqrt2$. Since $k^2-2l^2$ is a nonzero integer, we have $|k\theta-l|=|k^2-2l^2|/(2|k\theta+l|)\ge1/(4k)$ for all integers $k\ge1$ and $l$ with $|k\theta-l|<1/2$, so that $|k\theta-l|\ge1/(4k)$ for all integers $k\ge1$ and $l$, and \cref{prop:ch-regular-pairs} shows that the pair is regular and satisfies (QN). However, since both states are far from pure ($a\approx0.46$ and $b\approx0.61$), the commutator in~\eqref{eq:ch-commutator} is small,
{$\Delta_*\approx9.05\times10^{-4}$, so that the factor $e^{-n\Delta_*}$ in the remainder estimate~\eqref{eq:ch-rn-ratio} becomes small only for $n$ well above $1/\Delta_*\approx1.1\times10^3$.}
\end{example}

\section{Concluding Remarks and Future Directions}
\label{sec:conclusion}

In this work, we proved that the noncommutative minimum $m(A,B)=(\Tr[A]+\Tr[B]-\lVert A-B\rVert_1)/2$ of positive semidefinite operators is super-multiplicative. The proof relies on an exact paired rank-one decomposition, which preserves the noncommutative minimum and reduces the problem to a scalar inequality. As a consequence, we established sub-additivity of the quantum hockey-stick divergence and of the Datta--Leditzky divergences on product states. We showed that the hockey-stick bound is tight, also for noncommuting states, that it is strict for faithful states whenever at least two of the divergences are positive, and that the parameters must be combined multiplicatively. 
Next, we derived a parallel composition theorem for quantum differential privacy, showing that the privacy parameters degrade at most linearly even when the adversary uses collective measurements, with privacy parameters that are optimal even for classical mechanisms. We also established a sharp sub-additivity property of the hypothesis-testing relative entropy normalized by the acceptance probability, which, through its interpretation as a likelihood ratio, bounds what collective tests can achieve in terms of the individual systems. Finally, we studied refined asymptotics of symmetric binary quantum hypothesis testing. By comparing the optimal error with the harmonic overlap, whose difference is exponentially small for noncommuting faithful states, we determined the exact prefactor of the quantum Chernoff bound, the expansion to every fixed order in $1/n$ under (QN), and the sample complexity up to one sample, and we showed that the required regularity conditions hold for generic pairs of faithful states, for instance after adding generic noise. 

There are several directions for future work. First, our composition results concern product neighboring inputs, and it is natural to ask how the privacy parameters degrade for correlated, entangled, or permutation-invariant inputs, for instance in quantum local differential privacy, where all pairs of joint input states are neighbors, and for adaptive (sequential) composition of quantum channels. Second, it would be interesting to extend the refined Chernoff analysis to multiple quantum hypothesis testing, to asymmetric hypothesis testing with exponentially small type-I error (the Hoeffding regime), and to the boundary cases discussed in \cref{sec:ch-outside}, where the minimizer of $\Phi$ lies at an endpoint or the log-likelihood ratio is lattice valued, as well as to obtain explicit finite-$n$ versions of the sample complexity in \cref{cor:ch-sample}. Lastly, the exact paired rank-one decomposition may be useful for other problems in which a quantity defined through the trace norm needs to be compared across tensor products, such as chain rules for one-shot conditional entropies beyond product states.

 \section*{Acknowledgments}
 We thank Marius Junge and Felix Leditzky for helpful discussions on the sub-additivity of the hockey-stick divergence. TN acknowledges support from the IQUIST Postdoctoral Fellowship from the Illinois Quantum Information Science and Technology Center at the University of Illinois Urbana-Champaign.  SB was supported by NSF Grant No. 2426103.
CH received funding from the Deutsche Forschungsgemeinschaft (DFG, German
Research Foundation) – 550206990, the Federal Ministry of Research, Technology and Space (BMFTR), Germany, under the QC service centre QUICS (grant no. 13N17418), QUANDA (grant no. 01MQ26002B), and by Germany's Excellence Strategy EXC-2123 Quantum Frontiers.
 
\section*{AI Statement}
We acknowledge the use of Claude Opus 5.5 and ChatGPT 6.1 Sol and 6.0 Astra in proof formulation, manuscript development, literature checks, figure generation, and numerical tests during the development of this manuscript. The authors take full responsibility for the content presented in the paper.

\bibliographystyle{IEEEtran}
\bibliography{reference}

\appendix

\section{Proof of Lemma~\ref{lem:rank-one}}
\label{Sec:Lemma_rank_one}

We reduce the rank-one inequality and its strictness claim to a scalar quadratic identity.

If one of $A,B,C,D$ is zero, then both sides of~\eqref{eq:rank-one} vanish, since $m(\cdot,0)=m(0,\cdot)=0$. Otherwise write $A=a uu^\dagger$, $B=b vv^\dagger$, $C=c ww^\dagger$, and $D=d zz^\dagger$, with unit vectors and $a,b,c,d>0$. Set $r=|\langle u,v\rangle|^2$ and $s=|\langle w,z\rangle|^2$. The eigenvalues of $A-B$ give
\begin{equation}
x\coloneqq m(A,B)=\frac{a+b-\sqrt{(a+b)^2-4abr}}{2},
\label{eq:xformula}
\end{equation}
and hence
\begin{equation}
x(a+b-x)=abr,\qquad 0\le x\le\min\{a,b\}.
\label{eq:xquad}
\end{equation}
The same calculation for $y\coloneqq m(C,D)$ and $\zeta\coloneqq m(A\otimes C,B\otimes D)$ yields
\[
y(c+d-y)=cds,\qquad \zeta(ac+bd-\zeta)=abcd\,rs,
\]
with $0\le y\le\min\{c,d\}$ and $0\le \zeta\le\min\{ac,bd\}$. For $f(t)\coloneqq t(ac+bd-t)$ these identities give
\begin{equation}
f(\zeta)-f(xy)=xy\bigl[(a-x)(d-y)+(b-x)(c-y)\bigr]\ge0.
\label{eq:key-scalar}
\end{equation}
Both $\zeta$ and $xy$ belong to $[0,\min\{ac,bd\}]$, since $xy\le\min\{a,b\}\min\{c,d\}\le\min\{ac,bd\}$, and $f$ is strictly increasing on this interval, since $\min\{ac,bd\}\le(ac+bd)/2$.
Thus $\zeta\ge xy$, proving~\eqref{eq:rank-one}.

If both pairs satisfy the strictness assumptions, then $r,s<1$ and $x,y>0$. Equation~\eqref{eq:xformula} gives $x<\min\{a,b\}$ and, likewise, $y<\min\{c,d\}$. The right-hand side of~\eqref{eq:key-scalar} is therefore positive, so $\zeta>xy$.

\section{Proof of Lemma~\ref{lem:decomposition_general_to_rank}}
\label{sec:proof_lemma_decomposition_general_to_rank}

We explicitly remove a common positive part, then decompose the residual pair into rank-one pairs whose differences have aligned positive and negative parts.

\emph{Removing a common part.} Work on $\mathcal T\coloneqq\supp(A+B)$; if $\mathcal T=\{0\}$, the claim is immediate. Define
\[
L\coloneqq(A+B)^{1/2},\qquad W\coloneqq L^{-1}(A-B)L^{-1},
\]
where the inverses are restricted to $\mathcal T$. Since $-\id\le W\le\id$, the operators
\begin{equation}
C_0\coloneqq\tfrac12L(\id-|W|)L,\qquad
A_0\coloneqq LW_+L,\qquad B_0\coloneqq LW_-L
\label{eq:explicit-common-part}
\end{equation}
are positive and satisfy $A=C_0+A_0$, $B=C_0+B_0$. Moreover, $\ran(A_0)=L\ran(W_+)$ and $\ran(B_0)=L\ran(W_-)$, so invertibility of $L$ and orthogonality of the spectral subspaces of $W$ imply
\begin{equation}
\ran(A_0)\cap\ran(B_0)=\{0\}.
\label{eq:disjoint-range}
\end{equation}
Set $H\coloneqq A_0-B_0=A-B$ and $\cS\coloneqq\ran(A_0)+\ran(B_0)$. By~\eqref{eq:disjoint-range}, $Hx=0$ implies $A_0x=B_0x=0$, whence $\ker H=\ker A_0\cap\ker B_0=\cS^\perp$. If $\cS=\{0\}$, then $A=B=C_0$, and pairing the spectral summands of $C_0$ with themselves gives the required decomposition. Otherwise, all inverses below are on $\cS$.

The operator $A_0H^{-1}B_0=A_0H^{-1}A_0-A_0$ is Hermitian, so it also equals $B_0H^{-1}A_0$. Its range lies in both residual ranges and is therefore zero. Consequently,
\begin{equation}
A_0H^{-1}A_0=A_0.
\label{eq:AHinvA}
\end{equation}

\emph{Rank-one blocks.} Define $G\coloneqq|H|^{1/2}$, $J\coloneqq\sgn(H)$, and $K\coloneqq G^{-1}A_0G^{-1}$ on $\cS$. Then $G$ commutes with $J$, $H^{-1}=G^{-1}JG^{-1}$, and~\eqref{eq:AHinvA} gives
\begin{equation}
K\ge0,\qquad K-J=G^{-1}B_0G^{-1}\ge0,\qquad KJK=K.
\label{eq:KJK}
\end{equation}
Relative to the positive and negative spectral subspaces $\cS=\cS_+\oplus\cS_-$ of $H$, write
\[
J=\begin{pmatrix}\id&0\\0&-\id\end{pmatrix},\qquad
K=\begin{pmatrix}P&Z\\Z^\dagger&Q\end{pmatrix}.
\]
Equation~\eqref{eq:KJK} gives $P\ge\id$, $Q\ge0$, $P^2-P=ZZ^\dagger$, and $Q^2+Q=Z^\dagger Z$. Taking the positive square roots yields
\begin{equation}
P=\frac{\id+\sqrt{\id+4ZZ^\dagger}}2,\qquad
Q=\frac{-\id+\sqrt{\id+4Z^\dagger Z}}2.
\label{eq:PQ}
\end{equation}
Choose singular vectors $u_j\in\cS_+$, $v_j\in\cS_-$ for the positive singular values $\tau_j$ of $Z$. Set $\cS_j=\operatorname{span}\{u_j, v_j\}$. Then
\[
\cS=\left( \bigoplus_j\cS_j \right)\oplus\ker Z^\dagger\oplus\ker Z
\]
is an orthogonal decomposition. By~\eqref{eq:PQ}, $Pu_j=p_ju_j$ and $Qv_j=q_jv_j$, where
\[
p_j\coloneqq\frac{1+\sqrt{1+4\tau_j^2}}2,\qquad q_j\coloneqq p_j-1.
\]
Together with $Zv_j=\tau_ju_j$ and $Z^\dagger u_j=\tau_jv_j$, these identities show that $K$ and $K-J$ preserve $\cS_j$. In the ordered basis $(u_j, v_j)$, the restrictions are
\[
K\vert_{\cS_j}=\begin{pmatrix}p_j&\tau_j\\\tau_j&q_j\end{pmatrix}
=\alpha_j\alpha_j^\dagger,\qquad
(K-J)\vert_{\cS_j}=\begin{pmatrix}q_j&\tau_j\\\tau_j&p_j\end{pmatrix}
=\beta_j\beta_j^\dagger,
\]
where $\alpha_j=\sqrt{p_j}\,u_j+\sqrt{q_j}\,v_j$ and $\beta_j=\sqrt{q_j}\,u_j+\sqrt{p_j}\,v_j$; here we use $p_jq_j=\tau_j^2$ and $p_j-q_j=1$. Thus
\begin{equation}
\alpha_j\alpha_j^\dagger-\beta_j\beta_j^\dagger=u_ju_j^\dagger-v_jv_j^\dagger.
\label{eq:block-diff}
\end{equation}
Choose orthonormal bases ${\{\hat u_a\}}_a$ of $\ker Z^\dagger\subset\cS_+$ and ${\{\hat v_b\}}_b$ of $\ker Z\subset\cS_-$. By~\eqref{eq:PQ} and the action of $J$ on $\cS_+$ and $\cS_-$,
\begin{align*}
    K\hat u_a&=\hat u_a,& (K-J)\hat u_a&=0,\\
    K\hat v_b&=0,& (K-J)\hat v_b&=\hat v_b.
\end{align*}
Thus the two operators decompose as
\begin{align*}
    K&=\sum_j\alpha_j\alpha_j^\dagger+\sum_a\hat u_a\hat u_a^\dagger,\\
    K-J&=\sum_j\beta_j\beta_j^\dagger+\sum_b\hat v_b\hat v_b^\dagger.
\end{align*}
Pair the $j$-th terms with each other and each kernel term with zero. Multiplying each entry on the left and right by $G$ gives the pairs
\[
\begin{aligned}
    &(G\alpha_j\alpha_j^\dagger G, G\beta_j\beta_j^\dagger G) &&\text{for each }j,\\
    &(G\hat u_a\hat u_a^\dagger G, 0) &&\text{for each }a,\\
    &(0, G\hat v_b\hat v_b^\dagger G) &&\text{for each }b.
\end{aligned}
\]
Indexing these pairs by $i$, each $A_i$ and $B_i$ is positive semidefinite and has rank at most one, and
\[
\sum_i A_i=GKG=A_0,\quad\sum_i B_i=G(K-J)G=B_0.
\]

\emph{Exactness and the additional assertion.} Since $G$ preserves $\cS_+$ and $\cS_-$, equation~\eqref{eq:block-diff} and the one-sided blocks imply that the positive and negative parts of each $\Delta_i\coloneqq A_i-B_i$ lie in these respective subspaces. Therefore
\begin{equation}
\sum_i\|\Delta_i\|_1=\sum_i\Tr(J\Delta_i)=\Tr(JH)=\|H\|_1.
\label{eq:exact-tracenorm}
\end{equation}
Add identical pairs $(C_\ell,C_\ell)$ from a spectral decomposition $C_0=\sum_\ell C_\ell$. They have zero difference, so the resulting sums equal $A,B$ and~\eqref{eq:exact-tracenorm} remains valid. Substitution into~\eqref{eq:m-def} gives
\[
\sum_i m(A_i,B_i)=\frac{\Tr A+\Tr B-\|A-B\|_1}{2}=m(A,B).
\]

Suppose now that $(A-B)_+\nleq A$. If $Z=0$, then~\eqref{eq:PQ} gives $P=I$ and $Q=0$, so $K$ is the projection onto $\cS_+$. Hence $A_0=GKG=H_+$, which gives $(A-B)_+=H_+=A_0\leq A$, a contradiction. Thus $Z\ne0$, and we can choose a positive singular value $\tau_j$. Then $q_j>0$, the vectors $G\alpha_j,G\beta_j$ are nonzero and linearly independent, and $Gu_j\perp Gv_j$. Hence the corresponding pair satisfies
\[
m(A_j,B_j)=q_j\bigl(\|Gu_j\|^2+\|Gv_j\|^2\bigr)>0.
\]
This proves the additional assertion.
Finally, there are $\dim\cS-\rank Z$ residual pairs and $\rank C_0$ identical pairs, at most $2d$ in total on a $d$-dimensional space.

\section{Proof of \cref{prop:ch-regular-pairs}: Regular Pairs}
\label{sec:ch-prop-regular-pairs}
\noindent\underline{First item}: Since the states are faithful, $\Phi(0)=\ln\Tr[\sigma]=0$ and $\Phi(1)=\ln\Tr[\rho]=0$. Moreover, $e^{\Phi(1/2)}=\Tr[\rho^{1/2}\sigma^{1/2}]\le\lVert\rho^{1/2}\rVert_2\lVert\sigma^{1/2}\rVert_2=1$ by the Cauchy--Schwarz inequality, with equality only if $\rho^{1/2}=\sigma^{1/2}$. Since $\rho\ne\sigma$, we have $\Phi(1/2)<0$. By~\eqref{eq:ch-tilt-moments}, $\Phi''(\alpha)=\operatorname{Var}_\alpha[Z]\ge0$. Suppose that $\operatorname{Var}_\alpha[Z]=0$ for some $\alpha$. Since the states are faithful, the support of $\Pr_\alpha$ is the set of pairs $(i,j)$ with $\langle u_i,v_j\rangle\ne0$, which does not depend on $\alpha$, so that $Z$ is equal to a constant $z$ on this set. Since $Q_{\NS}$ sums to one on this set, $\Phi(\alpha)=\ln\sum_{i,j}Q_{\NS}(i,j)e^{\alpha Z(i,j)}=\alpha z$ is affine, and $\Phi(0)=\Phi(1)=0$ gives $\Phi\equiv0$, which contradicts $\Phi(1/2)<0$. Hence $\Phi''>0$ on $[0,1]$, so that $\Phi$ is strictly convex, and $\Phi(1/2)<0=\Phi(0)=\Phi(1)$ implies that its minimizer is unique and lies in $(0,1)$.

\medskip
\noindent\underline{Second item}: Let $r_1>r_2$ and $s_1>s_2$ be the eigenvalues of $\rho$ and $\sigma$, and set $A\coloneqq\ln(r_1/r_2)=2\operatorname{artanh}(a)$ and $B\coloneqq\ln(s_1/s_2)=2\operatorname{artanh}(b)$. Since the Bloch vectors are not collinear, all four overlaps $|\langle u_i,v_j\rangle|^2$ are positive, and the values of $Z$ are $z_0$, $z_0+B$, $z_0-A$, and $z_0-A+B$, where $z_0\coloneqq\ln(r_1/s_1)$. These values lie in a set of the form $z_0'+d\mathbb Z$ if and only if $A$ and $B$ are integer multiples of a common $d>0$, i.e., if and only if $\theta=A/B$ is rational. Together with the first item, this proves the first claim.

For (QN), let $|u|$ be large, and write $uA=2\pi l+e_1$ and $uB=2\pi k+e_2$ with integers $k,l$ and $|e_1|,|e_2|\le\pi$. Computing $uAB$ in two ways gives $2\pi(kA-lB)=Be_1-Ae_2$. Since $1\le|k|\le|u|B/\pi$ for large $|u|$, the assumption gives
\begin{equation}
c\left(\frac{|u|B}{\pi}\right)^{-\mu}\le c\,|k|^{-\mu}\le|k\theta-l|=\frac{|Be_1-Ae_2|}{2\pi B}\le\frac{1+\theta}{2\pi}\max\{|e_1|,|e_2|\},
\end{equation}
so that $\max\{|e_1|,|e_2|\}\ge c'|u|^{-\mu}$ for some $c'>0$. Let $z_m$ be the values of $Z$ and $w_m(\alpha)>0$ their tilted weights, so that $Q_{\alpha+iu}^{\NS}/Q_{\alpha}^{\NS}=\sum_mw_m(\alpha)e^{iuz_m}$. Then
\begin{align}
1-\frac{|Q_{\alpha+iu}^{\NS}|}{Q_{\alpha}^{\NS}}
&\ge\frac12\left(1-\frac{|Q_{\alpha+iu}^{\NS}|^2}{(Q_{\alpha}^{\NS})^2}\right)
=\frac12\sum_{m,m'}w_m(\alpha)w_{m'}(\alpha)\left[1-\cos(u(z_m-z_{m'}))\right]\\
&\ge w\left[(1-\cos e_1)+(1-\cos e_2)\right]
\ge\frac{2w}{\pi^2}\max\{e_1^2,e_2^2\}
\ge c''|u|^{-2\mu},
\end{align}
where the first inequality follows from $1-x\ge(1-x^2)/2$ for $x\in[0,1]$, the equality from $\sum_mw_m(\alpha)=1$, the second inequality from keeping only the pairs of values that differ by $A$ and by $B$, with $w>0$ the smallest product of two tilted weights and $\cos(uA)=\cos e_1$, $\cos(uB)=\cos e_2$, and the third inequality from $1-\cos e\ge2e^2/\pi^2$ for $|e|\le\pi$. This lower bound is at least $e^{-\eta|u|}$ for every $\eta>0$ and all sufficiently large $|u|$, which proves (QN). By Khinchin's theorem~\cite{khintchine1924}, for almost every $\theta$, the assumption holds with every $\mu>1$. If $a$ and $b$ are algebraic, then $e^A=(1+a)/(1-a)$ and $e^B=(1+b)/(1-b)$ are algebraic, and Baker's theorem gives $|kA-lB|\ge c\max\{|k|,|l|\}^{-\mu}$ for some $c,\mu>0$ and all $(k,l)\ne(0,0)$, since $kA-lB\ne0$ for irrational $\theta$.

For these qubit pairs, (QN) is equivalent to $\frac{\ln q_{j+1}}{q_j}\to0$, where $q_j$ are the continued-fraction denominators of $\theta$. Indeed, the phase estimates above and the evaluation at $u=2\pi k/B$ show that (QN) is equivalent to $\operatorname{dist}(k\theta,\mathbb Z)\ge e^{-\eta k}$ for every $\eta>0$ and all sufficiently large $k$, and the standard best-approximation bounds for continued fractions give the stated condition, which is condition (R) of~\cite[p.~343]{bounemoura2019optimality}.

\medskip
\noindent\underline{Third item}: The states $\rho_\eps$ and $\sigma_{\eps'}$ are faithful, and $[\rho_\eps,\sigma_{\eps'}]=(1-\eps)(1-\eps')[\rho,\sigma]\ne0$; in particular, $\rho_\eps\ne\sigma_{\eps'}$. Let $u_i$ be an orthonormal eigenbasis of $\rho$ with eigenvalues $r_i\ge0$, so that $\rho_\eps$ has the eigenvalues $r_i(\eps)\coloneqq(1-\eps)r_i+\eps/d$, and similarly for $\sigma_{\eps'}$. If every eigenvector $v_j$ of $\sigma$ were contained in an eigenspace of $\rho$, then $\rho$ and $\sigma$ would commute. Hence there are $j$ and $i,i'$ with $r_i\ne r_{i'}$ and $\langle u_i,v_j\rangle\ne0\ne\langle u_{i'},v_j\rangle$, and, similarly, there are $i''$ and $j',j''$ with $s_{j'}\ne s_{j''}$ and $\langle u_{i''},v_{j'}\rangle\ne0\ne\langle u_{i''},v_{j''}\rangle$. Then the values of $Z$ for $(\rho_\eps,\sigma_{\eps'})$ include two pairs with the differences
\[
g(\eps)\coloneqq\ln\frac{r_i(\eps)}{r_{i'}(\eps)}=\ln\frac{r_i+x}{r_{i'}+x},\quad x\coloneqq\frac{\eps}{d(1-\eps)},
\qquad\text{and}\qquad
h\coloneqq\ln\frac{s_{j'}(\eps')}{s_{j''}(\eps')}\ne0,
\]
where $g$ is smooth and strictly monotone in $\eps\in(0,1)$. If the pair is lattice, then $g(\eps)/h$ is rational. Since $g$ is strictly monotone, $g(\eps)/h$ takes each rational value for at most one $\eps$, so that the pair is lattice for at most countably many $\eps$; for all other $\eps$, the pair is regular by the first item. For (QN), the argument of the second item applies to the differences $g(\eps)$ and $h$, so that (QN) holds whenever $\theta(\eps)\coloneqq g(\eps)/h$ satisfies the Diophantine condition of the second item. This condition fails for every $\mu$ only on a set of $\theta$ of Lebesgue measure zero~\cite{khintchine1924}, and since $\theta(\eps)$ is smooth with nonvanishing derivative, its preimage also has measure zero.

\section{Auxiliary Results for the Chernoff Analysis}
\label{sec:ch-auxiliary}

In this appendix, we collect the auxiliary lemmas used in \cref{sec:chernoff}, together with their proofs. First, we show that the two quasi-entropies in~\eqref{eq:ch-Qalpha}--\eqref{eq:ch-Qalpha2} are super-additive on paired sums. This is used in the proof of \cref{thm:ch-three-bounds}.

\begin{lemma}[Super-Additivity of Quasi-Entropies]
\label{lem:ch-trace-concavity}
For finite families $A_j,B_j\ge0$ and $\alpha\in(0,1)$, the quantities in~\eqref{eq:ch-Qalpha}--\eqref{eq:ch-Qalpha2} satisfy
\begin{equation}
\sum_jQ_\alpha(A_j,B_j)\le Q_\alpha\!\left(\sum_jA_j,\sum_jB_j\right),
\qquad
\sum_jQ_{\alpha,2}(A_j,B_j)\le Q_{\alpha,2}\!\left(\sum_jA_j,\sum_jB_j\right).
\label{eq:ch-flag-concavity}
\end{equation}
\end{lemma}

\begin{proof}
Let $F$ be either $Q_\alpha$ or $Q_{\alpha,2}$. By~\cite[Theorem~1.1(1)]{hiai2013concavity} with identity maps and $(p,q,s)=(\alpha,1-\alpha,1)$ or $(\alpha/2,(1-\alpha)/2,2)$, respectively, $F$ is jointly concave on pairs of positive definite operators, and, by continuity, on pairs of positive semidefinite operators. Since $F$ is also homogeneous of degree one, for a family of $N$ pairs,
\begin{equation}
F\!\left(\sum_jA_j,\sum_jB_j\right)=N\,F\!\left(\frac1N\sum_jA_j,\frac1N\sum_jB_j\right)\ge\sum_jF(A_j,B_j),
\end{equation}
where the equality follows from homogeneity, and the inequality from joint concavity.
\end{proof}

\begin{proof}[Proof of \cref{lem:ch-harmonic}]
For $s\in(0,1)$, let $\mathrm{HM}_s(A,B)\coloneqq\sum_{i,j}\frac{a_ib_j}{sa_i+(1-s)b_j}|\langle u_i,v_j\rangle|^2$ denote the $s$-weighted Petz--Nussbaum--Szko{\l}a harmonic mean of~\cite[Definition~4.1]{cheng2026multiple}. By~\cite[Proposition~C.1(iii), Eq.~(202)]{cheng2026multiple} with $s=1/2$, our variational definition~\eqref{eq:ch-Har-def} satisfies $\Har(A,B)=\tfrac12\mathrm{HM}_{1/2}(A,B)$, which gives~\eqref{eq:ch-Har-spectral}, including singular inputs. The bound $\Har(A,B)\le m(A,B)$ is the case $s=1/2$ of~\cite[Theorem~4.2, Eq.~(65)]{cheng2026multiple}. Finally,
for super-additivity, the objective in~\eqref{eq:ch-Har-def}, $F_X(A,B)\coloneqq\Tr[(\id-X)^\dagger A(\id-X)]+\Tr[XBX^\dagger]$, is linear in $(A,B)$ for fixed $X$, so that
\begin{equation}
\Har\!\left(\sum_jA_j,\sum_jB_j\right)=\min_X\sum_jF_X(A_j,B_j)\ge\sum_j\min_XF_X(A_j,B_j)=\sum_j\Har(A_j,B_j).
\end{equation}
\end{proof}

We need the following uniform local limit theorem of Stone~\cite{stone1965local}, with the Gaussian density $\phi_v$ as in \cref{lem:ch-LLT-harmonic}.

\begin{lemma}[Local Limit Theorem~\cite{stone1965local}]
\label{lem:ch-LLT}
Let $S_n$ be a sum of $n$ independent copies of a nonlattice random variable $Z$ that takes finitely many values, with mean $\mu$ and variance $v>0$. Let $h\ge0$ be continuous and satisfy $h(z)\le Ce^{-c|z|}$ for some $C,c>0$. Then
\begin{equation}
\lim_{n\to\infty}\sup_{x\in\mathbb R}\sqrt n\left|\mathbb E\!\left[h(S_n-x)\right]-\phi_{nv}(x-n\mu)\int_{\mathbb R}h(z)\,dz\right|=0.
\label{eq:ch-LLT}
\end{equation}
\end{lemma}

\begin{proof}
Stone's interval theorem~\cite{stone1965local} gives~\eqref{eq:ch-LLT} for indicator functions of bounded intervals; hence, by linearity, for step functions with finitely many steps. It also gives $C_0>0$ such that $\sup_z\Pr(S_n\in[z,z+1])\le C_0/\sqrt n$ for all sufficiently large $n$. Fix $R\ge2$ and split $h=h\mathbf1_{[-R,R]}+h\mathbf1_{\{|z|>R\}}$.

\underline{Tails}: Covering $\{z:|z|>R\}$ by the unit intervals $[k,k+1]$ with $|k|\ge R-1$ and using $h(z)\le Ce^ce^{-c|k|}$ on each of them, we have that
\begin{equation}
\sup_x\sqrt n\,\mathbb E\!\left[h(S_n-x)\mathbf1_{\{|S_n-x|>R\}}\right]\le C\,C_0e^c\sum_{|k|\ge R-1}e^{-c|k|}.
\end{equation}
The corresponding Gaussian term is at most $(2\pi v)^{-1/2}\int_{|z|>R}h(z)\,dz$, since $\sqrt n\,\phi_{nv}\le(2\pi v)^{-1/2}$. Both bounds tend to zero as $R\to\infty$, uniformly in $n$ and $x$.

\underline{Bounded part}: Since $h$ is uniformly continuous on $[-R,R]$, for every $\epsilon>0$, there are step functions $h_-\le h\mathbf1_{[-R,R]}\le h_+$ with $\int(h_+-h_-)\le\epsilon$. Applying~\eqref{eq:ch-LLT} to $h_\pm$ and using monotonicity of the expectation, the $\limsup$ in~\eqref{eq:ch-LLT} for $h\mathbf1_{[-R,R]}$ is at most $(2\pi v)^{-1/2}\epsilon$. Letting $\epsilon\to0$ and then $R\to\infty$ proves the claim.
\end{proof}

\begin{proof}[Proof of \cref{prop:ch-local-expansion}]
By~\eqref{eq:ch-MB}, the left-hand side of~\eqref{eq:ch-local-expansion} is the logarithm of
\begin{equation}
\frac1{2\pi}\int_{\mathbb R}e^{\Lambda(\alpha+iu)}\left(\frac{Q_{\alpha+iu}^{\NS}}{Q_\alpha^{\NS}}\right)^ne^{-iux}\,du.
\label{eq:ch-MB-normalized}
\end{equation}
Fix $u_0\in(0,1)$ so small that $\beta\mapsto\ln Q_\beta^{\NS}$ is analytic on a neighborhood of the disk $\{|\beta-\alpha|\le u_0\}$ and $\operatorname{Re}\ln(Q_{\alpha+iu}^{\NS}/Q_\alpha^{\NS})\le-\Phi''(\alpha)u^2/4$ for $|u|\le u_0$. We show that the frequencies $|u|\ge u_0$ contribute $O(n^{-L})$ for every $L>0$, and that the frequencies $|u|\le u_0$ give the expansion.

\underline{Large frequencies}: Fix $L>0$, choose $K>L/\pi$, and set $\eta\coloneqq1/(2K)$. Since (QN) does not depend on the choice of $\alpha$ (see the discussion after \cref{def:ch-QN}), \eqref{eq:ch-QN} gives $U>u_0$ such that $1-|Q_{\alpha+iu}^{\NS}|/Q_\alpha^{\NS}\ge e^{-\eta|u|}$ for $|u|\ge U$. We bound the integrand of~\eqref{eq:ch-MB-normalized} in three ranges, using $|e^{\Lambda(\alpha+iu)}|\le\pi/\sinh(\pi|u|)$, which is integrable on $\{|u|\ge u_0\}$:
\begin{itemize}
\item For $u_0\le|u|\le U$, nonlatticeness and continuity give $\sup_{u_0\le|u|\le U}|Q_{\alpha+iu}^{\NS}|/Q_\alpha^{\NS}<1$, so that this range contributes an exponentially small term.
\item For $U\le|u|\le K\ln n$, which is a nonempty range for large $n$, we have $(|Q_{\alpha+iu}^{\NS}|/Q_\alpha^{\NS})^n\le\exp(-ne^{-\eta|u|})\le\exp(-ne^{-\eta K\ln n})=e^{-\sqrt n}$, so that this range contributes $O(e^{-\sqrt n})$.
\item For $|u|\ge K\ln n$, we use $|Q_{\alpha+iu}^{\NS}|\le Q_\alpha^{\NS}$ and $\pi/\sinh(\pi|u|)\le4\pi e^{-\pi|u|}$, whose integral over this range is $O(n^{-\pi K})=O(n^{-L})$.
\end{itemize}
Hence the frequencies $|u|\ge u_0$ contribute $O(n^{-L})$ to~\eqref{eq:ch-MB-normalized}, uniformly in $x\in\mathbb R$.

\underline{Small frequencies}: Set $x=n\Phi'(\alpha)+y$ and $v\coloneqq\Phi''(\alpha)$. For $|u|\le u_0$, the integrand of~\eqref{eq:ch-MB-normalized} is equal to $a_y(u)e^{-n\varphi(u)}$, where
\[
\varphi(u)\coloneqq-\ln\!\left(\frac{Q_{\alpha+iu}^{\NS}}{Q_\alpha^{\NS}}\right)+iu\Phi'(\alpha),\qquad
a_y(u)\coloneqq e^{\Lambda(\alpha+iu)-iuy}.
\]
These functions are analytic on a complex neighborhood of $[-u_0,u_0]$, and, since $\frac{d}{du}\ln Q_{\alpha+iu}^{\NS}=i\Phi'(\alpha+iu)$, they satisfy $\varphi(0)=\varphi'(0)=0$, $\varphi''(0)=v>0$, and $\operatorname{Re}\varphi(u)\ge vu^2/4$ for $|u|\le u_0$. The analytic simple-saddle expansion~\cite[\S2.4(iv), Eq.~(2.4.15)]{olver2010asymptotic} therefore gives
\[
\frac1{2\pi}\int_{-u_0}^{u_0}a_y(u)e^{-n\varphi(u)}\,du=\frac{e^{\Lambda(\alpha)}}{\sqrt{2\pi nv}}\left[1+\sum_{j=1}^M\frac{d_j(y)}{n^j}+O(n^{-M-1})\right]
\]
for some coefficients $d_j(y)$, uniformly for $|y|\le Y$, since the phase $\varphi$ does not depend on $y$ and the amplitudes $a_y$ and their derivatives are uniformly bounded on a common complex neighborhood for bounded $y$. The coefficients $d_j(y)$ are obtained by expanding $a_y(u)$ and $n[\varphi(u)-vu^2/2]$ around $u=0$, substituting $u=w/\sqrt n$, and integrating term by term against the Gaussian weight $e^{-vw^2/2}$; in particular, each $d_j(y)$ is a polynomial in $y$.

Choose $L=M+2$, add the large-frequency contribution, which is $O(n^{-M-3/2})$ relative to the leading term, and then take the logarithm. Expanding $\ln(1+\sum_jd_j(y)n^{-j})$ gives~\eqref{eq:ch-local-expansion} with the coefficients $c_j(y)$ of the rule~\eqref{eq:app-coefficient} in Appendix~\ref{sec:ch-coefficients}.
\end{proof}

\section{Computation of the Coefficients}
\label{sec:ch-coefficients}

In this appendix, we derive the coefficients in \cref{thm:ch-main,prop:ch-local-expansion}. Fix $\alpha\in(0,1)$ with $\Phi''(\alpha)>0$, and write $\Phi^{(k)}\coloneqq\Phi^{(k)}(\alpha)$, $\Lambda^{(k)}\coloneqq\Lambda^{(k)}(\alpha)$, and $s\coloneqq1/\Phi''$, with $\Phi$ in~\eqref{eq:ch-Phi}, $\Lambda$ in~\eqref{eq:ch-kernel}, and $\tau(\alpha)$ in~\eqref{eq:ch-tau}.

\underline{General rule}: In the small-frequency integral in the proof of \cref{prop:ch-local-expansion}, the Taylor expansions of $\Lambda(\alpha+iu)$ and $\ln Q_{\alpha+iu}^{\NS}=\Phi(\alpha+iu)$ around $u=0$ give
\begin{equation}
\Lambda(\alpha+iu)-iuy-n\varphi(u)=\Lambda(\alpha)-\frac{n\Phi''u^2}2+\sum_{k\ge1}\frac{\Lambda^{(k)}}{k!}(iu)^k+n\sum_{k\ge3}\frac{\Phi^{(k)}}{k!}(iu)^k-iuy.
\end{equation}
After the substitution $u=w/\sqrt n$, the quadratic term becomes the Gaussian weight $e^{-w^2/(2s)}$, and the remaining terms become $\sum_{k\ge1}n^{-k/2}R^{(k)}$ with
\begin{equation}
R^{(k)}\coloneqq\frac{\Lambda^{(k)}}{k!}(iW)^k+\frac{\Phi^{(k+2)}}{(k+2)!}(iW)^{k+2}-\mathbf1_{\{k=1\}}iyW
\label{eq:app-perturbations}
\end{equation}
evaluated at $W=w$. Hence, with $W$ Gaussian with mean zero and variance $s$, the small-frequency integral is equal to $e^{\Lambda(\alpha)}(2\pi n\Phi'')^{-1/2}\,\mathbb E[\exp(\sum_{k\ge1}n^{-k/2}R^{(k)})]$, as a formal power series in $n^{-1/2}$. Writing $t$ for $n^{-1/2}$ and comparing with~\eqref{eq:ch-local-expansion} and the proof of \cref{thm:ch-main}, for every integer $j\ge1$,
\begin{equation}
c_j(y)=[t^{2j}]\ln\mathbb E\!\left[\exp\!\left(\sum_{k=1}^{2j}t^kR^{(k)}\right)\right],
\qquad
C_j(p)=-c_j\!\left(\ln\frac{1-p}{p}\right)\bigg|_{\alpha=\alpha_*}.
\label{eq:app-coefficient}
\end{equation}
The expectation and the logarithm are evaluated as formal power series, coefficient by coefficient, so that no exponential-integrability assumption on the polynomial perturbations is needed. The coefficients of odd powers of $t$ vanish, since they involve only odd moments of $W$. For $c_j$, only derivatives of $\Phi$ through order $2j+2$, derivatives of $\Lambda$ through order $2j$, and Gaussian moments through order $6j$ are used, so that the rule is a finite computation at each order.

\underline{First coefficient}: The first four perturbation terms are
\begin{align*}
R^{(1)}&=i\left[(\Lambda'-y)W-\frac{\Phi'''}{6}W^3\right],&
R^{(2)}&=-\frac{\Lambda''}{2}W^2+\frac{\Phi^{(4)}}{24}W^4,\\
R^{(3)}&=-i\left[\frac{\Lambda'''}{6}W^3-\frac{\Phi^{(5)}}{120}W^5\right],&
R^{(4)}&=\frac{\Lambda^{(4)}}{24}W^4-\frac{\Phi^{(6)}}{720}W^6.
\end{align*}

Using the Gaussian moments $\mathbb E[W^{2r}]=(2r-1)!!\,s^r$, where $(2r-1)!!\coloneqq1\cdot3\cdot5\cdots(2r-1)$, and $\mathbb E[W^{2r+1}]=0$, the coefficient of $t^2$ is
\begin{align*}
c_1(y)&=\mathbb E\!\left[R^{(2)}\right]+\frac12\mathbb E\!\left[(R^{(1)})^2\right]\\
&=-\frac{\Lambda''}{2\Phi''}+\frac{\Phi^{(4)}}{8(\Phi'')^2}-\frac{(\Lambda'-y)^2}{2\Phi''}+\frac{(\Lambda'-y)\Phi'''}{2(\Phi'')^2}-\frac{5(\Phi''')^2}{24(\Phi'')^3}.
\end{align*}
Completing the square with $\tau(\alpha)=\Lambda'-\Phi'''/(2\Phi'')$ and using $\Lambda''=\pi^2/\sin^2(\pi\alpha)$, we obtain
\begin{equation}
c_1(y)=b_1(\alpha)-\frac{(y-\tau(\alpha))^2}{2\Phi''},
\label{eq:ch-c1}
\end{equation}
where $b_j(\alpha)\coloneqq c_j(\tau(\alpha))$ denotes the centered coefficients, and
\begin{equation}
b_1(\alpha)=\frac{\Phi^{(4)}}{8(\Phi'')^2}-\frac{(\Phi''')^2}{12(\Phi'')^3}-\frac{\pi^2}{2\Phi''\sin^2(\pi\alpha)}.
\label{eq:ch-b1}
\end{equation}
At $\alpha=\alpha_*$, $C_1(p)=-c_1(\ln\frac{1-p}p)$ with $\ln\frac{1-p}p-\tau(\alpha_*)=\delta_p$ gives~\eqref{eq:ch-C1}.

\underline{Second coefficient}: The coefficient of $t^4$ is obtained in the same way from the cumulant expansion
\[
c_2(y)=\mathbb E\!\left[R^{(4)}\right]+\mathbb E\!\left[R^{(1)}R^{(3)}\right]+\frac12\mathbb E\!\left[(R^{(2)})^2\right]+\frac12\mathbb E\!\left[(R^{(1)})^2R^{(2)}\right]+\frac1{24}\mathbb E\!\left[(R^{(1)})^4\right]-\frac12\left(c_1(y)\right)^2,
\]
where we used $\mathbb E[R^{(1)}]=\mathbb E[R^{(3)}]=0$ and $\mathbb E[R^{(2)}]+\frac12\mathbb E[(R^{(1)})^2]=c_1(y)$. Evaluating the Gaussian moments up to $\mathbb E[W^{12}]=10395s^6$ and writing the result in terms of $y-\tau(\alpha)$ gives
\begin{equation}
c_2(y)=b_2(\alpha)+\frac{b_1'(\alpha)}{\Phi''}(y-\tau(\alpha))+\frac{\tau'(\alpha)}{2(\Phi'')^2}(y-\tau(\alpha))^2+\frac{\Phi'''}{6(\Phi'')^3}(y-\tau(\alpha))^3,
\label{eq:ch-c2}
\end{equation}
with
\begin{align}
b_2(\alpha)&=\frac{\Lambda^{(4)}+2(\Lambda'')^2}{8(\Phi'')^2}-\frac{\Phi^{(6)}+8\Phi'''\Lambda'''+12\Phi^{(4)}\Lambda''}{48(\Phi'')^3}\notag\\
&\qquad+\frac{(\Phi^{(4)})^2+\Phi'''\Phi^{(5)}+3(\Phi''')^2\Lambda''}{12(\Phi'')^4}-\frac{(\Phi''')^2\Phi^{(4)}}{4(\Phi'')^5}+\frac{5(\Phi''')^4}{48(\Phi'')^6}.
\label{eq:ch-b2}
\end{align}
We verified this computation with computer algebra. At $\alpha=\alpha_*$, with $V_*=\Phi''(\alpha_*)$, the coefficient $C_2(p)=-c_2(\ln\frac{1-p}p)$ is
\begin{equation}
C_2(p)=-b_2(\alpha_*)-\frac{b_1'(\alpha_*)}{V_*}\delta_p-\frac{\tau'(\alpha_*)}{2V_*^2}\delta_p^2-\frac{\Phi'''(\alpha_*)}{6V_*^3}\delta_p^3.
\label{eq:ch-C2}
\end{equation}
These coefficients are also supported, independently of the harmonic overlap, by the numerical evaluation of the optimal error in \cref{ex:ch-numerics}.

\underline{Third coefficient}: The dependence of $c_{j+1}$ on $y$ is determined by $c_j$ through the identity
\begin{equation}
\partial_yc_{j+1}(y)=\frac1{\Phi''(\alpha)}\,\partial_\alpha c_j(y),\qquad j\ge1,
\label{eq:ch-coeff-identity}
\end{equation}
where $\partial_\alpha$ is taken at fixed $y$. To prove~\eqref{eq:ch-coeff-identity}, substitute $x=n\Phi'(\alpha)+y$ and $u=tw$ with $t=n^{-1/2}$ in the integrand of~\eqref{eq:ch-MB-normalized}, which gives
\[
F(w)\coloneqq\exp\!\left(\Lambda(\alpha+itw)-itwy+t^{-2}\left[\Phi(\alpha+itw)-\Phi(\alpha)-itw\Phi'(\alpha)\right]\right).
\]
Direct differentiation gives
\[
\left(\partial_\alpha-t^{-2}\Phi''(\alpha)\,\partial_y\right)F=yF+\frac1{it}\,\partial_wF.
\]
As a formal power series in $t$, every coefficient of $F$ is a polynomial in $w$ times $e^{-\Phi''(\alpha)w^2/2}$, so that $\int_{\mathbb R}\partial_wF\,dw=0$ coefficient by coefficient. Hence the formal integral $J\coloneqq\frac t{2\pi}\int_{\mathbb R}F\,dw$ satisfies $(\partial_\alpha-t^{-2}\Phi''(\alpha)\partial_y)\ln J=y$. By the general rule, $\ln J=\ln t+\Lambda(\alpha)-\frac12\ln(2\pi\Phi''(\alpha))+\sum_{j\ge1}c_j(y)t^{2j}$, and comparing the coefficients of $t^{2j}$ gives~\eqref{eq:ch-coeff-identity} for $j\ge1$; the coefficient of $t^0$ gives $\partial_yc_1(y)=(\tau(\alpha)-y)/\Phi''(\alpha)$, in agreement with~\eqref{eq:ch-c1}.

To give the next coefficient explicitly, abbreviate $v\coloneqq\Phi''(\alpha)$, $f_k\coloneqq\Phi^{(k)}(\alpha)$, and $\ell_k\coloneqq\Lambda^{(k)}(\alpha)$. Evaluating~\eqref{eq:app-coefficient} for $j=3$ at $y=\tau(\alpha)$ gives
\begin{align}
b_3(\alpha)
&=-\frac{\ell_6+10\ell_3^2+12\ell_2\ell_4+8\ell_2^3}{48v^3}\notag\\
&\quad+\frac{1}{384v^4}\Bigl(f_8+24f_6\ell_2+56f_5\ell_3+64f_4\ell_4
 +144f_4\ell_2^2\notag\\
&\hspace{39mm}+32f_3\ell_5+288f_3\ell_2\ell_3\Bigr)\notag\\
&\quad-\frac{1}{1920v^5}\Bigl(63f_5^2+100f_4f_6+640f_4^2\ell_2+40f_3f_7
 +760f_3f_5\ell_2\notag\\
&\hspace{39mm}+1360f_3f_4\ell_3+480f_3^2\ell_4+1200f_3^2\ell_2^2\Bigr)\notag\\
&\quad+\frac{33f_4^3+123f_3f_4f_5+31f_3^2f_6+420f_3^2f_4\ell_2+156f_3^3\ell_3}{288v^6}\notag\\
&\quad-\frac{271f_3^2f_4^2+117f_3^3f_5+216f_3^4\ell_2}{288v^7}
 +\frac{109f_3^4f_4}{96v^8}-\frac{97f_3^6}{288v^9}.
\label{eq:ch-b3}
\end{align}
Integrating~\eqref{eq:ch-coeff-identity} for $j=2$ from $y=\tau(\alpha)$, with $c_3(\tau(\alpha))=b_3(\alpha)$, and evaluating at $\alpha=\alpha_*$, so that $v=V_*$, and at $y=\ln\frac{1-p}p$, so that $y-\tau(\alpha_*)=\delta_p$, gives
\begin{align}
C_3(p)
&=-b_3-\left(\frac{b_2'}v-\frac{\tau'b_1'}{v^2}\right)\delta_p
 -\left(\frac{b_1''}{2v^2}-\frac{b_1'f_3+(\tau')^2}{2v^3}\right)\delta_p^2\notag\\
&\quad-\left(\frac{\tau''}{6v^3}-\frac{\tau'f_3}{2v^4}\right)\delta_p^3
 -\left(\frac{f_4}{24v^4}-\frac{f_3^2}{8v^5}\right)\delta_p^4,
\label{eq:ch-C3}
\end{align}
where primes on $b_1,b_2$, and $\tau$ denote derivatives with respect to $\alpha$. Direct expansion of~\eqref{eq:app-coefficient}, using Gaussian moments through $\mathbb E[W^{18}]=34459425s^9$, verifies both~\eqref{eq:ch-b3} and~\eqref{eq:ch-C3} by exact rational arithmetic.

\end{document}